\documentclass[final,5p,times,twocolumn,authoryear]{elsarticle}
\newif\ifreportversion
\reportversiontrue

\usepackage{microtype}
\usepackage{setspace}
\usepackage[skip=5pt plus 1pt,indent=0pt]{parskip}

\usepackage{mathtools}
\usepackage{amssymb}
\usepackage{graphicx}
\graphicspath{{figures/}}
\usepackage{booktabs}
\usepackage{array}
\usepackage{multirow}
\newcommand{\tablehead}[1]{\begin{tabular}[c]{@{}c@{}}#1\end{tabular}}
\newcommand{\bottomtablehead}[2][2]{\begin{tabular}[c]{@{}l@{}}\vphantom{X}\\\ifnum#1=3\vphantom{X}\\\fi #2\end{tabular}}
\newcommand{\systemhead}[1][2]{\bottomtablehead[#1]{System}}
\usepackage{tabularx}
\usepackage{longtable}
\usepackage{enumitem}
\setlist[itemize]{leftmargin=*,topsep=2pt,itemsep=1pt}
\setlist[enumerate]{leftmargin=*,topsep=2pt,itemsep=1pt}

\usepackage[font=small,labelfont=bf,skip=4pt]{caption}
\usepackage{subcaption}
\usepackage{placeins}
\usepackage[table]{xcolor}
\usepackage[normalem]{ulem}
\definecolor{revisionadd}{HTML}{DCFCE7}
\definecolor{revisiondel}{HTML}{FEE2E2}
\newif\ifshowrevisions
\showrevisionsfalse

\DeclareRobustCommand{\deleted}[1]{\ifshowrevisions\deletedmark{#1}\hspace{0.15em}\fi}
\DeclareRobustCommand{\addedmark}{%
  \bgroup
  \markoverwith{\textcolor{revisionadd}{\rule[-0.5ex]{2pt}{2.5ex}}}%
  \ULon}
\DeclareRobustCommand{\deletedmark}{%
  \bgroup
  \markoverwith{\textcolor{revisiondel}{\rule[-0.5ex]{2pt}{2.5ex}}}%
  \ULon}
\newcommand{\candidateseparator}[1]{%
  \noalign{\vskip2pt}%
  \multispan{#1}\leaders\hbox{\rule{1.5pt}{0.3pt}\hspace{2pt}}\hfill\cr
  \noalign{\vskip2pt}}
\definecolor{linkblue}{HTML}{1F4E79}
\definecolor{draftred}{HTML}{B91C1C}
\definecolor{draftbg}{HTML}{FEF2F2}
\definecolor{edittext}{HTML}{234E8A}
\definecolor{editemotion}{HTML}{6F3D78}
\definecolor{editpause}{HTML}{186C55}
\usepackage[colorlinks=true,
  linkcolor=linkblue,
  citecolor=linkblue,
  urlcolor=linkblue,
  breaklinks=true]{hyperref}
\makeatletter
\let\originalNAT@hyper@\NAT@hyper@
\renewcommand{\NAT@hyper@}[1]{\mbox{\originalNAT@hyper@{#1}}}
\makeatother
\newcommand{\cref}[1]{\autoref{#1}}
\newcommand{\Cref}[1]{\autoref{#1}}
\newcommand{\suppref}[1]{\ifreportversion\ref{#1}\else Supplementary Material, Section~\ref{#1}\fi}
\newcounter{pipelinealgorithm}

\newenvironment{pipelinealgorithm}[2]{%
  \par\smallskip\noindent\begin{minipage}{\linewidth}%
  \refstepcounter{pipelinealgorithm}\label{#2}%
  \hrule\smallskip
  \textbf{Algorithm \thepipelinealgorithm\quad #1}\par\smallskip\hrule\smallskip
  \linespread{1}\small
}{%
  \smallskip\hrule\end{minipage}\par\smallskip
}

\biboptions{round,authoryear}
\newcommand{\dotstts}{dots.tts}
\newcommand{\dotsttsedit}{dots.tts.edit}
\newcommand{\dotsttsedittitle}{dots.tts.edit}
\newcommand{\srcproj}[1]{g_{\mathrm{src}}\!\left(#1\right)}
\newcommand{\tgtproj}[1]{g_{\mathrm{tgt}}\!\left(#1\right)}
\newcommand{\audiosrc}{A_{\mathrm{src}}}
\newcommand{\audiotgt}{A_{\mathrm{tgt}}}

\newcommand{\textsrc}{T_{\mathrm{src}}}
\newcommand{\texttgt}{T_{\mathrm{tgt}}}

\newcommand{\latentsrc}{Z_{\mathrm{src}}}
\newcommand{\latenttgt}{Z_{\mathrm{tgt}}}

\usepackage{etoolbox}
="[fonts/resource-icons-solid.otf]" at 9pt
="[fonts/resource-icons-brands.otf]" at 9pt
\newcommand{\resourceicon}[2]{%
  {\csname resource#1font\endcsname\symbol{"#2}}\kern0.35em}
\newcommand{\resourcelink}[4]{%
  \href{#3}{\resourceicon{#1}{#2}#4}}
\newcommand{\paperresources}{%
  \par\medskip
  \begingroup\hypersetup{urlcolor=linkblue}%
  \noindent\makebox[\textwidth][c]{\normalfont
    \ifreportversion\fontsize{11}{13}\selectfont\else\normalsize\fi
    \resourcelink{solid}{F11B}{https://dots-studio-dots-tts-edit.hf.space}{Playground}\quad
    \resourcelink{solid}{F144}{https://huggingface.co/spaces/dots-studio/dots.tts.edit-demo}{Demo}\quad
    \resourcelink{solid}{F1B3}{https://huggingface.co/dots-studio/dots.tts.edit}{Models}\quad
    \resourcelink{brand}{F09B}{https://github.com/studio-dots-ai/dots.tts}{Code}\quad
    \resourcelink{solid}{F080}{https://github.com/studio-dots-ai/doteBench}{Benchmark}\quad
    \resourcelink{solid}{F542}{https://github.com/studio-dots-ai/dots.tts.edit-data-pipeline}{Data Pipeline}}%
  \par\endgroup}
\newcommand{\authorlinebreak}{%
  \gappto\elsauthors{\def\authorsep{\unskip\\}}}
\ifreportversion\else
  \patchcmd{\MaketitleBox}
    {\elsaddress\par\vskip36pt}
    {\elsaddress\par\vskip6pt\paperresources\vskip18pt}
    {}{\PackageError{resource-links}{Cannot insert journal resource links}{Check the elsarticle title layout.}}
\fi

\showrevisionsfalse
\input{title-page}

\begin{document}
\begin{frontmatter}
\title{\dotsttsedittitle{}: Precisely Controlled Speech Editing\texorpdfstring{\\}{ }
with a Continuous Autoregressive Model}
\author[xhs,sjtu]{Hankun Wang\fnref{intern}}
\ead{wanghankun@xiaohongshu.com}
\ead{wanghankun@sjtu.edu.cn}
\author[xhs,sjtu]{Bohan Li\fnref{intern}}
\author[xhs]{Shi Lian}
\author[xhs,sjtu]{Xiaoyu Gu\fnref{intern}}
\author[xhs,sjtu]{Jing Peng\fnref{intern}}
\author[xhs]{Da Zheng}
\authorlinebreak
\author[sjtu]{Yiwei Guo}
\author[xhs]{Colin Zhang}
\author[nju]{Shuai Wang}
\author[sjtu]{Kai Yu\corref{corresponding}}
\ead{kai.yu@sjtu.edu.cn}
\fntext[intern]{Work done during an internship at Xiaohongshu Inc.}
\cortext[corresponding]{Corresponding author.}
\affiliation[xhs]{organization={Xiaohongshu Inc.}}
\affiliation[sjtu]{organization={X-LANCE Lab, School of Computer Science, Shanghai Jiao Tong University}}
\affiliation[nju]{organization={School of Intelligence Science and Technology, Nanjing University}}

\begin{abstract}
Speech editing for content creation requires precise control over both what an
edit should do and where it should apply.  Free-form natural language provides
a flexible interface for expressing edit requests, but its ambiguity may leave
the intended operation, parameters, or target region underspecified.  We study
a semi-formal transcript-grounded edit instruction that specifies operation categories and parameters
and binds them to transcript spans or boundaries.  This
semantic timeline avoids explicit timestamp alignment and provides an
externally inspectable contract for compositional edits.  We
instantiate the interface in \dotsttsedit{}, a speech editing model built on a
continuous autoregressive speech generation architecture.  Four representative
speech-creation controls cover lexical content, affective expression,
pitch and speaking-rate delivery, and temporal phrasing through text,
emotion, prosody, and pause editing.  Task-specific data pipelines construct
operation- and scope-controlled pairs while retaining source-derived context
outside each target region.  We further introduce \emph{doteBench}, a
bilingual benchmark that measures precise instruction
following, local preservation, and audio quality across the four controls and
their composition.  Experiments show leading overall instruction following
and local preservation across its five editing categories, while audio quality
remains comparable to existing open-source systems.  Across three
Seed-TTS-Eval shards, the model shows negligible differences from the base
model in zero-shot TTS recognition error rate and speaker similarity.
\end{abstract}

\end{frontmatter}

\section{Introduction}
\label{sec:introduction}

Speech generation for content creation requires control over both what is
said and how it is delivered. In text-to-speech (TTS) synthesis, an input
transcript specifies the words to be spoken, while speech editing additionally
connects requested changes to an existing utterance. Early neural TTS systems,
including Tacotron and VQTTS, autoregressively generated continuous acoustic
features or discrete self-supervised speech tokens~\citep{wang2017tacotron,du2022vqtts}.
In parallel, an established line of speech editing research addressed text
editing: inserting, deleting, or replacing spoken words while preserving the
surrounding speech. VoCo combined synthesis, retrieval, voice conversion, and
waveform assembly for such changes~\citep{jin2017voco}. Neural approaches
including EditSpeech, CampNet, A$^3$T, and UniCATS subsequently used text and
acoustic context to generate edited regions, with the surrounding speech
providing context for voice consistency and natural transitions
~\citep{tan2021editspeech,wang2022campnet,bai2022a3t,du2023unicats}.

\ifreportversion
\begin{table*}[t]
  \centering
  \small
  \caption{Precise editing capabilities. $\checkmark$ denotes
  documented coverage, and --- denotes coverage not established in the cited work.
  Open-source denotes publicly available code and model weights.
  An explicit instruction interface binds operations, parameters, and scopes
  in a user-supplied instruction. The pitch/rate/pause column requires all three
  local controls. Compositional local editing combines operation categories
  in one output. The aligner- and duration-model-free column denotes
  executing edits without depending on an external aligner or explicit
  duration prediction.}
  \label{tab:precise-editing-capabilities}
  \setlength{\tabcolsep}{4pt}
  \renewcommand{\arraystretch}{1.2}
  \newcommand{\capabilityhead}[1]{{\renewcommand{\arraystretch}{1.0}\tablehead{#1}}}
  \resizebox{\textwidth}{!}{%
  \begin{tabular}{@{}lccccccc@{}}
    \toprule
    {\renewcommand{\arraystretch}{1.0}\systemhead} & Open-source & \capabilityhead{Explicit instruction\\interface} &
      \capabilityhead{Local text\\editing} & \capabilityhead{Local emotion\\editing} &
      \capabilityhead{Local pitch/rate/pause\\control} & \capabilityhead{Compositional\\local editing} &
      \capabilityhead{Aligner- and\\duration-model-free} \\
    \midrule
\rowcolor{black!4}
    VoiceCraft~\citep{peng2024voicecraft} & $\checkmark$ & --- & $\checkmark$ & --- & --- & --- & --- \\
    Step-Audio-EditX~\citep{yan2025stepaudioeditx} & $\checkmark$ & --- & --- & --- & --- & --- & $\checkmark$ \\
\rowcolor{black!4}
    AST~\citep{lv2026ast} & --- & --- & $\checkmark$ & $\checkmark$ & --- & $\checkmark$ & --- \\
    UniSAE~\citep{zhu2026unisae} & --- & --- & $\checkmark$ & --- & --- & --- & --- \\
\rowcolor{black!4}
    Ming-UniAudio~\citep{yan2025minguniaudio} & $\checkmark$ & --- & $\checkmark$ & --- & --- & --- & $\checkmark$ \\
    \begin{tabular}[c]{@{}l@{}}MiMo-Audio-Instruct\\\citep{xiaomi2025mimoaudio}\end{tabular} & $\checkmark$ & --- & --- & --- & --- & --- & $\checkmark$ \\
\rowcolor{black!4}
    \begin{tabular}[c]{@{}l@{}}FireRedTTS3-Instruct\\\citep{shen2026fireredtts3}\end{tabular} & $\checkmark$ & --- & $\checkmark$ & --- & --- & --- & $\checkmark$ \\
    AuK~\citep{ma2026auk} & $\checkmark$ & --- & $\checkmark$ & --- & --- & --- & --- \\
    \midrule
    \rowcolor{edittext!7}
    \textbf{\dotsttsedit{}} & $\checkmark$ & $\checkmark$ & $\checkmark$ & $\checkmark$ & $\checkmark$ & $\checkmark$ & $\checkmark$ \\
    \bottomrule
  \end{tabular}}
\end{table*}

\else
\ifreportversion
\captionsetup{type=figure,hypcap=false}
\else
\begin{figure*}[t]
\fi
  \centering
  \setstretch{1}
  \fontsize{9}{10}\selectfont
  \setlength{\tabcolsep}{6pt}
  \renewcommand{\arraystretch}{1.0}
  \renewcommand{\tabularxcolumn}[1]{m{#1}}
  \begin{tabularx}{\textwidth}{@{}>{\bfseries\raggedright\arraybackslash}m{0.28\textwidth} X@{}}
    \toprule
    Source transcript $\textsrc$ &
      ``I thought the meeting starts after lunch today.'' \\
    \midrule
    Edit instruction $u$ &
      {\fontsize{8}{9}\selectfont\ttfamily
      \begin{tabular}[c]{@{}l@{}}
        I thought the
        \textcolor{edittext}{<sub targ="concert">meeting</sub>} \\
        \textcolor{editemotion}{<emo desc="a happy and excited tone">starts after lunch</emo>} \\
        \textcolor{editpause}{<pause act="ins" level="2"/>} today.
      \end{tabular}} \\
    \midrule
    Expected target transcript $\texttgt$ &
      ``I thought the \textcolor{edittext}{concert}
      \textcolor{editemotion}{starts after lunch}
      \textcolor{editpause}{$\Vert$} today.'' \\
    \bottomrule
  \end{tabularx}
  \caption{A compositional editing example.  A semi-formal transcript-grounded edit instruction specifies operation
  categories and parameters while colored spans and a boundary localize their
  effects.  The transcript rendering replaces \emph{meeting} with
  \emph{concert}. Emotion and pause operations alter delivery without changing
  the remaining words.  Colors and $\Vert$ in the target rendering are visual
  annotations that mark edited content, delivery, and the inserted pause. They
  are not part of $\texttgt$.}
  \label{fig:instruction-example}
\ifreportversion\else
\end{figure*}
\fi

\fi

Following advances in large language models such as GPT-3 and LLaMA
~\citep{brown2020gpt3,touvron2023llama}, large-scale pretraining has also
broadened the capabilities of speech generation models. VALL-E formulates
TTS synthesis as conditional language modeling over audio codec tokens and
supports unseen speakers through short acoustic prompts~\citep{chen2025valle}.
CosyVoice and Spark-TTS further develop this approach through speech token
design and large-scale training~\citep{du2024cosyvoice,wang2025sparktts}.
A complementary route uses diffusion or flow matching to model continuous
speech representations directly, including non-autoregressive models such as
E2 TTS and F5-TTS~\citep{eskimez2024e2tts,chen2024f5tts} and autoregressive
models such as ARDiT, DiTAR, and \dotstts{}
~\citep{liu2024ardit,jia2025ditar,dotstts2026}.
Control has also expanded beyond the transcript: PromptTTS conditions
TTS synthesis on descriptions of vocal style~\citep{guo2023prompttts}, and
MiMo-Audio and proprietary systems such as GPT-4o mini TTS support
natural-language control over speech delivery
~\citep{xiaomi2025mimoaudio,openai2025audio}.
For speech editing, Voicebox uses speech infilling to support TTS synthesis,
text editing, and other transformations within one model~\citep{le2023voicebox}.
Recent general-purpose speech editing systems also change vocal attributes
such as emotion and prosody. Ming-UniAudio and AuK combine TTS synthesis
with multiple editing capabilities, using natural-language instructions to
express changes to spoken content and acoustic attributes
~\citep{yan2025minguniaudio,ma2026auk}.

Natural language is flexible, but its flexibility can also introduce ambiguity:
textual references may admit multiple interpretations, while the intended
operation category, parameters, or target region may remain underspecified.
Additionally, professional creation frequently requires repeatable controls
whose requested effect and scope can be checked independently. Such settings
benefit from a precise and explicit representation of editing instructions.
A machine-readable, inspectable, and composable representation, rather than open-ended semantics, is also suitable
as a controllable backend for audio-creation studios or as a callable tool in
reliable agentic systems~\citep{li2026xflow,yu2026das}.

We formalize this precision along two axes.  \emph{Precise operation
specification} makes the operation category and parameters
explicit---what to edit and how.  \emph{Precise localization} states where it
applies.  Absolute timestamps may appear to provide precise, unambiguous localization. They are used for fine-grained temporal grounding in audio-understanding models~\citep{ghosh2026audioflamingonext,sun2026spotsound}, but are less natural as an interface for speech editing: acoustic transitions between linguistic units can be gradual, and a time boundary alone does not specify the intended treatment of a partially covered word. For example, when a replacement interval begins in the middle of a word in the source speech, the timestamp alone leaves it unclear whether the entire word should be replaced. Anchoring edits to transcript spans and boundaries instead makes the intended linguistic scope explicit, reducing this ambiguity and providing a natural interface for speech-creation tasks. We therefore propose a semi-formal transcript-grounded edit instruction that specifies operations and parameters along the transcript. Its XML-style syntax uses tag names for operation categories, attributes for parameters, and tag positions for linguistic spans or boundaries. Parameter values can be numerical, such as a pitch offset in semitones, or descriptive, such as an emotion specification in a \texttt{type} or \texttt{desc} attribute. By combining a fixed, machine-readable syntax with flexible parameter values, this semi-formal representation makes the requested operation
and its scope inspectable while leaving their acoustic realization to the
speech editing model.
\Cref{fig:instruction-example} illustrates explicit operation specification and transcript-based localization in a compositional edit.

\ifreportversion\else

\fi

The space of speech-creation requests is broader than any fixed task list.  We
select four representative, recurring controls that exercise distinct
operation categories and localization patterns.  Text editing changes lexical
content at spans or boundaries.  Emotion editing changes affective expression
globally or over a selected span.  Prosody editing controls pitch or speaking
rate over a span. Pause editing requires naturally adjusting the duration of silence gaps at word boundaries.

Realizing this formulation remains difficult.  The output must execute every
requested operation, preserve non-target content and acoustics, and remain
coherent even when a local edit changes duration. Training such a model
requires large amounts of corresponding paired data. However, natural
recordings rarely provide paired utterances that differ only in one requested
property, while alignment, resynthesis, and stitching can introduce incidental
changes. In addition, accurately evaluating instruction following, local
preservation, and overall audio quality across editing tasks remains a challenge.

We instantiate the formulation in \dotsttsedit{}, a unified model for
precisely controlled speech editing. It adapts the continuous autoregressive
\dotstts{} TTS foundation model~\citep{dotstts2026}. The model conditions
on source speech and the edit instruction, whose
deterministic transcript renderings provide source and target transcripts,
then generates target speech in continuous latent space.
In addition to edits within a single category, the model can execute multiple
operations across categories from one instruction in a single pass.
To train this model, we develop tailored data pipelines that construct
large-scale paired data for the four representative editing categories,
with explicit operation and scope supervision under a shared instruction
representation.

We also introduce \emph{doteBench}, a bilingual benchmark for precise speech
editing. It uses the semi-formal transcript-grounded edit instruction as the
primary specification of the intended edits and provides fine-grained metrics
across editing tasks and evaluation dimensions.
Instruction Following tests whether the requested operation is realized,
Local Preservation tests whether speech outside the edited regions and
non-target attributes within them remain unchanged, and Audio Quality
evaluates the complete output.
Across text, emotion, prosody, pause, and compositional editing,
\dotsttsedit{} achieves leading overall instruction following and local
preservation among the evaluated open-source audio-generation and
speech-editing systems, while maintaining comparable audio quality.  Its
Seed-TTS-Eval recognition and speaker-similarity scores remain close to the
strongest \dotstts{} variants.

Our contributions are:

\begin{itemize}
  \item A semi-formal transcript-grounded edit instruction that combines typed operations,
  explicit parameters, and transcript-bound scopes, supporting inspectable
  compositional control.
  \item \dotsttsedit{}, a continuous autoregressive speech editing model trained
  on controlled pairs from tailored data pipelines, enabling four
  representative editing controls within one model and executing multiple
  operations within one generation pass.
  \item doteBench, a bilingual benchmark that evaluates precise
  instruction following, local preservation, and audio quality for individual
  and compositional edits.
\end{itemize}

\section{Related Work}
\label{sec:related-work}

\begin{table*}[t]
  \centering
  \small
  \caption{Evaluation of precise editing. $\checkmark$ denotes
  documented coverage, and --- denotes coverage not established.
  Open-source denotes publicly available benchmark data and evaluation code.
  Explainable task-specific measurements directly assess linguistic or acoustic
  properties. Within-edit preservation checks non-target attributes inside
  edited regions, such as pitch during a rate change.}
  \label{tab:precise-editing-evaluation}
  \setlength{\tabcolsep}{5pt}
  \renewcommand{\arraystretch}{1.2}
  \newcommand{\evaluationhead}[1]{{\renewcommand{\arraystretch}{0.9}\tablehead{#1}}}
  \resizebox{\textwidth}{!}{%
  \begin{tabular}{@{}lccccc@{}}
    \toprule
    {\renewcommand{\arraystretch}{0.9}\bottomtablehead{Benchmark}} & Open-source & \evaluationhead{Explainable task-specific\\measurements} &
      \evaluationhead{Compositional\\evaluation} &
      \evaluationhead{Word-level prosody\\preservation} &
      \evaluationhead{Within-edit attribute\\preservation} \\
    \midrule
\rowcolor{black!4}
    \begin{tabular}[c]{@{}l@{}}Step-Audio-Edit-Benchmark\\\citep{yan2025stepaudioeditx}\end{tabular} & $\checkmark$ & --- & --- & --- & --- \\
    \begin{tabular}[c]{@{}l@{}}Ming-Freeform-Audio-Edit\\\citep{yan2025minguniaudio}\end{tabular} & $\checkmark$ & $\checkmark$ & --- & --- & --- \\
\rowcolor{black!4}
    LibriSpeech-Edit~\citep{lv2026ast} & --- & $\checkmark$ & --- & $\checkmark$ & --- \\
    SpeechEditBench~\citep{zhang2026speecheditbench} & $\checkmark$ & $\checkmark$ & $\checkmark$ & --- & --- \\
\rowcolor{black!4}
    MMAE~\citep{ma2026mmae} & $\checkmark$ & --- & $\checkmark$ & --- & $\checkmark$ \\
    \midrule
    \rowcolor{edittext!7}
    \textbf{doteBench} & $\checkmark$ & $\checkmark$ & $\checkmark$ & $\checkmark$ & $\checkmark$ \\
    \bottomrule
  \end{tabular}}
\end{table*}

\mbox{\Cref{tab:precise-editing-capabilities}} compares the editing capabilities of recent systems.

\paragraph{Earlier text editing in speech}
Text editing changes the spoken content of an existing utterance through
insertion, deletion, or substitution while preserving the surrounding speech.
VoCo combines synthesis, retrieval, voice conversion, and stitching to replace
speech in an existing narration~\citep{jin2017voco}.  Neural speech editing models then
learned missing-region acoustics from text and surrounding speech: EditSpeech
uses partial inference and bidirectional fusion~\citep{tan2021editspeech},
CampNet predicts masked speech~\citep{wang2022campnet}, A$^3$T introduces
alignment-aware acoustic--text pretraining~\citep{bai2022a3t}, and
E$^3$TTS supports text-guided insertion, deletion, and replacement in a joint
TTS and editing model~\citep{liang2024e3tts}.
FluentEditor variants regularize boundary acoustics and global prosody
~\citep{liu2023fluenteditor,liu2024fluenteditor2}, whereas UniCATS uses
contextual VQ-diffusion over semantic tokens~\citep{du2023unicats}.
Foundation-scale systems extend infilling through flow matching in Voicebox
~\citep{le2023voicebox} and autoregressive codec generation in VoiceCraft
~\citep{peng2024voicecraft}. CosyEdit and AST adapt pretrained TTS models for
precise content edits~\citep{chen2026cosyedit,lv2026ast}.  Many such systems
rely on an explicit aligner module to map the edited transcript span to the
acoustic region that is masked or regenerated.  This progression
improves realization, boundary fluency, and contextual continuity, but
predominantly studies lexical edits rather than localized control of delivery.

\paragraph{General-purpose speech editing}
Beyond changes to spoken content, speech editing also concerns attributes
of delivery, speaker identity, and acoustic conditions. Recent general-purpose
models bring several generation and editing capabilities into a shared system.
SpeechX uses task-dependent prompt tokens to condition
one codec language model for TTS, enhancement, extraction, and
editing~\citep{wang2023speechx}.
Step-Audio-EditX performs iterative utterance-level editing of emotion,
speaking style, and paralinguistics~\citep{yan2025stepaudioeditx}, while
SpeechEdit selectively controls speaker, emotion, and prosody attributes during
TTS generation~\citep{pei2026speechedit}.  Ming-UniAudio supports free-form
content editing and utterance-level acoustic changes
~\citep{yan2025minguniaudio}.  UniSAE extends local content editing from
sub-phoneme to word level and composes it with speaker and emotion control
~\citep{zhu2026unisae}.  Beyond speech attributes, MMEdit localizes general-audio
events~\mbox{\citep{tao2025mmedit}}, while Audio-Omni and UNISON unify editing across speech, sound, music,
or audio scenes~\mbox{\citep{tian2026audioomni,li2026unison}}.
Recently, FireRedTTS3-Instruct unifies speech generation and instruction-guided
editing~\citep{shen2026fireredtts3}, and AuK likewise combines speech generation
and editing through natural-language instructions and audio
context~\citep{ma2026auk}.
Both models natively support local text-based speech editing without
an external aligner.
However, AuK still requires a separate, explicit estimate of the target
utterance duration before generation. Neither model supports localized
editing of paralinguistic attributes such as prosody and emotion.
To our knowledge,
\dotsttsedit{} is
the first end-to-end speech editing model to jointly support fine-grained local
content and paralinguistic editing while preserving unrequested regions and
attributes.

\paragraph{Speech editing benchmarks}
\mbox{\Cref{tab:precise-editing-evaluation}} compares benchmark availability
and evaluation coverage.
RealEdit targets zero-shot content editing in diverse acoustics
~\citep{peng2024voicecraft}.
Step-Audio-Edit-Benchmark evaluates emotion, speaking style, and
paralinguistic control through audio-language-model judgments
~\citep{yan2025stepaudioeditx}. Ming-Freeform-Audio-Edit covers lexical
operations and utterance-level acoustic transformations
~\citep{yan2025minguniaudio}. Neither benchmark evaluates local acoustic
editing with explicit preservation checks for unrequested acoustic attributes.
LibriSpeech-Edit adds controlled text and style
edits with temporal-consistency measures~\citep{lv2026ast}.  SpeechEditBench
covers content, emotion, prosody, and four other atomic editing tasks, together
with compositional editing~\citep{zhang2026speecheditbench}.
Its preservation-success gate for non-content tasks checks only speech
recognition error rates---word error rate (WER) for English and character
error rate (CER) for Chinese---and therefore does not assess whether source
prosody or other paralinguistic attributes remain preserved.
Recently, MMAE broadens coverage to multimodal and mixed
acoustic scenarios with complex instructions and local or global operations
~\citep{ma2026mmae}.  Its fine-grained instruction-following and consistency
rubrics rely on Qwen3-Omni judgments~\citep{xu2025qwen3omni}.  Three independent
queries, majority voting, and option shuffling mitigate variation in discrete
rubric decisions and positional bias, but the resulting scores remain
constrained by the judge's fine-grained perceptual capability and offer limited
interpretability. Our doteBench instead defines category-specific, scope-aware
protocols for text,
emotion, prosody, pause, and compositional editing under a common evaluation of
instruction following, local preservation, and audio quality, emphasizing interpretable
instruction following and preservation measurements for localized speech controls.

\section{Problem Definition and Benchmark}
\label{sec:problem-benchmark}

\subsection{Precisely Controlled Speech Editing}

We define an edit sample as the five-tuple
\begin{equation}
  \begin{aligned}
    d_{\mathrm{edit}}
      &=\left(\textsrc,\audiosrc,u,\texttgt,\audiotgt\right),\\
    \textsrc&=\srcproj{u},\qquad \texttgt=\tgtproj{u}.
  \end{aligned}
  \label{eq:edit-sample}
\end{equation}
Here $\textsrc$ and $\texttgt$ are the source and target transcripts,
$\audiosrc$ and $\audiotgt$ are the corresponding waveforms, and $u$ is a
semi-formal transcript-grounded edit instruction. Tag names and XML attributes encode operations and their parameters, with tag positions specifying the scope.
The deterministic
renderer $\srcproj{u}$ removes the tags while retaining source-side lexical
content, whereas $\tgtproj{u}$ applies lexical insertions, deletions, and
substitutions.  Attribute tags leave the transcript unchanged.
The speech editing model produces
\begin{equation}
  \hat{A}_{\mathrm{tgt}}
  =f_{\theta}\!\left(
    \textsrc,\audiosrc,u,\texttgt
  \right).
\end{equation}
Here $f_{\theta}$ is the speech editing model with trainable parameters $\theta$, and
$\hat{A}_{\mathrm{tgt}}$ is its predicted target waveform.
Each tag specifies an operation category and its parameters, while its position
in the transcript binds the operation to a span or boundary. This separates
operation specification from localization while making their binding explicit
in the transcript, avoiding the need for the generative model to infer either field from an
underspecified request.  A successful output must (i) realize every tagged
operation, (ii) preserve content and attributes outside the tagged locations,
and (iii) remain natural and coherent as a complete utterance.

\paragraph{Extensibility}
\begin{figure}[t]
  \centering
  \includegraphics[width=\columnwidth,trim=0bp 33bp 0bp 15bp,clip]{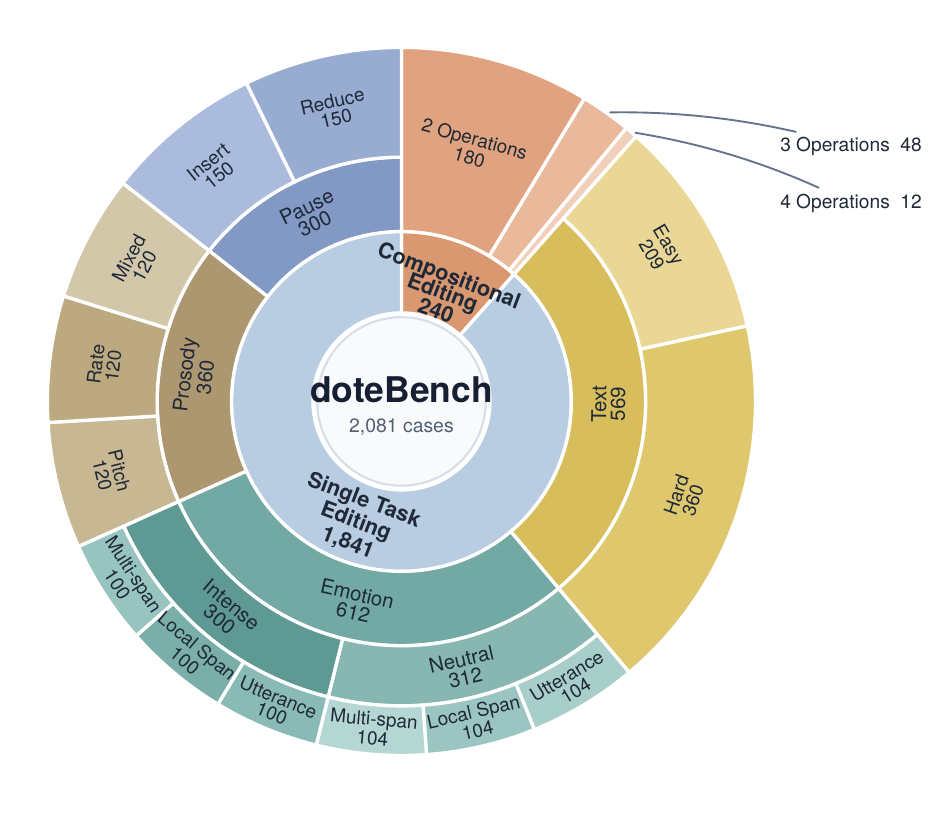}
  \caption{Hierarchical composition of doteBench.}
  \label{fig:doteb-distribution}
\end{figure}

\begin{table}[t]
  \centering
  \small
  \caption{doteBench categories, sizes, and evaluation dimensions.}
  \label{tab:benchmark-overview}
  \setlength{\tabcolsep}{2pt}
  \renewcommand{\arraystretch}{1.18}
  \renewcommand{\tabularxcolumn}[1]{m{#1}}
  \begin{tabularx}{\columnwidth}{@{}lc >{\raggedright\arraybackslash}X >{\raggedright\arraybackslash}X@{}}
    \toprule
    Category & Cases & Instruction following & Local preservation \\
    \midrule
\rowcolor{black!4}
    Text & 569 &
      Edit WER/CER &
      Non-edit WER/CER, WDTW, SpkSim \\
    \addlinespace[2pt]
    Emotion & 612 &
      Emotion accuracy &
      WER/CER, WDTW, SpkSim \\
    \addlinespace[2pt]
\rowcolor{black!4}
    Prosody & 360 &
      Duration/pitch error & WER/CER, WDTW, SpkSim \\
    \addlinespace[2pt]
    Pause & 300 &
      Direction accuracy & WER/CER, WDTW, SpkSim \\
    \addlinespace[2pt]
\rowcolor{black!4}
    Compositional & 240 &
      Component/all success & WER/CER, WDTW, SpkSim \\
    \bottomrule
  \end{tabularx}
\end{table}

\ifreportversion
\begin{figure*}[!t]
  \centering
  \includegraphics[width=\textwidth]{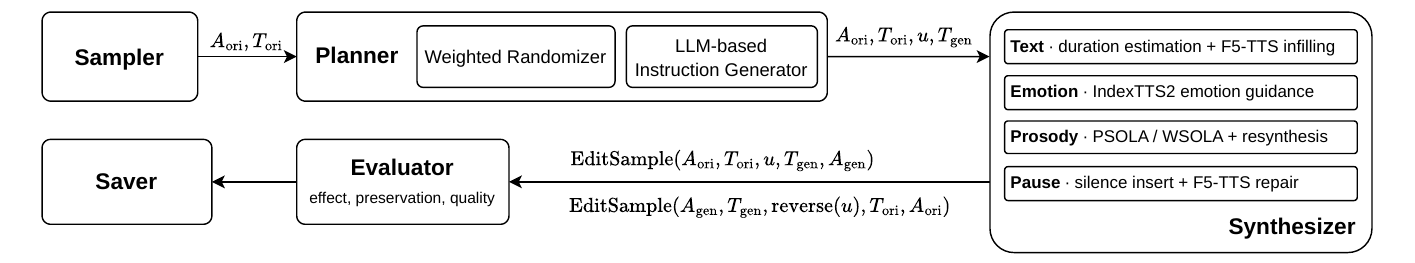}
  \caption{\textbf{Data pipeline.}  A sampler selects an
  original utterance, a planner produces an instruction and target
  transcript, and a task-specific synthesizer constructs the generated
  counterpart.  The evaluator checks edit effect (instruction following), preservation, and quality
  before the saver materializes accepted examples.  For text, prosody, and pause
  editing, we construct supervision in both directions.}
  \label{fig:data-pipeline}
\end{figure*}

\fi
Within speech, the typed vocabulary can incorporate additional attributes and
edit actions.  Its explicit schema also provides a stable tool boundary through
which a studio frontend, planner, or agent can construct, validate, and invoke
an edit instruction.  More broadly, the same instruction design could extend to
non-verbal events, general audio, or music given a suitable symbolic timeline
and paired supervision. This work evaluates speech only.

\subsection{doteBench}
\label{sec:benchmark}

Evaluating precisely controlled speech editing requires distinguishing requested
changes from unintended alterations according to the instruction's operation,
parameters, and scope. We therefore introduce doteBench, a bilingual benchmark
that measures instruction following, local preservation, and audio quality
across editing categories. It comprises five categories: text editing, emotion editing,
prosody editing, pause editing, and compositional editing.  The first four
evaluate one type of edit at a time on single-speaker utterances, with explicit
target and preserved regions. \Cref{fig:doteb-distribution} shows the category
hierarchy and case counts.
Text editing has an Easy shard for smaller local changes and a Hard
shard with longer contexts, larger replacements, and multiple text operations.
Emotion editing uses Neutral and Intense shards, starting from neutral speech
and strongly emotional speech, respectively.
Prosody and pause editing
each use one primary set.  The compositional editing
category covers two-, three-, and four-operation instructions. Its instruction
following metrics report
component and all-operation success rates. Together,
the three evaluation dimensions match the control contract: Instruction
Following tests execution of the specified operation, Local Preservation tests
content and acoustic attributes that the instruction leaves unchanged, and
Audio Quality tests the complete output audio.
We assign operation-specific metrics for instruction following and local
preservation to each editing category, as summarized in
\cref{tab:benchmark-overview} and detailed below.

\ifreportversion\else
\begin{figure*}[!t]
  \centering
  \includegraphics[width=\textwidth]{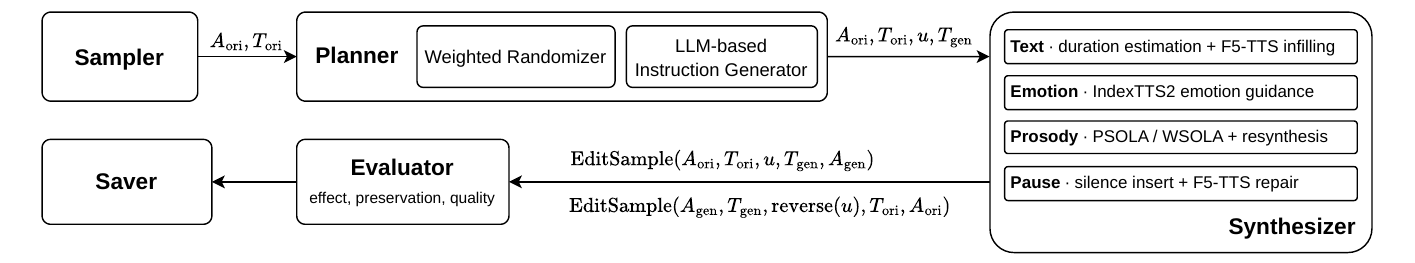}
  \caption{\textbf{Data pipeline.}  A sampler selects an
  original utterance, a planner produces an instruction and target
  transcript, and a task-specific synthesizer constructs the generated
  counterpart.  The evaluator checks edit effect (instruction following), preservation, and quality
  before the saver materializes accepted examples.  For text, prosody, and pause
  editing, we construct supervision in both directions.}
  \label{fig:data-pipeline}
\end{figure*}

\begin{figure*}[!t]
  \centering
  \ifreportversion
  \includegraphics[width=\textwidth]{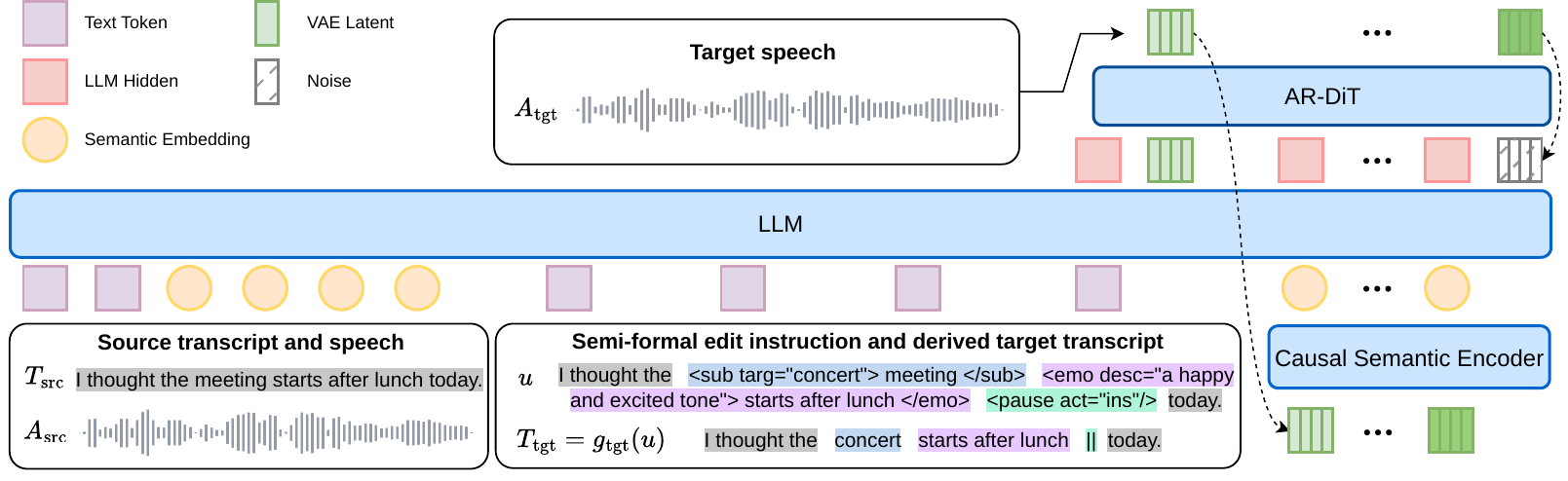}
  \else
  \includegraphics[width=0.95\textwidth]{dots.tts.edit.pdf}
  \fi
  \caption{\textbf{\dotsttsedit{} overview.}  Source transcript and speech,
  together with the semi-formal transcript-grounded edit instruction and its target-transcript rendering, condition the retained
  continuous autoregressive \dotstts{} backbone. Only target speech is
  generated. The pause symbol in the target-transcript illustration
  is a visual marker. It is not part of the transcript supplied to the model.}
  \label{fig:overview}
\end{figure*}

\fi

\paragraph{Instruction following}
Text editing uses edited-region WER/CER. Emotion editing uses
Qwen3-Omni-30B-A3B-Instruct~\citep{xu2025qwen3omni} to classify the requested
emotion from the edited region under a fixed prompt and label set.
Prosody editing measures absolute duration error in seconds and pitch-shift
error in semitones.
Pause editing requires a gap change of at least 160\,ms in the requested direction.
For a compact comparison across heterogeneous operation families,
compositional instruction following uses binary success criteria for content, emotion, the direction
and magnitude of prosody changes, and the direction of pause changes.
Component success is the fraction of requested operations meeting their
criteria. All-operation success is the fraction of examples in which every
requested operation meets its criterion.
\suppref{app:compositional-criteria} defines the criteria.

\paragraph{Local preservation}
For text editing, Qwen3-ASR~\citep{shi2026qwen3asr} measures recognition error
outside the instruction-derived edit neighborhood.  Because the other tasks
retain the transcript, they use full-utterance WER/CER.  We additionally
compare matched, preserved text units (English words and Chinese characters)
acoustically. WDTW-Dur inherits the
word-level duration comparison introduced by AST~\citep{lv2026ast}.
Intuitively, it force-aligns instruction-selected units and measures the
normalized change in their duration sequences as a percentage, where lower
values indicate better timing preservation. This differs from the absolute
Duration L1 error in seconds used to assess requested prosody changes.
Duration alone cannot expose pitch drift, so we
introduce WDTW-F0, which compares the minimum, maximum, and mean voiced F0 of
eligible preserved units in semitones. Lower values indicate less pitch drift.
We expect both duration and F0 to be preserved in regions with no edit.
Within an edited region, a duration-only change should preserve F0, and a
pitch-only change should preserve duration. Accordingly, WDTW-F0 includes
duration-only regions, and WDTW-Dur includes pitch-only regions.
We also assess whether editing preserves speaker identity using speaker
similarity (SpkSim), computed as the cosine similarity between source and
edited speech embeddings from WavLM-large/ECAPA-TDNN
~\citep{chen2022wavlm,desplanques2020ecapa}.
\suppref{app:evaluation-criteria} details the preservation metrics and
their aggregation.

\paragraph{Audio quality}
Beyond instruction following and preservation of unedited regions, evaluation
must also measure the overall acoustic quality of the generated speech.
Audio quality is measured by UTMOS~\citep{saeki2022utmos}, an automatic predictor
of mean opinion scores for speech quality.

\section{Data Pipelines for Precisely Controlled Speech Editing}
\label{sec:data}

\paragraph{Challenges}
Precisely controlled speech editing requires paired utterances that realize a
specified change while preserving the remaining content and acoustics. Such
pairs are rare in natural recordings. Independently synthesizing both sides
can introduce unrelated differences in timing, voice, or rendering, while
replacing only a local segment can create abrupt, incoherent transitions at
its boundaries.

\paragraph{Principle}
Our data pipeline in \Cref{fig:data-pipeline} constructs pairs around explicit
interventions, reuses surrounding speech where possible, and repairs local
transitions. Context-conditioned F5-TTS~\mbox{\citep{chen2024f5tts}} regeneration repairs short boundary
neighborhoods to make local joins natural. The planner specifies the operation, parameters, and location
through a semi-formal transcript-grounded edit instruction.
Dozens of speech models and signal-processing tools synthesize and
assemble the paired utterances, which share the five-tuple representation in
\Cref{eq:edit-sample}. The evaluator filters pairs for edit realization,
intelligibility, preservation, speaker consistency, and quality. Text,
prosody, and pause construction also supplies
reverse supervision by exchanging the original and generated waveforms and
inverting the instruction. When the original waveform is human-recorded speech, the reverse direction uses it as the target. Construction procedures and parameter
settings for each editing task are selected through evaluation using objective
metrics and subjective preferences.
\suppref{app:data-pipeline-details} provides construction algorithms and parameters.

\ifreportversion
\begin{figure*}[!t]
  \centering
  \ifreportversion
  \includegraphics[width=\textwidth]{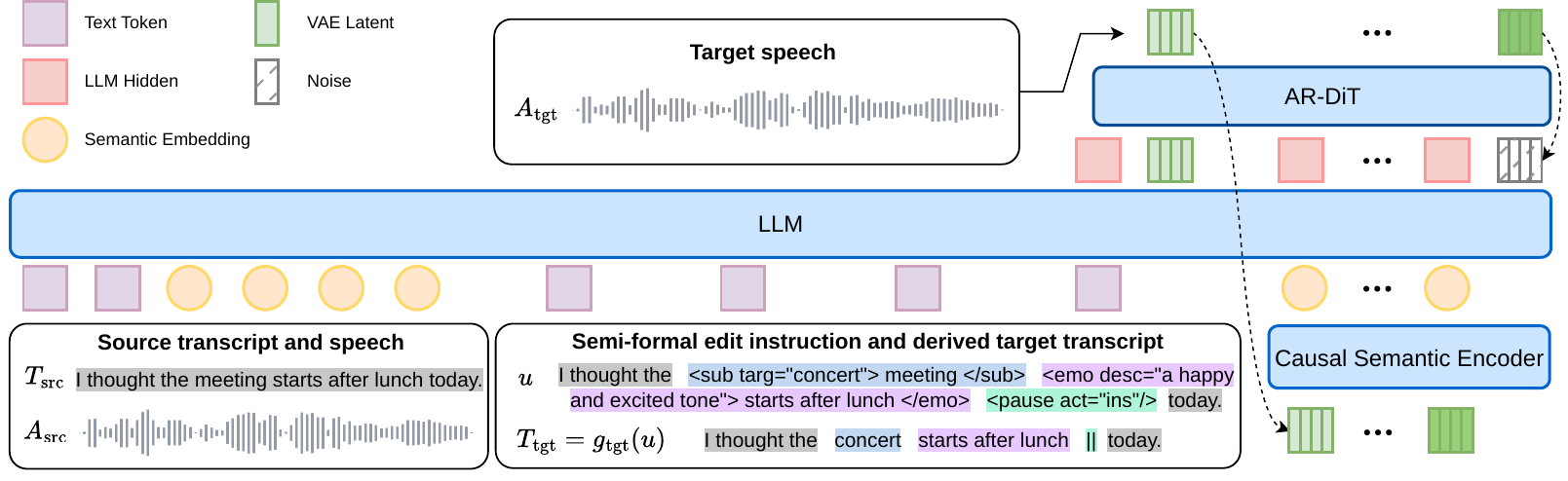}
  \else
  \includegraphics[width=0.95\textwidth]{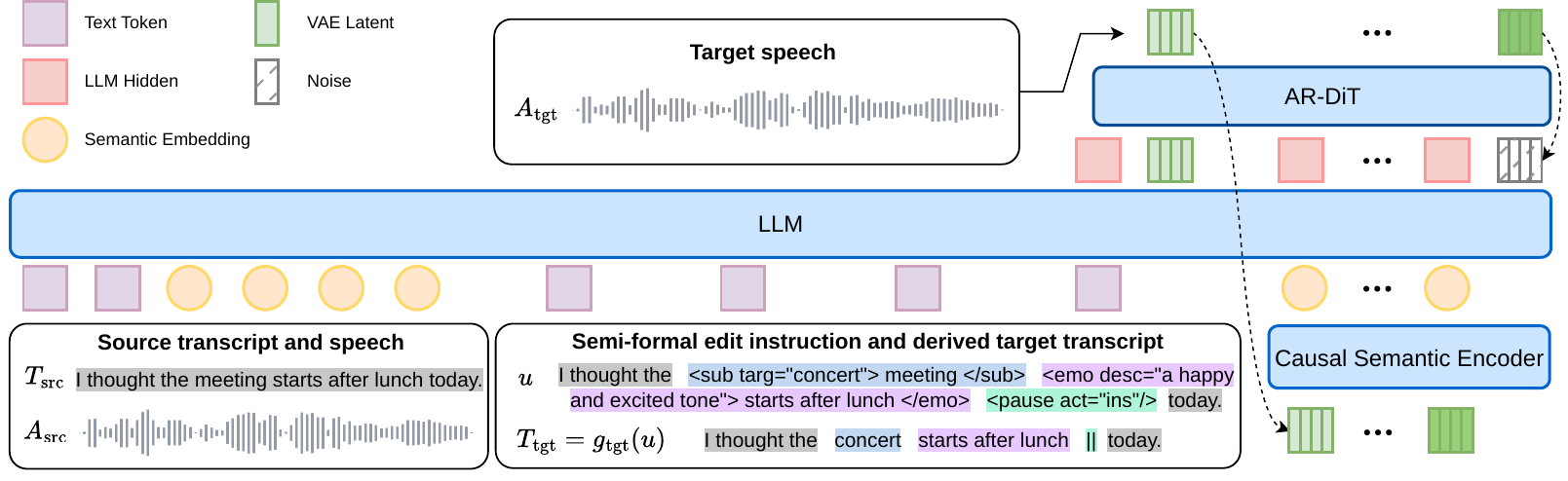}
  \fi
  \caption{\textbf{\dotsttsedit{} overview.}  Source transcript and speech,
  together with the semi-formal transcript-grounded edit instruction and its target-transcript rendering, condition the retained
  continuous autoregressive \dotstts{} backbone. Only target speech is
  generated. The pause symbol in the target-transcript illustration
  is a visual marker. It is not part of the transcript supplied to the model.}
  \label{fig:overview}
\end{figure*}

\fi

\section{\dotsttsedit{}: Continuous Autoregressive Speech Editing}
\label{sec:method}

\subsection{Continuous Autoregressive Backbone}

With the paired data in place, we train a model for the conditional
sequence-generation task formulated above, which naturally admits an
autoregressive architecture. The close relationship between speech editing
and TTS motivates adapting a TTS model pretrained on large-scale speech data.
We adopt \dotstts{}~\mbox{\citep{dotstts2026}}: continuous latents offer greater acoustic fidelity potential,
while autoregressive generation avoids separate duration prediction and reuses
pretrained LLM capabilities for semantic modeling. This architecture also
provides a basis for scalable training and future streaming interaction.

\dotstts{} is a 2B-parameter continuous autoregressive TTS foundation
model.  It represents 48\,kHz speech with an AudioVAE
based on HoliTok~\citep{li2026holitok}, producing a 25\,Hz latent stream.  It
reduces each 4-frame patch to a 6.25\,Hz semantic
representation. The LLM autoregressively produces hidden states from which a
flow-matching head renders continuous latent patches.  A frozen CAM++ speaker
encoder~\citep{wang2023campplus} provides global identity conditioning.
\dotsttsedit{} retains this architecture and generation interface.  We change
the conditioning sequence and paired training examples rather than adding
task-specific inpainting networks.    \Cref{fig:overview} shows the retained
backbone and the editing conditioning sequence.

\subsection{Editing and TTS Training}

Training mixes ordinary TTS examples with the paired editing examples from
\Cref{sec:data}.  Let
$\latentsrc=\mathcal{E}_{\mathrm{VAE}}(\audiosrc)$ and
$\latenttgt=\mathcal{E}_{\mathrm{VAE}}(\audiotgt)$ denote the source and target
AudioVAE latent sequences produced by the frozen encoder
$\mathcal{E}_{\mathrm{VAE}}$, and let $\mathbf{e}_{\mathrm{spk}}$ denote the
frozen CAM++ speaker embedding.  Editing examples follow the sequence
$[\,\textsrc,\latentsrc,u,\texttgt,\latenttgt\,]$.  The target-audio positions
carry flow-matching and stopping supervision, and $\latentsrc$ provides the
complete source utterance as acoustic context. The model generates the complete
edited utterance, learning to preserve unrequested content and acoustics from
the source-conditioned training pairs. TTS and editing retain distinct
conditioning contexts but share the same generative model for target audio latents:
\begin{equation}
 \begin{aligned}
 p_{\theta}^{\mathrm{TTS}}
   &=p_{\theta}\!\left(Z\mid T,\mathbf{e}_{\mathrm{spk}}\right),\\
 p_{\theta}^{\mathrm{edit}}
   &=p_{\theta}\!\left(
     \latenttgt\mid\textsrc,\latentsrc,u,\texttgt,
     \mathbf{e}_{\mathrm{spk}}\right).
 \end{aligned}
\label{eq:tts-edit-conditioning}
\end{equation}
Here $T$ denotes a generic transcript, $Z$ a target AudioVAE latent sequence,
and $p_{\theta}$ the shared conditional generative model.
Thus TTS is conditioned on the requested transcript and speaker identity,
whereas editing additionally observes the source transcript, source speech,
instruction, and target transcript.  All editing tasks use $p_{\theta}^{\mathrm{edit}}$ with shared model parameters $\theta$.

For each supervised target patch $x_1$, flow matching~\citep{lipman2023flow}
draws standard Gaussian noise $x_0\sim\mathcal{N}(0,I)$, where $I$ is the
identity covariance, and time $t\sim\mathcal{U}(0,1)$.  With zero terminal-noise scale, the interpolated patch is $x_t=t x_1+(1-t)x_0$ and the target velocity is $v^\star=x_1-x_0$.  The autoregressive backbone supplies
the causal semantic history and conditioning context for this patch-level
velocity field.  The training objective is
\begin{equation}
 \mathcal{L}
 =\lambda_{\mathrm{FM}}\,
   \mathbb{E}\!\left[\left\|
   v_{\theta}(x_t,t;c)-v^\star\right\|_2^2\right]
 +\lambda_{\mathrm{EOS}}\mathcal{L}_{\mathrm{EOS}}.
\label{eq:training-objective}
\end{equation}
Here $v_{\theta}$ is the predicted velocity. The conditioning variable $c$ comprises the causal history of preceding audio patches and either $(T,\mathbf{e}_{\mathrm{spk}})$ for TTS or $(\textsrc,\latentsrc,u,\texttgt,\mathbf{e}_{\mathrm{spk}})$ for editing, as specified in \Cref{eq:tts-edit-conditioning}. The expectation is
over training examples, target patches, Gaussian noise, and interpolation
times. $\mathcal{L}_{\mathrm{EOS}}$ is the two-class cross-entropy of the
audio-stop projection at supervised target-audio positions. The flow-matching error and $\mathcal{L}_{\mathrm{EOS}}$ are each computed as a mask-weighted mean: the sum of weighted errors divided by the sum of the corresponding supervision weights. Flow-matching weights select target latent patches; EOS weights select target-audio positions for stop prediction. Source-audio, text, and padding positions contribute to neither objective.

\section{Experiments}
\label{sec:experiments}

Our experiments ask four questions: whether the speech editing model executes explicitly
specified operations, whether it preserves speech outside their localizations,
whether several operations can be composed in one generation pass, and
whether editing post-training retains the zero-shot TTS capability of the base
model.  doteBench answers the first three through Instruction Following, Local
Preservation, and Audio Quality comparisons against existing open-source
audio-generation and speech-editing models and task-specific pipelines.
SpeechEditBench~\citep{zhang2026speecheditbench} provides a complementary bilingual, multi-attribute
comparison (\suppref{app:additional-results}), and Seed-TTS-Eval~\citep{seedtts2024} tests
TTS capability retention. A listening study (Section~\ref{sec:listening})
assesses agreement between doteBench metrics and human judgments.

\subsection{Setup}
\label{sec:exp_setup}

\subsubsection{Training}
\paragraph{Model initialization}
\dotsttsedit{} initializes all speech editing model parameters from the public 2B-parameter
\dotstts{} checkpoint~\citep{dotstts2026}.\footnote{\scriptsize\url{https://huggingface.co/collections/rednote-hilab/dotstts}}
The backbone comprises a Qwen2.5-1.5B language model
~\citep{qwen2024qwen25}, a 24-layer semantic encoder, and an 18-layer
autoregressive flow-matching DiT~\citep{peebles2023dit} operating on
4-frame patches of a frozen 48\,kHz AudioVAE.  A frozen 512-dimensional
CAM++ speaker encoder
~\citep{wang2023campplus} supplies the voice condition. We use the
VoxCeleb-trained checkpoint.\footnote{\scriptsize\url{https://modelscope.cn/models/iic/speech_campplus_sv_en_voxceleb_16k}}

\paragraph{Objectives and optimization}
\begin{figure*}[!t]
  \centering
  \includegraphics[width=\textwidth]{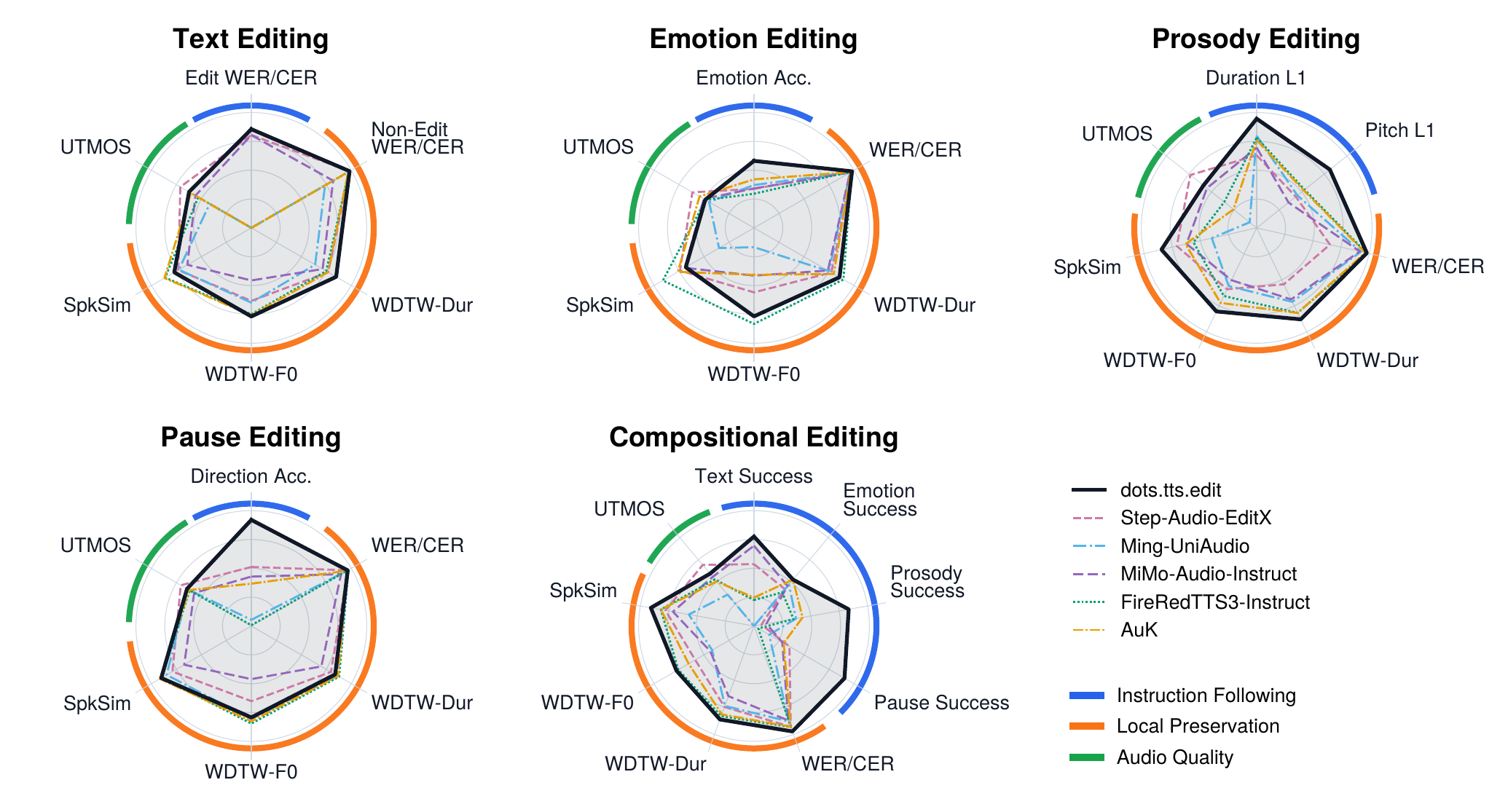}
  \caption{\textbf{One-take model comparison on doteBench.}  Five panels
  summarize the text, emotion, prosody, pause, and compositional editing
  categories.  Axes form contiguous groups for instruction following, local
  preservation, and audio quality.  Fixed model-independent semantic bounds
  map every metric to $[0,1]$, with outward indicating better performance.
  Text uses Hard. Emotion averages the Neutral and Intense shard metrics
  equally before normalization.
  The outer ring corresponds to 40\% on the Emotion Acc. and Emotion Success axes.}
  \label{fig:radar-models}
\end{figure*}

We optimize the flow-matching and audio-stop objectives end-to-end while
keeping the AudioVAE and speaker encoder fixed.
We set $\lambda_{\mathrm{FM}}=\lambda_{\mathrm{EOS}}=1$ in \Cref{eq:training-objective}.
  Training
uses bfloat16 arithmetic with FP32 parameters, AdamW~\citep{loshchilov2019adamw} with learning rate
$2.0\times10^{-5}$, moment coefficients $\beta_1=0.9$ and $\beta_2=0.99$,
weight decay $0.01$, and gradient clipping at 2.  The WSD
schedule~\citep{hu2024minicpm} uses 1,000 linear warm-up updates and a
15,000-update linear decay to 15\% of the peak rate.

\paragraph{Training mixture}
TTS replay is sampled from the 1.5M-hour multilingual corpus used by the base
model~\citep{dotstts2026}.  Before the online quality filters in the training
loader, the edit manifests contain approximately 11M text, 10M emotion, 8M
prosody, and 5M pause pairs.
The base sampling weights for TTS:text:emotion:prosody:pause are $24{:}2{:}2{:}1.2{:}0.7$.
A shared multiplier on the edit weights increases linearly from 0.2 at initialization
to 1.0 after 20\% of training and 1.5 after 60\%, then remains fixed. TTS replay maintains
exposure to synthesis examples alongside localized editing supervision.

\paragraph{Augmentation}
For 60\% of edit examples, we concatenate two or three independently sampled
segments, up to 40\,s total, creating longer, compositional, and potentially
multi-speaker contexts that train preservation of each speaker identity.
All concatenated examples drop the global CAM++ speaker embedding.
Noise augmentation at 10--20\,dB SNR supports speech enhancement and
background-noise preservation.  TTS replay is not augmented.

\subsubsection{Evaluation}
\paragraph{Evaluation protocol}
We primarily use doteBench to evaluate how accurately systems execute requested edits while preserving the remaining speech and maintaining audio quality.
One-take is the primary comparison and permits one model call per case,
with the request expressed through each candidate's supported interface.
Sequential execution applies text operations first and then acoustic
operations, using each output as the next input. For each case, both modes start from the same source audio, and the final output is evaluated against the complete editing instruction.
The compositional data pipeline uses multiple passes.
Generation failures remain in metric denominators with fixed penalties.
\suppref{app:aggregation} details metric eligibility, aggregation, and failure
handling. To assess editing on an independently constructed benchmark, we also evaluate on SpeechEditBench~\citep{zhang2026speecheditbench}, which covers a broader range of speech attributes. Its protocol and results are given in \suppref{sec:speecheditbench_results}. To measure retention of zero-shot TTS capability, we use Seed-TTS-Eval~\citep{seedtts2024}, with 1,088 English, 2,020 Chinese, and 400 Chinese Hard examples, following its WER/CER and speaker-similarity protocol.

\paragraph{Baselines}
The open-source model comparison (\cref{fig:radar-models}) includes Step-Audio-EditX~\citep{yan2025stepaudioeditx},
Ming-UniAudio~\citep{yan2025minguniaudio}, MiMo-Audio-Instruct~\citep{xiaomi2025mimoaudio},
FireRedTTS3-Instruct~\citep{shen2026fireredtts3}, and AuK~\citep{ma2026auk}.
AuK and FireRedTTS3-Instruct provide native local text-editing baselines.
Both perform text edits without an external aligner. AuK uses its prompt enhancer (PE), backed by
DeepSeek-v4.1-flash-1017~\citep{deepseekai2026v41flash}, to convert editing requests into native controls.
Adapters follow each model's native editing or generation interface.
In preliminary adapter validation, adding unsupported local-span or
multi-operation instructions could destabilize speech generation and
linguistic content. We therefore use supported controls and simplify
requests when necessary to obtain each system's usable editing output.
\suppref{app:candidate-adaptation} gives the
candidate-specific mappings and operation selection rules.
The tables additionally report an Identity reference that copies the source
audio and the training data pipelines as baselines, grouped separately from learned systems.
For TTS capability retention, we compare with \dotstts{}~\citep{dotstts2026}, Qwen3-TTS~\citep{hu2026qwen3tts}, and VoxCPM2~\citep{zhou2026voxcpm2}.

\begin{table*}[!t]
\begin{minipage}[t]{\columnwidth}
  \vspace{0pt}
  \captionsetup{type=table}
  \centering
  \footnotesize
  \setlength{\tabcolsep}{2pt}
  \caption{Text editing results under one-take execution. Bold and underline mark the best and second-best learned systems within each mode and shard.}
  \label{tab:text-results}
  \resizebox{\columnwidth}{!}{%
  \begin{tabular}{@{}l c c c c c c @{}}
    \toprule
    \systemhead[2] & \tablehead{Edit WER/\\CER (\%)\,$\downarrow$} & \tablehead{Non-edit WER/\\CER (\%)\,$\downarrow$} & \tablehead{WDTW\\Dur (\%)\,$\downarrow$} & \tablehead{WDTW\\F0 (st)\,$\downarrow$} & SpkSim\,$\uparrow$ & UTMOS\,$\uparrow$ \\
    \midrule
    \addlinespace[2pt]
    \multicolumn{7}{l}{\rmfamily\bfseries Shard: Easy} \\
    Identity & 43.68 & 0.21 & 1.45 & 0.16 & 1.000 & 3.37 \\
    Data pipeline & 1.17 & 0.46 & 3.76 & 0.97 & 0.918 & 3.22 \\
    \midrule[0.3pt]
    AuK & 24.34 & \underline{0.85} & \textbf{5.65} & \textbf{1.46} & \textbf{0.902} & 3.36 \\
    \addlinespace[1pt]
    Step-Audio-EditX & 15.38 & 1.71 & 7.92 & 2.88 & 0.768 & \textbf{3.68} \\
    Ming-UniAudio & 40.13 & 7.33 & 9.30 & 2.18 & 0.777 & 3.11 \\
    MiMo-Audio-Instruct & \underline{5.85} & 5.62 & 13.05 & 4.12 & 0.636 & 3.27 \\
    FireRedTTS3-Instruct & 26.81 & 0.96 & 7.09 & \underline{1.77} & 0.820 & 3.17 \\
    \rowcolor{blue!5}
    \dotsttsedit{} & \textbf{1.74} & \textbf{0.78} & \underline{6.51} & 2.12 & \underline{0.852} & \underline{3.39} \\
    \addlinespace[3pt]\midrule[0.65pt]
    \multicolumn{7}{l}{\rmfamily\bfseries Shard: Hard} \\
    Identity & 178.52 & 0.17 & 7.75 & 0.80 & 1.000 & 3.49 \\
    Data pipeline & 24.87 & 0.25 & 6.84 & 1.32 & 0.766 & 3.11 \\
    \midrule[0.3pt]
    AuK & 103.67 & 1.53 & 11.60 & \underline{1.94} & \textbf{0.870} & 3.42 \\
    \addlinespace[1pt]
    Step-Audio-EditX & \underline{19.47} & \textbf{0.85} & \underline{11.56} & 2.94 & 0.732 & \textbf{3.84} \\
    Ming-UniAudio & 145.97 & 13.14 & 18.08 & 2.82 & 0.725 & 2.67 \\
    MiMo-Audio-Instruct & 19.56 & 9.07 & 14.34 & 4.35 & 0.640 & 3.24 \\
    FireRedTTS3-Instruct & 152.22 & 1.19 & 12.55 & 2.01 & \underline{0.862} & 3.12 \\
    \rowcolor{blue!5}
    \dotsttsedit{} & \textbf{14.45} & \textbf{0.85} & \textbf{7.57} & \textbf{1.86} & 0.771 & \underline{3.49} \\
    \bottomrule
  \end{tabular}
  }
\end{minipage}
\hfill
\begin{minipage}[t]{\columnwidth}
  \vspace{0pt}
  \captionsetup{type=table}
  \centering
  \footnotesize
  \setlength{\tabcolsep}{2pt}
  \caption{Emotion editing results under one-take execution.}
  \label{tab:emotion-results}
  \resizebox{\columnwidth}{!}{%
  \begin{tabular}{@{}l c c c c c c @{}}
    \toprule
    \systemhead[2] & \tablehead{Emotion\\Acc. (\%)\,$\uparrow$} & \tablehead{WER/\\CER (\%)\,$\downarrow$} & \tablehead{WDTW\\Dur (\%)\,$\downarrow$} & \tablehead{WDTW\\F0 (st)\,$\downarrow$} & SpkSim\,$\uparrow$ & UTMOS\,$\uparrow$ \\
    \midrule
    \addlinespace[2pt]
    \multicolumn{7}{l}{\rmfamily\bfseries Shard: Neutral} \\
    Identity & 20.43 & 2.76 & 0.00 & 0.00 & 1.000 & 3.07 \\
    Data pipeline & 36.78 & 3.14 & 3.19 & 0.74 & 0.870 & 2.77 \\
    \midrule[0.3pt]
    AuK & \underline{27.40} & 5.62 & 9.87 & 4.84 & \underline{0.728} & \underline{3.18} \\
    \addlinespace[1pt]
    Step-Audio-EditX & 23.08 & \textbf{2.46} & 11.28 & 3.58 & 0.701 & \textbf{3.54} \\
    Ming-UniAudio & 25.72 & 6.82 & 12.73 & 6.68 & 0.325 & 2.86 \\
    MiMo-Audio-Instruct & 23.08 & 4.59 & 13.49 & 4.88 & 0.644 & 2.95 \\
    FireRedTTS3-Instruct & 20.43 & 3.70 & \textbf{5.65} & \textbf{1.55} & \textbf{0.906} & 2.92 \\
    \rowcolor{blue!5}
    \dotsttsedit{} & \textbf{37.98} & \underline{3.19} & \underline{6.96} & \underline{2.09} & 0.657 & 3.00 \\
    \addlinespace[3pt]\midrule[0.65pt]
    \multicolumn{7}{l}{\rmfamily\bfseries Shard: Intense} \\
    Identity & 3.50 & 0.62 & 0.00 & 0.00 & 1.000 & 3.11 \\
    Data pipeline & 9.00 & 0.80 & 3.06 & 0.60 & 0.905 & 2.75 \\
    \midrule[0.3pt]
    AuK & \underline{6.25} & 0.96 & 10.31 & 4.65 & \underline{0.795} & \underline{3.18} \\
    \addlinespace[1pt]
    Step-Audio-EditX & 4.00 & \textbf{0.63} & 11.28 & 3.48 & 0.780 & \textbf{3.39} \\
    Ming-UniAudio & 4.00 & 0.94 & 12.22 & 6.64 & 0.375 & 2.74 \\
    MiMo-Audio-Instruct & 4.25 & 3.43 & 12.56 & 4.49 & 0.740 & 3.07 \\
    FireRedTTS3-Instruct & 3.25 & 0.68 & \textbf{4.86} & \textbf{1.14} & \textbf{0.918} & 2.94 \\
    \rowcolor{blue!5}
    \dotsttsedit{} & \textbf{8.50} & \underline{0.67} & \underline{7.46} & \underline{1.63} & 0.709 & 2.90 \\
    \bottomrule
  \end{tabular}
  }
\end{minipage}

\ifreportversion\else
\par\medskip
\ifreportversion
\begin{table}[!tbp]
\else
\begin{minipage}[t]{\columnwidth}
  \vspace{0pt}
  \captionsetup{type=table}
\fi
  \centering
  \footnotesize
  \setlength{\tabcolsep}{2pt}
  \caption{Prosody editing results under one-take execution.}
  \label{tab:prosody-results}
  \resizebox{\columnwidth}{!}{%
  \begin{tabular}{@{}l c c c c c c c @{}}
    \toprule
    \systemhead[2] & \tablehead{Duration\\L1 (s)\,$\downarrow$} & \tablehead{Pitch\\L1 (st)\,$\downarrow$} & \tablehead{WER/\\CER (\%)\,$\downarrow$} & \tablehead{WDTW\\Dur (\%)\,$\downarrow$} & \tablehead{WDTW\\F0 (st)\,$\downarrow$} & SpkSim\,$\uparrow$ & UTMOS\,$\uparrow$ \\
    \midrule
    Identity & 0.212 & 4.39 & 1.65 & 0.01 & 0.00 & 1.000 & 3.45 \\
    Data pipeline & 0.045 & 1.53 & 2.07 & 3.39 & 0.93 & 0.901 & 3.15 \\
    \midrule[0.3pt]
    AuK & 0.240 & 3.83 & 3.46 & \underline{9.06} & \underline{2.23} & 0.632 & 2.03 \\
    \addlinespace[1pt]
    Step-Audio-EditX & 0.371 & 4.53 & 34.87 & 22.86 & 3.27 & \underline{0.710} & \textbf{3.94} \\
    Ming-UniAudio & \underline{0.201} & 4.39 & 7.36 & 14.43 & 3.53 & 0.399 & 1.31 \\
    MiMo-Audio-Instruct & 0.304 & 5.18 & 8.31 & 15.59 & 3.98 & 0.623 & 3.21 \\
    FireRedTTS3-Instruct & 0.218 & \underline{3.75} & \underline{2.32} & 9.11 & 2.76 & 0.562 & 2.44 \\
    \rowcolor{blue!5}
    \dotsttsedit{} & \textbf{0.054} & \textbf{1.51} & \textbf{2.04} & \textbf{6.00} & \textbf{1.57} & \textbf{0.846} & \underline{3.36} \\
    \bottomrule
  \end{tabular}
  }
\ifreportversion
\end{table}
\else
\end{minipage}
\fi
\hfill
\ifreportversion
\begin{table}[!tbp]
\else
\begin{minipage}[t]{\columnwidth}
  \vspace{0pt}
  \captionsetup{type=table}
\fi
  \centering
  \footnotesize
  \setlength{\tabcolsep}{2pt}
  \caption{Pause editing results under one-take execution.}
  \label{tab:pause-results}
  \resizebox{\columnwidth}{!}{%
  \begin{tabular}{@{}l c c c c c c @{}}
    \toprule
    \systemhead[2] & \tablehead{Direction\\Acc. (\%)\,$\uparrow$} & \tablehead{WER/\\CER (\%)\,$\downarrow$} & \tablehead{WDTW\\Dur (\%)\,$\downarrow$} & \tablehead{WDTW\\F0 (st)\,$\downarrow$} & SpkSim\,$\uparrow$ & UTMOS\,$\uparrow$ \\
    \midrule
    Identity & 0.00 & 2.30 & 0.00 & 0.00 & 1.000 & 3.67 \\
    Data pipeline & 92.83 & 3.96 & 7.19 & 1.10 & 0.971 & 3.53 \\
    \midrule[0.3pt]
    AuK & 36.67 & 4.19 & \underline{6.57} & \underline{1.39} & \textbf{0.915} & 3.49 \\
    \addlinespace[1pt]
    Step-Audio-EditX & \underline{51.17} & 3.40 & 10.17 & 2.79 & 0.789 & \textbf{3.82} \\
    Ming-UniAudio & 5.33 & \textbf{2.87} & 6.74 & 1.54 & 0.848 & 3.54 \\
    MiMo-Audio-Instruct & 42.83 & 10.23 & 14.97 & 4.32 & 0.672 & 3.29 \\
    FireRedTTS3-Instruct & 0.67 & \underline{2.89} & \textbf{6.05} & \textbf{1.23} & 0.897 & 3.47 \\
    \rowcolor{blue!5}
    \dotsttsedit{} & \textbf{92.00} & 3.53 & 8.03 & 1.66 & \underline{0.901} & \underline{3.58} \\
    \bottomrule
  \end{tabular}
  }
\ifreportversion
\end{table}
\else
\end{minipage}
\fi

\fi
\end{table*}

\ifreportversion

\begin{table*}[!t]
  \centering
  \footnotesize
  \setlength{\tabcolsep}{3pt}
  \caption{Compositional editing results. Component and All report operation-level and case-level success rates using category-specific criteria. The pipeline baseline uses multiple passes.}
  \label{tab:compositional-results}
  \resizebox{\textwidth}{!}{%
  \begin{tabular}{l c c c c c c c c c c c }
    \toprule
    \systemhead[2] & \tablehead{Component\\(\%)\,$\uparrow$} & \tablehead{All\\(\%)\,$\uparrow$} & \tablehead{Text\\(\%)\,$\uparrow$} & \tablehead{Emotion\\(\%)\,$\uparrow$} & \tablehead{Prosody\\(\%)\,$\uparrow$} & \tablehead{Pause\\(\%)\,$\uparrow$} & \tablehead{WER/CER\\(\%)\,$\downarrow$} & \tablehead{WDTW\\Dur (\%)\,$\downarrow$} & \tablehead{WDTW\\F0 (st)\,$\downarrow$} & SpkSim\,$\uparrow$ & UTMOS\,$\uparrow$ \\
    \midrule
    Identity (source audio) & 3.62 & 0.00 & 0.00 & 13.04 & 0.00 & 1.45 & 8.58 & 2.25 & 0.23 & 1.000 & 3.54 \\
    \midrule[0.3pt]
    \multicolumn{12}{l}{\rmfamily\itshape Execution: One-take} \\
    AuK & 29.53 & 2.08 & 24.64 & \textbf{21.01} & \underline{42.75} & 29.71 & 7.02 & 9.24 & 2.77 & \underline{0.820} & 3.01 \\
    \addlinespace[1pt]
    Step-Audio-EditX & 28.08 & 7.50 & 53.62 & 16.67 & 6.52 & \underline{35.51} & 7.04 & 12.66 & 3.29 & 0.779 & \textbf{3.76} \\
    Ming-UniAudio & 17.75 & 1.67 & 0.00 & 18.84 & 36.96 & 15.22 & 12.00 & 13.12 & 4.65 & 0.573 & 2.42 \\
    MiMo-Audio-Instruct & \underline{31.70} & \underline{9.58} & \underline{69.57} & 18.12 & 10.87 & 28.26 & 12.80 & 17.64 & 4.50 & 0.713 & 3.23 \\
    FireRedTTS3-Instruct & 19.20 & 1.67 & 22.46 & 15.22 & 34.78 & 4.35 & \underline{6.71} & \underline{8.44} & \underline{1.94} & \underline{0.820} & 3.11 \\
    \rowcolor{blue!5}
    \dotsttsedit{} & \textbf{68.12} & \textbf{37.92} & \textbf{77.54} & \textbf{21.01} & \textbf{83.33} & \textbf{90.58} & \textbf{2.73} & \textbf{6.88} & \textbf{1.80} & \textbf{0.908} & \underline{3.37} \\
    \midrule[0.3pt]
    \multicolumn{12}{l}{\rmfamily\itshape Execution: Sequential} \\
    Data pipeline & 72.10 & 45.83 & 78.99 & 31.88 & 84.78 & 92.75 & 2.88 & 5.00 & 1.02 & 0.953 & 3.24 \\
    \candidateseparator{12}
    AuK & \underline{43.66} & \underline{18.33} & 42.03 & \underline{21.01} & 57.97 & \underline{53.62} & 6.13 & 11.80 & 4.28 & 0.663 & 2.73 \\
    \addlinespace[1pt]
    Step-Audio-EditX & 34.60 & 7.50 & \underline{62.32} & 19.57 & 10.14 & 46.38 & 7.18 & 13.58 & 4.03 & \underline{0.672} & \textbf{3.73} \\
    Ming-UniAudio & 24.46 & 5.83 & 23.19 & 17.39 & 32.61 & 24.64 & 18.23 & 16.95 & 6.30 & 0.410 & 2.06 \\
    MiMo-Audio-Instruct & 30.43 & 7.50 & 60.87 & 18.12 & 13.04 & 29.71 & 12.90 & 17.98 & 5.35 & 0.605 & 3.20 \\
    FireRedTTS3-Instruct & 35.14 & 9.17 & 51.45 & 15.22 & \underline{65.22} & 8.70 & \underline{4.41} & \underline{9.82} & \underline{3.04} & 0.657 & 2.74 \\
    \rowcolor{blue!5}
    \dotsttsedit{} & \textbf{68.48} & \textbf{42.92} & \textbf{74.64} & \textbf{33.33} & \textbf{77.54} & \textbf{88.41} & \textbf{2.96} & \textbf{7.29} & \textbf{1.98} & \textbf{0.839} & \underline{3.29} \\
    \bottomrule
  \end{tabular}
  }
\end{table*}

\ifreportversion
\begin{table}[!tbp]
\else
\begin{minipage}[t]{\columnwidth}
  \vspace{0pt}
  \captionsetup{type=table}
\fi
  \centering
  \footnotesize
  \setlength{\tabcolsep}{2pt}
  \caption{Prosody editing results under one-take execution.}
  \label{tab:prosody-results}
  \resizebox{\columnwidth}{!}{%
  \begin{tabular}{@{}l c c c c c c c @{}}
    \toprule
    \systemhead[2] & \tablehead{Duration\\L1 (s)\,$\downarrow$} & \tablehead{Pitch\\L1 (st)\,$\downarrow$} & \tablehead{WER/\\CER (\%)\,$\downarrow$} & \tablehead{WDTW\\Dur (\%)\,$\downarrow$} & \tablehead{WDTW\\F0 (st)\,$\downarrow$} & SpkSim\,$\uparrow$ & UTMOS\,$\uparrow$ \\
    \midrule
    Identity & 0.212 & 4.39 & 1.65 & 0.01 & 0.00 & 1.000 & 3.45 \\
    Data pipeline & 0.045 & 1.53 & 2.07 & 3.39 & 0.93 & 0.901 & 3.15 \\
    \midrule[0.3pt]
    AuK & 0.240 & 3.83 & 3.46 & \underline{9.06} & \underline{2.23} & 0.632 & 2.03 \\
    \addlinespace[1pt]
    Step-Audio-EditX & 0.371 & 4.53 & 34.87 & 22.86 & 3.27 & \underline{0.710} & \textbf{3.94} \\
    Ming-UniAudio & \underline{0.201} & 4.39 & 7.36 & 14.43 & 3.53 & 0.399 & 1.31 \\
    MiMo-Audio-Instruct & 0.304 & 5.18 & 8.31 & 15.59 & 3.98 & 0.623 & 3.21 \\
    FireRedTTS3-Instruct & 0.218 & \underline{3.75} & \underline{2.32} & 9.11 & 2.76 & 0.562 & 2.44 \\
    \rowcolor{blue!5}
    \dotsttsedit{} & \textbf{0.054} & \textbf{1.51} & \textbf{2.04} & \textbf{6.00} & \textbf{1.57} & \textbf{0.846} & \underline{3.36} \\
    \bottomrule
  \end{tabular}
  }
\ifreportversion
\end{table}
\else
\end{minipage}
\fi

\ifreportversion
\begin{table}[!tbp]
\else
\begin{minipage}[t]{\columnwidth}
  \vspace{0pt}
  \captionsetup{type=table}
\fi
  \centering
  \footnotesize
  \setlength{\tabcolsep}{2pt}
  \caption{Pause editing results under one-take execution.}
  \label{tab:pause-results}
  \resizebox{\columnwidth}{!}{%
  \begin{tabular}{@{}l c c c c c c @{}}
    \toprule
    \systemhead[2] & \tablehead{Direction\\Acc. (\%)\,$\uparrow$} & \tablehead{WER/\\CER (\%)\,$\downarrow$} & \tablehead{WDTW\\Dur (\%)\,$\downarrow$} & \tablehead{WDTW\\F0 (st)\,$\downarrow$} & SpkSim\,$\uparrow$ & UTMOS\,$\uparrow$ \\
    \midrule
    Identity & 0.00 & 2.30 & 0.00 & 0.00 & 1.000 & 3.67 \\
    Data pipeline & 92.83 & 3.96 & 7.19 & 1.10 & 0.971 & 3.53 \\
    \midrule[0.3pt]
    AuK & 36.67 & 4.19 & \underline{6.57} & \underline{1.39} & \textbf{0.915} & 3.49 \\
    \addlinespace[1pt]
    Step-Audio-EditX & \underline{51.17} & 3.40 & 10.17 & 2.79 & 0.789 & \textbf{3.82} \\
    Ming-UniAudio & 5.33 & \textbf{2.87} & 6.74 & 1.54 & 0.848 & 3.54 \\
    MiMo-Audio-Instruct & 42.83 & 10.23 & 14.97 & 4.32 & 0.672 & 3.29 \\
    FireRedTTS3-Instruct & 0.67 & \underline{2.89} & \textbf{6.05} & \textbf{1.23} & 0.897 & 3.47 \\
    \rowcolor{blue!5}
    \dotsttsedit{} & \textbf{92.00} & 3.53 & 8.03 & 1.66 & \underline{0.901} & \underline{3.58} \\
    \bottomrule
  \end{tabular}
  }
\ifreportversion
\end{table}
\else
\end{minipage}
\fi

\fi

\paragraph{Inference parameters}
At inference time, \dotsttsedit{} uses Euler integration with 10 flow steps,
a classifier-free guidance~\mbox{\citep{ho2022classifierfree}} scale of
1.2, speaker guidance of 1.5, bfloat16 inference,
and a maximum generation length of 500 latent steps.
Speaker guidance is enabled for every evaluation category except Emotion.

\subsection{doteBench Results}
\label{sec:radar_results}

The radar plot summarizes the performance of speech editing systems
under one-take execution. Sequential single-task results and the corresponding radar plot are provided in
\suppref{app:sequential-results}. In the tables reporting results on doteBench, bold
and underline mark the best and second-best learned systems within each
execution mode and shard. Across the five doteBench categories,
\dotsttsedit{} combines leading instruction execution with consistent local
preservation and audio quality among the evaluated learned systems. Its
advantage is the ability to carry out different precisely localized edits,
including their compositions, within one model and one generation pass.

\paragraph{Single-task editing}
On Text Hard (\cref{tab:text-results}), \dotsttsedit{} obtains
the lowest edited-region WER/CER among the evaluated learned systems at
14.45\%, with non-edit WER/CER of 0.85\%. It also achieves the lowest
WDTW-Dur and WDTW-F0. These results show that the model realizes difficult
transcript changes while preserving surrounding speech content, timing, and pitch.

On emotion editing (\cref{tab:emotion-results}), \dotsttsedit{} reaches 37.98\%
accuracy on Neutral and 8.50\% on Intense, the highest among the evaluated learned
systems on both shards. It combines these gains with low recognition error and strong
timing and F0 preservation. Changing an already emotional source remains challenging
across the evaluated systems, but \dotsttsedit{} still provides the strongest
target-emotion matching on Intense.

\ifreportversion\else

\fi

On prosody editing (\cref{tab:prosody-results}), \dotsttsedit{} achieves the lowest
duration and pitch errors, at 0.054 s and 1.51 semitones. Its WDTW-Dur and
WDTW-F0 are 6.00\% and 1.57 semitones, respectively. The model therefore improves
control of the requested rate and pitch changes while preserving the timing and pitch of
speech outside those targets.

On pause editing (\cref{tab:pause-results}), \dotsttsedit{} reaches 92.00\%
direction accuracy, compared with 51.17\% for the best open-source baseline. These
results show that the model can modify a specified pause boundary while retaining voice
and overall acoustic quality.

\paragraph{Compositional editing}
In one-take compositional editing (\cref{tab:compositional-results}), \dotsttsedit{}
reaches 68.12\% component success and 37.92\% all-operation success, compared with
31.70\% and 9.58\% for the best open-source baseline. It also achieves the best or joint-best success rate among the evaluated learned systems in every operation family and has the lowest recognition error. This
advantage across both individual operations and complete requests demonstrates the
model's ability to combine precisely scoped controls in one pass.
Sequential execution increases all-operation success from 37.92\% to 42.92\%, with
emotion success rising from 21.01\% to 33.33\%. One-take execution retains higher
speaker similarity (0.908 versus 0.839) and UTMOS (3.37 versus 3.29) while completing
the request in a single generation pass. The two settings show that the unified speech editing model
supports both direct composition and operation-wise execution, with one-take editing
favoring preservation and deployment efficiency.

The data pipelines offer a reference for precise local expert
operations: each isolates the requested region, applies a task-specific
method, and retains the remaining waveform. As expected, this gives them
strong instruction-following and preservation scores. Their compositional
baseline reaches 72.10\% component success and 45.83\% all-operation success
through multiple passes. In contrast, \dotsttsedit{} handles difficult Text
Hard target transcripts more reliably than the text pipeline (14.45\% versus
24.87\% edit error) and achieves higher UTMOS across the five categories.
It performs the complete edit with one model and one generation pass,
retaining a simpler deployment path for combined controls.

\subsection{Human Listening Evaluation}
\label{sec:listening}

\ifreportversion\else
\begin{table}[!t]
\centering
\caption{Objective--subjective directional agreement. $n$ denotes the number of valid comparison pairs.}
\label{tab:listening-agreement}
\footnotesize
\setlength{\tabcolsep}{4pt}
\renewcommand{\tabularxcolumn}[1]{m{#1}}
\begin{tabularx}{\columnwidth}{lXcc}
\toprule
Dimension & Objective metric & Agreement (\%) & $n$ \\
\midrule
\multirow[c]{5}{*}{IF} & WER/CER (\%) & 94.64 & 14 \\
 & Pitch error & 86.36 & 11 \\
 & Duration error & 100.00 & 11 \\
 & Pause direction & 91.67 & 10 \\
 & Emotion accuracy & 75.00 & 7 \\
\midrule
\multirow[c]{3}{*}{LP} & WDTW-Dur & 77.88 & 52 \\
 & WDTW-F0 & 82.69 & 52 \\
 & Non-edit WER/CER (\%) & 62.50 & 16 \\
\midrule
AQ & UTMOS & 80.48 & 70 \\
\bottomrule
\end{tabularx}
\end{table}

\begin{table}[!t]
\centering
\caption{Mean system preferences against sampled opponents
(\textminus{}3 to +3, higher is better). Bold and underline mark the highest and second-highest means.}
\label{tab:listening-systems}
\footnotesize
\setlength{\tabcolsep}{5pt}
\begin{tabularx}{\columnwidth}{Xccc}
\toprule
System & IF & LP & AQ \\
\midrule
AuK & \textminus{}0.671 & \underline{0.463} & \textminus{}0.418 \\
FireRedTTS3-Instruct & \textminus{}0.774 & 0.244 & \textminus{}0.571 \\
MiMo-Audio-Instruct & \underline{0.222} & \textminus{}0.619 & 0.120 \\
Ming-UniAudio & \textminus{}1.279 & \textminus{}1.292 & \textminus{}1.597 \\
Step-Audio-EditX & \textminus{}0.208 & \textminus{}0.106 & \textbf{1.010} \\
\rowcolor{blue!5}
\dotsttsedit{} & \textbf{1.538} & \textbf{0.735} & \underline{0.598} \\
\bottomrule
\end{tabularx}
\end{table}

\fi

We assessed agreement between doteBench metrics and human judgments,
following the core paired-comparison framework of ITU-R BS.1284
{\citep{itu2019bs1284}} and ITU-T P.808 quality-control principles for
equipment checks and listener screening {\citep{itu2021p808}}.
The study involved 24 bilingual volunteers, who evaluated 70 cases across text, emotion, pitch, rate, and pause editing, with four listeners per case. Each completed 10--12 comparisons
in approximately 10--15 minutes. After practice, listeners rated IF, LP,
and AQ from $-3$ (A much better) to $+3$ (B much better), with 0 indicating
a tie. {\suppref{app:listening-protocol}} gives prompts and scoring details.

{\Cref{tab:listening-agreement}} reports directional agreement.
For each valid comparison pair, we compute the fraction of valid directional
listener preferences favoring the output with the better objective score,
then average these fractions equally across pairs. WDTW-Dur reaches 77.88\%
(95\% CI: 66.83--88.46\%) and WDTW-F0 reaches 82.69\%
(72.12--92.31\%), supporting their use in assessing perceptual preservation
of timing and pitch.
Among the learned systems,
{\cref{tab:listening-systems}} shows that {\dotsttsedit{}} has the highest
mean IF and LP preferences in the sampled comparisons, and the second-highest mean AQ preference.

\ifreportversion
\begin{table}[!t]
\centering
\caption{Objective--subjective directional agreement. $n$ denotes the number of valid comparison pairs.}
\label{tab:listening-agreement}
\footnotesize
\setlength{\tabcolsep}{4pt}
\renewcommand{\tabularxcolumn}[1]{m{#1}}
\begin{tabularx}{\columnwidth}{lXcc}
\toprule
Dimension & Objective metric & Agreement (\%) & $n$ \\
\midrule
\multirow[c]{5}{*}{IF} & WER/CER (\%) & 94.64 & 14 \\
 & Pitch error & 86.36 & 11 \\
 & Duration error & 100.00 & 11 \\
 & Pause direction & 91.67 & 10 \\
 & Emotion accuracy & 75.00 & 7 \\
\midrule
\multirow[c]{3}{*}{LP} & WDTW-Dur & 77.88 & 52 \\
 & WDTW-F0 & 82.69 & 52 \\
 & Non-edit WER/CER (\%) & 62.50 & 16 \\
\midrule
AQ & UTMOS & 80.48 & 70 \\
\bottomrule
\end{tabularx}
\end{table}

\begin{table}[!t]
\centering
\caption{Mean system preferences against sampled opponents
(\textminus{}3 to +3, higher is better). Bold and underline mark the highest and second-highest means.}
\label{tab:listening-systems}
\footnotesize
\setlength{\tabcolsep}{5pt}
\begin{tabularx}{\columnwidth}{Xccc}
\toprule
System & IF & LP & AQ \\
\midrule
AuK & \textminus{}0.671 & \underline{0.463} & \textminus{}0.418 \\
FireRedTTS3-Instruct & \textminus{}0.774 & 0.244 & \textminus{}0.571 \\
MiMo-Audio-Instruct & \underline{0.222} & \textminus{}0.619 & 0.120 \\
Ming-UniAudio & \textminus{}1.279 & \textminus{}1.292 & \textminus{}1.597 \\
Step-Audio-EditX & \textminus{}0.208 & \textminus{}0.106 & \textbf{1.010} \\
\rowcolor{blue!5}
\dotsttsedit{} & \textbf{1.538} & \textbf{0.735} & \underline{0.598} \\
\bottomrule
\end{tabularx}
\end{table}

\fi

\subsection{TTS Capability Retention}
\label{sec:tts_retention}

\cref{tab:seed-tts-retention} shows that editing post-training largely retains the
foundation model's zero-shot TTS capability, with negligible differences in WER/CER and
speaker similarity relative to the pretrained \dotstts{} model.

\begin{table}[!t]
  \centering
  \footnotesize
  \setlength{\tabcolsep}{2pt}
  \caption{Retention of zero-shot TTS capability on Seed-TTS-Eval.}
  \label{tab:seed-tts-retention}
  \resizebox{\columnwidth}{!}{\begin{tabular}{@{}l cc cc cc@{}}
    \toprule
    & \multicolumn{2}{c}{en} & \multicolumn{2}{c}{zh} & \multicolumn{2}{c}{zh-hard} \\
    \cmidrule(lr){2-3}\cmidrule(lr){4-5}\cmidrule(lr){6-7}
    System & WER (\%)\,$\downarrow$ & SIM\,$\uparrow$ & CER (\%)\,$\downarrow$ & SIM\,$\uparrow$ & CER (\%)\,$\downarrow$ & SIM\,$\uparrow$ \\
    \midrule
    Qwen3-TTS 1.7B & \textbf{1.23} & 0.717 & 1.22 & 0.770 & 6.76 & 0.748 \\
    VoxCPM2 2B & 1.84 & 0.753 & \underline{0.97} & 0.795 & 8.13 & 0.753 \\
    \midrule
    \dotstts{} & 1.34 & \textbf{0.768} & \textbf{0.96} & \textbf{0.805} & \textbf{6.46} & \underline{0.792} \\
    \candidateseparator{7}
    \rowcolor{blue!5}
    \dotsttsedit{} & \underline{1.26} & \underline{0.764} & \underline{0.97} & \textbf{0.805} & \underline{6.48} & \textbf{0.794} \\
    \bottomrule
  \end{tabular}}
\end{table}

\subsection{Ablation Studies}
\label{sec:ablations}

\paragraph{Editing versus zero-shot TTS}
\cref{tab:edit-zero-shot-ablation} compares source-conditioned editing with
full-utterance zero-shot TTS on the same Text Hard cases. The Editing row uses the same outputs as the main Text Hard comparison, with WER/CER computed over the full target transcript. Editing reduces target WER/CER
from 13.39\% to 12.50\%, WDTW-Dur from 9.38\% to 7.57\%, and WDTW-F0 from 2.79 to 1.86
semitones. Thus, conditioning on the source improves both realization of the target
transcript and preservation of its local timing and pitch, supporting a dedicated
editing mode for precisely localized changes.

\begin{table}[!t]
  \centering
  \footnotesize
  \setlength{\tabcolsep}{2pt}
  \caption{Source-conditioned editing versus full-utterance zero-shot TTS on Text Hard. WER/CER measures the target transcript of the entire utterance.}
  \label{tab:edit-zero-shot-ablation}
  \resizebox{\columnwidth}{!}{%
  \begin{tabular}{@{}l c c c c c @{}}
    \toprule
    \systemhead[2] & \tablehead{WER/\\CER (\%)\,$\downarrow$} & \tablehead{WDTW\\Dur (\%)\,$\downarrow$} & \tablehead{WDTW\\F0 (st)\,$\downarrow$} & SpkSim\,$\uparrow$ & UTMOS\,$\uparrow$ \\
    \midrule
    Editing & \textbf{12.50} & \textbf{7.57} & \textbf{1.86} & \underline{0.771} & \underline{3.49} \\
    Zero-shot TTS & \underline{13.39} & \underline{9.38} & \underline{2.79} & \textbf{0.818} & \textbf{3.65} \\
    \bottomrule
  \end{tabular}
  }
\end{table}

\paragraph{Instruction representation}
We next examine how instruction representation affects the model's ability to
realize the requested edits.
We compare the semi-formal transcript-grounded edit instruction
against a canonical bilingual natural-language rendering that preserves the same
operations, parameters, and target locations. Initialization, training data,
sampling, optimization, seed, and update budgets are matched; the details of
constructing equivalent natural-language instructions are given in \suppref{app:instruction-ablations}. In \cref{tab:report-ablation-summary}, XML lowers Text Hard edit error, prosody duration error, and pitch error, and raises Pause
accuracy from 69.33\% to 92.00\%. Compositional component/all-operation success
increases from 55.80\%/26.67\% to 68.12\%/37.92\%, with similar speaker similarity
and UTMOS. These results support explicit operation-to-location binding,
particularly when a request combines multiple edits.
\ifreportversion

\begin{table}[!t]
  \centering\footnotesize
  \setlength{\tabcolsep}{2pt}
  \caption{Instruction and category-expert ablation on the principal execution metrics. Bold and underline mark the best and second-best results. Text uses the Hard shard and Emotion uses the Intense shard. Single-category results come from the corresponding category experts. All settings retain the same proportion of TTS replay.}
  \label{tab:report-ablation-summary}
  \resizebox{\columnwidth}{!}{\begin{tabular}{@{}ll *{7}{c}@{}}
    \toprule
    & & Text & Emotion & \multicolumn{2}{c}{Prosody} & Pause & \multicolumn{2}{c}{Compositional} \\
    \cmidrule[0.8pt](lr){3-3}\cmidrule[0.8pt](lr){4-4}\cmidrule[0.8pt](lr){5-6}\cmidrule[0.8pt](lr){7-7}\cmidrule[0.8pt](lr){8-9}
    \tablehead{Instruction} & \tablehead{Editing\\data} & \tablehead{Edit WER/\\CER (\%)\,$\downarrow$} & \tablehead{Emotion\\Acc. (\%)\,$\uparrow$} & \tablehead{Duration\\L1 (s)\,$\downarrow$} & \tablehead{Pitch\\L1 (st)\,$\downarrow$} & \tablehead{Direction\\Acc. (\%)\,$\uparrow$} & \tablehead{Component\\(\%)\,$\uparrow$} & \tablehead{All\\(\%)\,$\uparrow$} \\
    \midrule
    \rowcolor{black!4}
\begin{tabular}[c]{@{}l@{}}XML\end{tabular} & \begin{tabular}[c]{@{}l@{}}All\\categories\end{tabular} & \begin{tabular}[c]{@{}c@{}}\textbf{14.45}\end{tabular} & \begin{tabular}[c]{@{}c@{}}\textbf{8.50}\end{tabular} & \begin{tabular}[c]{@{}c@{}}\underline{0.054}\end{tabular} & \begin{tabular}[c]{@{}c@{}}\underline{1.51}\end{tabular} & \begin{tabular}[c]{@{}c@{}}\textbf{92.00}\end{tabular} & \begin{tabular}[c]{@{}c@{}}\textbf{68.12}\end{tabular} & \begin{tabular}[c]{@{}c@{}}\textbf{37.92}\end{tabular} \\
    \addlinespace[4pt]
\begin{tabular}[c]{@{}l@{}}\shortstack[l]{Natural\\[1pt]language}\end{tabular} & \begin{tabular}[c]{@{}l@{}}All\\categories\end{tabular} & \begin{tabular}[c]{@{}c@{}}\underline{14.73}\end{tabular} & \begin{tabular}[c]{@{}c@{}}\textbf{8.50}\end{tabular} & \begin{tabular}[c]{@{}c@{}}0.056\end{tabular} & \begin{tabular}[c]{@{}c@{}}1.66\end{tabular} & \begin{tabular}[c]{@{}c@{}}69.33\end{tabular} & \begin{tabular}[c]{@{}c@{}}\underline{55.80}\end{tabular} & \begin{tabular}[c]{@{}c@{}}\underline{26.67}\end{tabular} \\
    \addlinespace[4pt]
    \rowcolor{black!4}
\begin{tabular}[c]{@{}l@{}}XML\end{tabular} & \begin{tabular}[c]{@{}l@{}}Single\\category\end{tabular} & \begin{tabular}[c]{@{}c@{}}15.42\end{tabular} & \begin{tabular}[c]{@{}c@{}}\textbf{8.50}\end{tabular} & \begin{tabular}[c]{@{}c@{}}\textbf{0.049}\end{tabular} & \begin{tabular}[c]{@{}c@{}}\textbf{1.36}\end{tabular} & \begin{tabular}[c]{@{}c@{}}\underline{77.17}\end{tabular} & \begin{tabular}[c]{@{}c@{}}---\end{tabular} & \begin{tabular}[c]{@{}c@{}}---\end{tabular} \\
    \bottomrule
  \end{tabular}}
\end{table}

\fi

\paragraph{Unified and category-specific training}
We further examine whether a unified model can match category-specific models
on individual editing tasks.
We train category experts by concentrating editing samples on a single category while keeping
the proportion of TTS replay unchanged. The unified model reduces
Text Hard edit error from 15.42\% to 14.45\% and raises Pause accuracy from 77.17\% to
92.00\%. It exceeds the Emotion expert's accuracy on Neutral and matches it on Intense while better
preserving content, timing, and pitch. On the prosody editing task, the unified model better preserves content, timing, pitch, and speaker identity while achieving comparable pitch and duration errors to the Prosody expert. Unlike the category experts, the unified model also supports compositional editing across categories in a single pass. These results support a single speech editing model that combines accurate control with
local preservation across categories.

\ifreportversion\else

\fi

\section{Conclusion}
\label{sec:conclusion}

We presented \dotsttsedit{}, a precisely controlled speech editing model built on a continuous
autoregressive TTS model. Its semi-formal transcript-grounded edit instruction combines explicit operation specifications and transcript-defined scopes, serving as a common interface for text, emotion, prosody, and pause editing.
Task-specific data pipelines provide operation- and scope-controlled supervision, and
doteBench evaluates requested changes together with preservation and audio quality.
Across its five categories, \dotsttsedit{} combines leading instruction following among
the evaluated learned systems with consistent local preservation and audio quality.
 The instruction comparison favors the XML form rather than the natural-language form for compositional execution.
Human subjective listening tests further favor \dotsttsedit{} in instruction following and local
preservation.
Editing post-training also retains the base model's zero-shot TTS capability.
These results establish a practical foundation for precise individual and compositional
edits through an explicit, inspectable interface.

\ifreportversion
\onecolumn
\raggedbottom
\setlength{\parskip}{3pt plus 1pt}
\setlength{\abovedisplayskip}{4pt plus 1pt minus 1pt}
\setlength{\belowdisplayskip}{4pt plus 1pt minus 1pt}
\setlength{\abovedisplayshortskip}{2pt plus 1pt}
\setlength{\belowdisplayshortskip}{3pt plus 1pt minus 1pt}
\setlength{\intextsep}{12pt plus 2pt minus 2pt}
\makeatletter
\global\@colht=\textheight
\global\@colroom=\textheight
\global\vsize=\textheight
\setlength{\@fptop}{0pt}
\setlength{\@dblfptop}{0pt}
\makeatother
\fi
\section*{Limitations}

This work studies four representative controls for precisely controlled speech editing.
Extending the same interface to speaker conversion, acoustic-scene editing, and dialect
or accent conversion will require corresponding paired supervision and evaluation.
Future data pipelines could use
stronger foundation TTS or caption-to-speech models to construct fluent, locally
controlled pairs with fewer synthesis and boundary-repair stages.

\ifreportversion
\appendix
\ifreportversion\else
\renewcommand{\thesection}{\Alph{section}}
\fi
\section{Online Resources}
\label{app:resources}
The online resources for the model, demonstrations, evaluation, and data
construction are listed below.

\begingroup
\small
\renewcommand{\arraystretch}{1.2}
\begin{longtable}{@{}>{\raggedright\arraybackslash}m{0.16\textwidth}>{\raggedright\arraybackslash}m{0.34\textwidth}>{\scriptsize\raggedright\arraybackslash}m{0.44\textwidth}@{}}
\caption{Online resources.}\label{tab:online-resources}\\
\toprule
Resource & Description & {\small URL} \\
\midrule
\endfirsthead
\toprule
Resource & Description & {\small URL} \\
\midrule
\endhead
\rowcolor{black!4}
Playground & Interactive speech editing interface. &
\url{https://dots-studio-dots-tts-edit.hf.space} \\
Demo & Audio examples of speech editing. &
\url{https://huggingface.co/spaces/dots-studio/dots.tts.edit-demo} \\
\rowcolor{black!4}
Model checkpoints & Released \dotsttsedit{} model weights. &
\url{https://huggingface.co/dots-studio/dots.tts.edit} \\
Code & Model training and inference implementation. &
\url{https://github.com/studio-dots-ai/dots.tts} \\
\rowcolor{black!4}
doteBench & Benchmark data, evaluator, and candidate adapters. &
\url{https://github.com/studio-dots-ai/doteBench} \\
Data pipeline & Paired editing data construction implementation. &
\url{https://github.com/studio-dots-ai/dots.tts.edit-data-pipeline} \\
\bottomrule
\end{longtable}
\endgroup

\section{Additional Results}
\label{app:additional-results}
\subsection{Sequential Editing Results}
\label{app:sequential-results}

Sequential execution applies edits through successive model calls and evaluates
the final waveform against the complete instruction. The tables below report
the four single-task categories; the main compositional table retains both
one-take and sequential results. Bold and underline indicate the best and
second-best learned systems within each shard. Data pipeline baselines and
Identity are listed separately from learned systems.

\begin{figure}[!htbp]
  \centering
  \includegraphics[width=\textwidth]{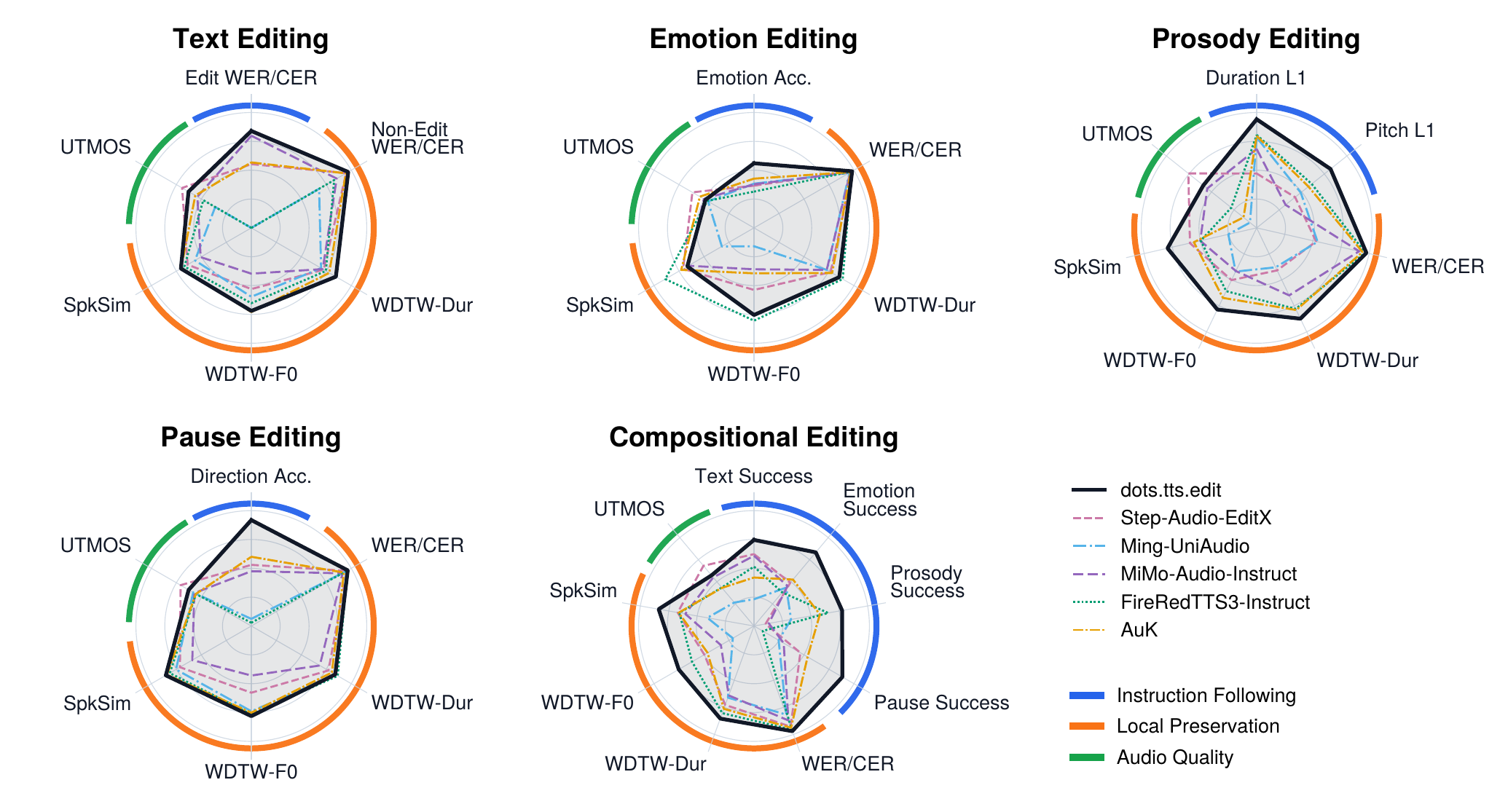}
  \caption{Sequential model comparison on doteBench. All six learned systems
  are evaluated on their final outputs. Axes, fixed normalization, and shard
  aggregation match the one-take radar plot; Text uses Hard and Emotion averages
  the Neutral and Intense shards.}
  \label{fig:radar-sequential}
\end{figure}

\noindent
\begin{minipage}[t]{0.49\textwidth}
\vspace{0pt}

\begingroup
  \centering
  \footnotesize
  \setlength{\tabcolsep}{2pt}
  \captionof{table}{Text editing results under sequential execution. Bold and underline mark the best and second-best learned systems within each mode and shard.}
  \label{tab:text-results-sequential}
  \resizebox{\linewidth}{!}{%
  \begin{tabular}{@{}l c c c c c c @{}}
    \toprule
    \systemhead[2] & \tablehead{Edit WER/\\CER (\%)\,$\downarrow$} & \tablehead{Non-edit WER/\\CER (\%)\,$\downarrow$} & \tablehead{WDTW\\Dur (\%)\,$\downarrow$} & \tablehead{WDTW\\F0 (st)\,$\downarrow$} & SpkSim\,$\uparrow$ & UTMOS\,$\uparrow$ \\
    \midrule
    \addlinespace[2pt]
    \multicolumn{7}{l}{\rmfamily\bfseries Shard: Easy} \\
    Identity & 43.68 & 0.21 & 1.45 & 0.16 & 1.000 & 3.37 \\
    \midrule[0.3pt]
    \multicolumn{7}{l}{\rmfamily\itshape Execution: Sequential} \\
    Data pipeline & 1.17 & 0.46 & 3.76 & 0.97 & 0.918 & 3.22 \\
    \candidateseparator{7}
    AuK & 14.74 & 1.46 & \textbf{6.48} & \textbf{1.68} & \textbf{0.859} & 3.23 \\
    \addlinespace[1pt]
    Step-Audio-EditX & 25.15 & 2.10 & 8.04 & 3.09 & 0.732 & \textbf{3.62} \\
    Ming-UniAudio & 41.46 & 10.40 & 9.87 & 2.34 & 0.721 & 3.00 \\
    MiMo-Audio-Instruct & \underline{5.77} & 6.44 & 13.85 & 4.58 & 0.583 & 3.28 \\
    FireRedTTS3-Instruct & 24.71 & \underline{1.25} & 7.71 & \underline{2.04} & 0.764 & 3.10 \\
    \rowcolor{blue!5}
    \dotsttsedit{} & \textbf{1.49} & \textbf{0.96} & \underline{6.76} & 2.15 & \underline{0.826} & \underline{3.38} \\
    \addlinespace[3pt]\midrule[0.65pt]
    \multicolumn{7}{l}{\rmfamily\bfseries Shard: Hard} \\
    Identity & 178.52 & 0.17 & 7.75 & 0.80 & 1.000 & 3.49 \\
    \midrule[0.3pt]
    \multicolumn{7}{l}{\rmfamily\itshape Execution: Sequential} \\
    Data pipeline & 24.87 & 0.25 & 6.84 & 1.32 & 0.766 & 3.11 \\
    \candidateseparator{7}
    AuK & 43.24 & 2.71 & \underline{11.00} & \textbf{2.17} & \underline{0.703} & 3.24 \\
    \addlinespace[1pt]
    Step-Audio-EditX & 44.77 & \underline{2.54} & 13.81 & 3.76 & 0.638 & \textbf{3.77} \\
    Ming-UniAudio & 106.86 & 16.10 & 15.09 & 3.22 & 0.564 & 2.45 \\
    MiMo-Audio-Instruct & \underline{20.12} & 7.29 & 14.29 & 4.83 & 0.507 & 3.16 \\
    FireRedTTS3-Instruct & 117.31 & 8.56 & 12.69 & 2.78 & 0.670 & 2.92 \\
    \rowcolor{blue!5}
    \dotsttsedit{} & \textbf{15.99} & \textbf{1.53} & \textbf{7.70} & \underline{2.24} & \textbf{0.705} & \underline{3.51} \\
    \bottomrule
  \end{tabular}
  }
\endgroup

\end{minipage}\hfill
\begin{minipage}[t]{0.49\textwidth}
\vspace{0pt}

\begingroup
  \centering
  \footnotesize
  \setlength{\tabcolsep}{2pt}
  \captionof{table}{Emotion editing results under sequential execution.}
  \label{tab:emotion-results-sequential}
  \resizebox{\linewidth}{!}{%
  \begin{tabular}{@{}l c c c c c c @{}}
    \toprule
    \systemhead[2] & \tablehead{Emotion\\Acc. (\%)\,$\uparrow$} & \tablehead{WER/\\CER (\%)\,$\downarrow$} & \tablehead{WDTW\\Dur (\%)\,$\downarrow$} & \tablehead{WDTW\\F0 (st)\,$\downarrow$} & SpkSim\,$\uparrow$ & UTMOS\,$\uparrow$ \\
    \midrule
    \addlinespace[2pt]
    \multicolumn{7}{l}{\rmfamily\bfseries Shard: Neutral} \\
    Identity & 20.43 & 2.76 & 0.00 & 0.00 & 1.000 & 3.07 \\
    \midrule[0.3pt]
    \multicolumn{7}{l}{\rmfamily\itshape Execution: Sequential} \\
    Data pipeline & 36.78 & 3.14 & 3.19 & 0.74 & 0.870 & 2.77 \\
    \candidateseparator{7}
    AuK & \underline{27.64} & 5.65 & 10.64 & 4.93 & \underline{0.695} & \underline{3.18} \\
    \addlinespace[1pt]
    Step-Audio-EditX & 24.52 & \textbf{2.28} & 11.60 & 3.79 & 0.665 & \textbf{3.54} \\
    Ming-UniAudio & 26.92 & 7.79 & 13.82 & 6.76 & 0.294 & 2.90 \\
    MiMo-Audio-Instruct & 25.24 & 4.42 & 14.02 & 5.37 & 0.610 & 2.95 \\
    FireRedTTS3-Instruct & 21.15 & 3.98 & \textbf{5.97} & \textbf{1.82} & \textbf{0.881} & 2.86 \\
    \rowcolor{blue!5}
    \dotsttsedit{} & \textbf{36.06} & \underline{2.92} & \underline{7.06} & \underline{2.14} & 0.642 & 2.99 \\
    \addlinespace[3pt]\midrule[0.65pt]
    \multicolumn{7}{l}{\rmfamily\bfseries Shard: Intense} \\
    Identity & 3.50 & 0.62 & 0.00 & 0.00 & 1.000 & 3.11 \\
    \midrule[0.3pt]
    \multicolumn{7}{l}{\rmfamily\itshape Execution: Sequential} \\
    Data pipeline & 9.00 & 0.80 & 3.06 & 0.60 & 0.905 & 2.75 \\
    \candidateseparator{7}
    AuK & \underline{6.50} & 0.92 & 11.15 & 4.77 & \underline{0.762} & \underline{3.15} \\
    \addlinespace[1pt]
    Step-Audio-EditX & 4.75 & \textbf{0.53} & 11.74 & 3.61 & 0.753 & \textbf{3.39} \\
    Ming-UniAudio & 3.75 & 1.23 & 13.06 & 6.69 & 0.349 & 2.78 \\
    MiMo-Audio-Instruct & 5.00 & 3.07 & 13.22 & 4.89 & 0.703 & 3.06 \\
    FireRedTTS3-Instruct & 4.00 & \underline{0.69} & \textbf{5.25} & \textbf{1.31} & \textbf{0.893} & 2.88 \\
    \rowcolor{blue!5}
    \dotsttsedit{} & \textbf{8.75} & 0.72 & \underline{7.57} & \underline{1.78} & 0.693 & 2.90 \\
    \bottomrule
  \end{tabular}
  }
\endgroup

\end{minipage}
\par\medskip
\noindent
\begin{minipage}[t]{0.49\textwidth}
\vspace{0pt}

\begingroup
  \centering
  \footnotesize
  \setlength{\tabcolsep}{2pt}
  \captionof{table}{Prosody editing results under sequential execution.}
  \label{tab:prosody-results-sequential}
  \resizebox{\linewidth}{!}{%
  \begin{tabular}{@{}l c c c c c c c @{}}
    \toprule
    \systemhead[2] & \tablehead{Duration\\L1 (s)\,$\downarrow$} & \tablehead{Pitch\\L1 (st)\,$\downarrow$} & \tablehead{WER/\\CER (\%)\,$\downarrow$} & \tablehead{WDTW\\Dur (\%)\,$\downarrow$} & \tablehead{WDTW\\F0 (st)\,$\downarrow$} & SpkSim\,$\uparrow$ & UTMOS\,$\uparrow$ \\
    \midrule
    Identity & 0.212 & 4.39 & 1.65 & 0.01 & 0.00 & 1.000 & 3.45 \\
    \midrule[0.3pt]
    \multicolumn{8}{l}{\rmfamily\itshape Execution: Sequential} \\
    Data pipeline & 0.045 & 1.53 & 2.07 & 3.39 & 0.93 & 0.901 & 3.15 \\
    \candidateseparator{8}
    AuK & 0.203 & 3.40 & 5.86 & \underline{10.49} & \underline{2.61} & 0.558 & 1.58 \\
    \addlinespace[1pt]
    Step-Audio-EditX & 0.528 & 4.59 & 47.16 & 29.56 & 3.98 & \underline{0.593} & \textbf{4.03} \\
    Ming-UniAudio & 0.216 & 4.09 & 45.88 & 31.04 & 4.64 & 0.252 & 1.30 \\
    MiMo-Audio-Instruct & 0.316 & 5.44 & 8.93 & 17.54 & 4.68 & 0.501 & 3.19 \\
    FireRedTTS3-Instruct & \underline{0.194} & \underline{3.12} & \underline{3.44} & 11.04 & 3.13 & 0.497 & 2.12 \\
    \rowcolor{blue!5}
    \dotsttsedit{} & \textbf{0.059} & \textbf{1.45} & \textbf{2.43} & \textbf{6.31} & \textbf{1.71} & \textbf{0.793} & \underline{3.36} \\
    \bottomrule
  \end{tabular}
  }
\endgroup

\end{minipage}\hfill
\begin{minipage}[t]{0.49\textwidth}
\vspace{0pt}

\begingroup
  \centering
  \footnotesize
  \setlength{\tabcolsep}{2pt}
  \captionof{table}{Pause editing results under sequential execution.}
  \label{tab:pause-results-sequential}
  \resizebox{\linewidth}{!}{%
  \begin{tabular}{@{}l c c c c c c @{}}
    \toprule
    \systemhead[2] & \tablehead{Direction\\Acc. (\%)\,$\uparrow$} & \tablehead{WER/\\CER (\%)\,$\downarrow$} & \tablehead{WDTW\\Dur (\%)\,$\downarrow$} & \tablehead{WDTW\\F0 (st)\,$\downarrow$} & SpkSim\,$\uparrow$ & UTMOS\,$\uparrow$ \\
    \midrule
    Identity & 0.00 & 2.30 & 0.00 & 0.00 & 1.000 & 3.67 \\
    \midrule[0.3pt]
    \multicolumn{7}{l}{\rmfamily\itshape Execution: Sequential} \\
    Data pipeline & 92.83 & 3.96 & 7.19 & 1.10 & 0.971 & 3.53 \\
    \candidateseparator{7}
    AuK & \underline{60.00} & 7.36 & 9.70 & 1.99 & \underline{0.845} & 3.22 \\
    \addlinespace[1pt]
    Step-Audio-EditX & 53.00 & 4.41 & 11.43 & 3.37 & 0.714 & \textbf{3.83} \\
    Ming-UniAudio & 6.17 & 4.62 & \underline{7.61} & 2.09 & 0.754 & 3.33 \\
    MiMo-Audio-Instruct & 47.50 & 8.72 & 15.75 & 4.56 & 0.591 & 3.28 \\
    FireRedTTS3-Instruct & 2.83 & \underline{3.65} & \textbf{6.77} & \underline{1.87} & 0.816 & 3.21 \\
    \rowcolor{blue!5}
    \dotsttsedit{} & \textbf{91.83} & \textbf{3.55} & 8.01 & \textbf{1.75} & \textbf{0.856} & \underline{3.51} \\
    \bottomrule
  \end{tabular}
  }
\endgroup

\end{minipage}

\subsection{SpeechEditBench Results}
\label{sec:speecheditbench_results}

SpeechEditBench~\citep{zhang2026speecheditbench} is an English--Chinese
benchmark covering content, speaker, emotion, style, prosody, paralinguistic,
and acoustic editing. It evaluates whether requested changes succeed while
preserving the required content. Its compositional setting combines two or
three changes across semantic content, speaker identity, expressive delivery,
and acoustic environment in a single instruction, such as a content edit
combined with speaker conversion and an emotion change. This tests
cross-attribute control, including utterance-level changes to voice and
environment. In doteBench, compositional editing instead combines two to four
explicit text, emotion, prosody, or pause operations localized to transcript
spans or boundaries, testing their joint execution and preservation outside
the specified scopes. The two settings therefore provide complementary
views of compositional editing, with different operation coverage and
localization requirements.

We follow the official joint-success protocol and use Qwen3-ASR transcripts for
\dotsttsedit{}. We evaluate the 3,680 requests whose components map to the model's
controls. A category unsupported in full is marked N/T.
Unsupported operations include speaker transfer and the insertion of coughs,
laughter, and sound effects. For partially supported categories, we evaluate only
the operations supported by \dotsttsedit{}. These unsupported operations also occur
in compositional editing, so we evaluate only requests whose components are all
supported, such as content with emotion editing or content with prosody editing.
If the text span specified by an editing instruction cannot be located in the ASR
transcript, we count the request as a failure without using an LLM or another
prompt enhancer for approximate matching.
For fair comparison, the table scores still use the full benchmark denominators,
with unsupported requests counted as failures.

On SpeechEditBench (\cref{tab:speecheditbench-results}), \dotsttsedit{} achieves the
highest reported open-source joint success on content, emotion, and prosody editing. Its 94.00\% content success and 91.50\% prosody joint success
demonstrate transfer of lexical and prosodic control to an independently constructed
benchmark, with prosody also outperforming the reported closed-source systems.
Compositional joint success is comparable to the strongest reported open-source baseline (7.25\% versus 7.30\%). As noted above, the lower compositional score relative to doteBench primarily reflects the many SpeechEditBench requests containing operations outside the model's current coverage, which remain in the denominator, rather than a limitation of the precise editing interface itself.

\begin{table*}[t]
  \centering
  \scriptsize
  \setlength{\tabcolsep}{2.2pt}
  \caption{Results on SpeechEditBench. Content reports joint success. Other atomic categories report joint success with target success in parentheses. Compositional reports joint success with component success in parentheses. N/T denotes a category not supported by the evaluated interface. Best and second-best results are bolded and underlined, respectively. Sources: $^{\dagger}$ \citet{zhang2026speecheditbench}, $^{\ddagger}$ \citet{ma2026auk}.}
  \label{tab:speecheditbench-results}
  \resizebox{\textwidth}{!}{%
  \begin{tabular}{l c c c c c c c c}
    \toprule
    System & Content & Speaker & Emotion & Style & Prosody & Paraling. & Acoustic & Compositional \\
    \midrule
    \multicolumn{9}{l}{\textit{Open-source systems}} \\
    Ming-UniAudio$^{\dagger}$ & 76.46 & N/T & 3.43 (5.29) & 22.17 (32.50) & 26.50 (28.00) & 11.25 (29.25) & 25.85 (29.66) & 1.76 (14.81) \\
    Step-Audio-EditX$^{\dagger}$ & 16.50 & N/T & 7.71 (9.29) & 49.67 (54.00) & 20.13 (51.51) & 31.25 (61.75) & 22.89 (40.96) & 2.01 (16.17) \\
    MiMo-Audio-Base$^{\dagger}$ & 31.67 & N/T & 0.21 (8.79) & 5.83 (42.17) & 5.67 (36.17) & 1.00 (49.75) & 4.44 (43.35) & 1.75 (15.00) \\
    MiMo-Audio-Instruct$^{\dagger}$ & 64.17 & N/T & 0.86 (\textbf{42.36}) & 0.50 (77.50) & 0.67 (42.83) & 2.00 (\underline{67.50}) & 0.80 (\textbf{45.69}) & 7.30 (32.16) \\
    AuK$^{\ddagger}$ & 91.83 & --- & 9.94 & --- & \underline{71.33} & \underline{38.50} & \textbf{37.07} & --- \\
    \midrule
    \multicolumn{9}{l}{\textit{Closed-source systems}} \\
    Gemini-Live$^{\dagger}$ & 93.17 & N/T & \textbf{27.79} (\underline{34.43}) & \underline{63.67} (\textbf{84.00}) & 65.17 (69.67) & 26.50 (61.75) & \underline{36.69} (41.53) & \textbf{11.03} (\textbf{38.57}) \\
    GPT-Realtime$^{\dagger}$ & \textbf{96.67} & N/T & 14.57 (21.00) & \textbf{68.67} (\underline{82.33}) & 63.94 (\underline{70.12}) & \textbf{47.00} (\textbf{81.50}) & 27.60 (\underline{43.60}) & \underline{10.05} (\underline{34.97}) \\
    \midrule
    \multicolumn{9}{l}{\textit{Ours}} \\
    \rowcolor{blue!5}
    \dotsttsedit{} & \underline{94.00} & N/T & \underline{15.07} (16.93) & 43.50 (54.17) & \textbf{91.50} (\textbf{96.33}) & 11.25 (11.75) & 31.80 (37.20) & 7.25 (12.05) \\
    \bottomrule
  \end{tabular}%
  }
\end{table*}

\begin{figure}[htbp]
\centering
\begin{minipage}[t]{0.32\textwidth}
\vspace{0pt}
\includegraphics[width=\linewidth]{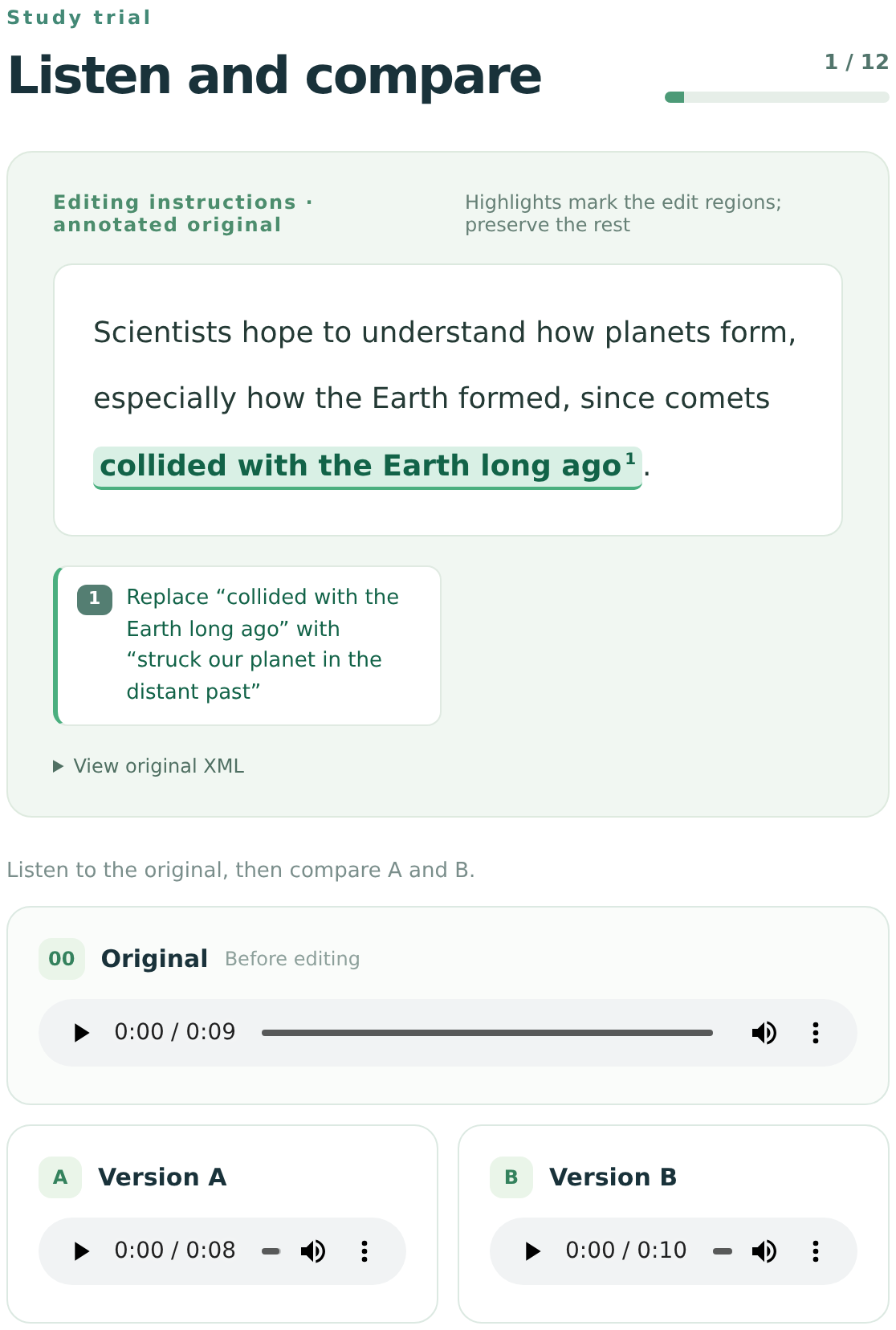}
\end{minipage}\hspace{0.025\textwidth}
\begin{minipage}[t]{0.32\textwidth}
\vspace{0pt}
\includegraphics[width=\linewidth]{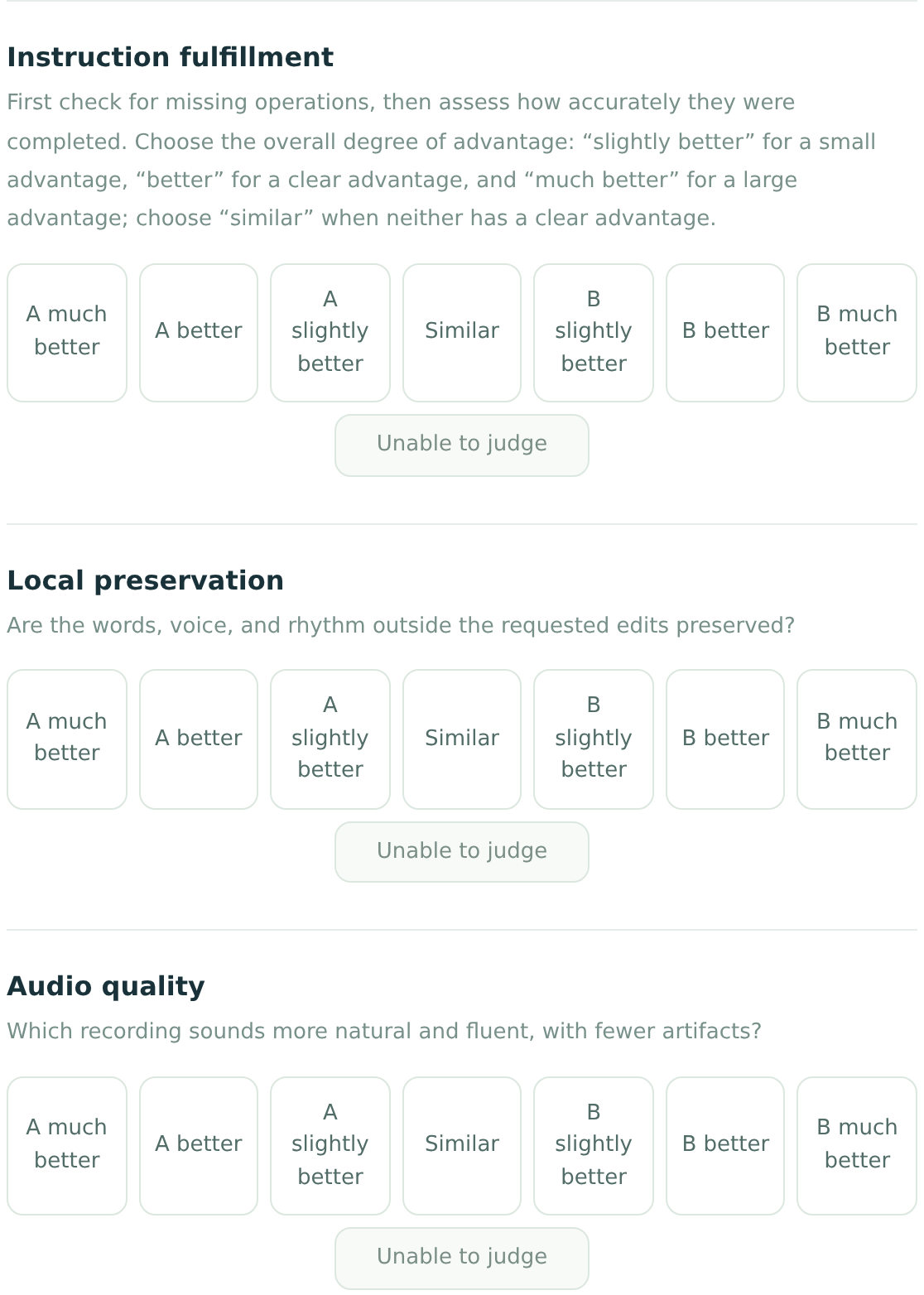}
\end{minipage}
\caption{Listening interface for an English text-editing trial.
Left: the annotated request and audio players. Right: IF, LP, and AQ
comparison prompts and response options.}
\label{fig:listening-interface}
\end{figure}

\subsection{Listening Instructions and Scoring}
\label{app:listening-protocol}

To make instruction-following preferences interpretable against the
corresponding objective metrics, we selected samples containing either a
single edit operation or multiple operations of the same type, such as
pitch edits at several locations or multiple text edits. This keeps each
trial focused on one editing objective and reduces the need for listeners
to weigh the completion of different types of edits against one another.
\Cref{fig:listening-interface} shows an English listening trial.
Listeners heard the source, then compared anonymous A/B outputs,
with each pair presented twice in each order. The dimension-specific
prompts were as follows. IF: ``First check for missing operations, then assess how
accurately they were completed.'' LP: ``Are the words, voice, and rhythm
outside the requested edits preserved?'' AQ: ``Which recording sounds more
natural and fluent, with fewer artifacts?''

For rating $r$, A receives $-r$ and B receives $r$. Mean preference scores for each system
include ties and omit unable-to-judge responses. Agreement uses applicable
cases with both objective values, omitting objective ties
and listener ties or unable-to-judge responses. Cases with no directional
votes are omitted.

\subsection{Instruction Representation Ablation}
\label{app:instruction-ablations}
\paragraph{Motivation and design principles}
This ablation examines whether explicit structure helps precise editing when
natural-language instructions already convey complete and accurate edit
specifications. We aim to give natural-language control a strong basis for
accurate execution while keeping requests concise. We therefore use fixed
bilingual templates that preserve every operation, parameter, and
target location while minimizing variation in wording and ambiguity in
localization. The comparison thus focuses on how the two representations
organize the same editing requirements.

\paragraph{Natural-language instruction construction}
\begin{table}[!t]
\centering
\small
\caption{Examples of equivalent instruction representations. The same
conversion also provides Chinese templates. Repeated text is indexed in
the source transcript, and combined requests retain each operation's scope.}
\label{tab:nl-instruction-examples}
\renewcommand{\tabularxcolumn}[1]{m{#1}}
\begin{tabularx}{\linewidth}{@{}>{\raggedright\arraybackslash}m{0.11\linewidth}>{\raggedright\arraybackslash}m{0.32\linewidth}>{\raggedright\arraybackslash}X@{}}
\toprule
Control & XML-style instruction & Natural-language instruction \\
\midrule
\rowcolor{black!4}
Replacement & \texttt{<sub targ="blue">red</sub>} & Replace ``red'' with ``blue''. \\
Deletion & \texttt{<del>red</del>} & Delete ``red''. \\
\rowcolor{black!4}
Emotion & \texttt{<emo type="happy">hello</emo>} & Render ``hello'' with happy emotion. \\
Rate & \texttt{<rate factor="0.8">hello</rate>} & Slow ``hello'' to $0.8\times$ speed. \\
\rowcolor{black!4}
Repeated span & \texttt{go <pitch semitones="3">go</pitch>} & Raise the pitch of occurrence 2 of ``go'' by 3 semitones. \\
Pause & \texttt{Hello<pause act="ins"/>world} & Insert a pause between ``Hello'' and ``world''. \\
\rowcolor{black!4}
Composition & \texttt{<pitch semitones="3">hello</pitch> <rate factor="0.8">world</rate>} & Raise the pitch of ``hello'' by 3 semitones. Slow ``world'' to $0.8\times$ speed. \\
\bottomrule
\end{tabularx}
\end{table}

We convert each semi-formal transcript-grounded edit instruction into a natural-language description using fixed rules. Text operations
retain the original and replacement content, emotion operations retain the
target category and any accompanying description, and prosody operations retain
the signed pitch change or speech-rate factor. Pause operations specify
insertion or reduction at a transcript boundary.

Quoted spans and occurrence numbers identify the intended text. Pause boundaries
use short adjacent text anchors, expanded when necessary to distinguish repeated
contexts, with occurrence numbers resolving any remaining ambiguity. This keeps
requests concise while identifying their target locations precisely. Multiple
operation clauses follow transcript order, and nested operations retain their
enclosing scope. The same conversion rules are used during training and
inference. \Cref{tab:nl-instruction-examples} illustrates the English templates,
with corresponding templates used for Chinese.

\begin{table*}[!htbp]
  \centering\footnotesize
  \setlength{\tabcolsep}{1.5pt}
  \renewcommand{\arraystretch}{1.12}
  \caption{Complete instruction and category-expert comparison. XML denotes the semi-formal transcript-grounded edit instruction. XML with all-category editing data is the shared unified reference for both comparisons. Single-category results come from the corresponding category experts. All settings retain the same proportion of TTS replay. Category experts are evaluated only on their own tasks, so compositional results are not applicable to them.}
  \label{tab:report-ablation-full}
  {\sffamily\bfseries Category: Text editing}\par\smallskip
  \setlength{\tabcolsep}{1.5pt}
  \begin{tabular*}{\linewidth}{@{\extracolsep{\fill}}ll c c c c c c @{}}
    \toprule[1pt]
    \tablehead{Instruction} & \tablehead{Editing\\data} & \tablehead{Edit WER/\\CER (\%)\,$\downarrow$} & \tablehead{Non-edit WER/\\CER (\%)\,$\downarrow$} & \tablehead{WDTW\\Dur (\%)\,$\downarrow$} & \tablehead{WDTW\\F0 (st)\,$\downarrow$} & SpkSim\,$\uparrow$ & UTMOS\,$\uparrow$ \\
    \midrule[0.3pt]
    \addlinespace[2pt]
    \multicolumn{8}{l}{\rmfamily\bfseries Shard: Hard} \\
    XML & All categories & \textbf{14.45} & \underline{0.85} & \underline{7.57} & \textbf{1.86} & \underline{0.771} & \textbf{3.49} \\
    Natural language & All categories & \underline{14.73} & \textbf{0.76} & \textbf{6.87} & \underline{1.94} & \textbf{0.772} & \textbf{3.49} \\
    XML & Single category & 15.42 & 0.93 & 8.28 & 2.31 & 0.761 & 3.44 \\
    \bottomrule
  \end{tabular*}
  \par\vspace{9pt}
  {\sffamily\bfseries Category: Emotion editing}\par\smallskip
  \setlength{\tabcolsep}{1.5pt}
  \begin{tabular*}{\linewidth}{@{\extracolsep{\fill}}ll c c c c c c @{}}
    \toprule[1pt]
    \tablehead{Instruction} & \tablehead{Editing\\data} & \tablehead{Emotion\\Acc. (\%)\,$\uparrow$} & \tablehead{WER/CER\\(\%)\,$\downarrow$} & \tablehead{WDTW\\Dur (\%)\,$\downarrow$} & \tablehead{WDTW\\F0 (st)\,$\downarrow$} & SpkSim\,$\uparrow$ & UTMOS\,$\uparrow$ \\
    \midrule[0.3pt]
    \addlinespace[2pt]
    \multicolumn{8}{l}{\rmfamily\bfseries Shard: Neutral} \\
    XML & All categories & \textbf{37.98} & \underline{3.19} & \underline{6.96} & \underline{2.09} & \underline{0.657} & \textbf{3.00} \\
    Natural language & All categories & \underline{37.02} & \textbf{2.92} & \textbf{6.75} & \textbf{2.02} & \textbf{0.666} & \textbf{3.00} \\
    XML & Single category & 36.06 & 4.57 & 7.99 & 2.27 & 0.616 & 2.94 \\
    \addlinespace[3pt]\midrule[0.65pt]
    \multicolumn{8}{l}{\rmfamily\bfseries Shard: Intense} \\
    XML & All categories & \textbf{8.50} & \textbf{0.67} & \textbf{7.46} & \textbf{1.63} & \textbf{0.709} & \underline{2.90} \\
    Natural language & All categories & \textbf{8.50} & \underline{0.74} & \textbf{7.46} & \underline{1.67} & 0.700 & \textbf{2.95} \\
    XML & Single category & \textbf{8.50} & 1.24 & 7.79 & 1.80 & \textbf{0.709} & 2.86 \\
    \bottomrule
  \end{tabular*}
  \par\vspace{9pt}
  {\sffamily\bfseries Category: Prosody editing}\par\smallskip
  \setlength{\tabcolsep}{1.5pt}
  \begin{tabular*}{\linewidth}{@{\extracolsep{\fill}}ll c c c c c c c @{}}
    \toprule[1pt]
    \tablehead{Instruction} & \tablehead{Editing\\data} & \tablehead{Duration\\L1 (s)\,$\downarrow$} & \tablehead{Pitch\\L1 (st)\,$\downarrow$} & \tablehead{WER/CER\\(\%)\,$\downarrow$} & \tablehead{WDTW\\Dur (\%)\,$\downarrow$} & \tablehead{WDTW\\F0 (st)\,$\downarrow$} & SpkSim\,$\uparrow$ & UTMOS\,$\uparrow$ \\
    \midrule[0.3pt]
    XML & All categories & \underline{0.054} & \underline{1.51} & \textbf{2.04} & \textbf{6.00} & \textbf{1.57} & \textbf{0.846} & 3.36 \\
    Natural language & All categories & 0.056 & 1.66 & \underline{2.12} & \underline{6.17} & 1.70 & \textbf{0.846} & \textbf{3.40} \\
    XML & Single category & \textbf{0.049} & \textbf{1.36} & 2.21 & 6.26 & \underline{1.66} & 0.830 & \textbf{3.40} \\
    \bottomrule
  \end{tabular*}
  \par\vspace{9pt}
  {\sffamily\bfseries Category: Pause editing}\par\smallskip
  \setlength{\tabcolsep}{1.5pt}
  \begin{tabular*}{\linewidth}{@{\extracolsep{\fill}}ll c c c c c c @{}}
    \toprule[1pt]
    \tablehead{Instruction} & \tablehead{Editing\\data} & \tablehead{Direction\\Acc. (\%)\,$\uparrow$} & \tablehead{WER/CER\\(\%)\,$\downarrow$} & \tablehead{WDTW\\Dur (\%)\,$\downarrow$} & \tablehead{WDTW\\F0 (st)\,$\downarrow$} & SpkSim\,$\uparrow$ & UTMOS\,$\uparrow$ \\
    \midrule[0.3pt]
    XML & All categories & \textbf{92.00} & \underline{3.53} & \underline{8.03} & \underline{1.66} & \textbf{0.901} & 3.58 \\
    Natural language & All categories & 69.33 & \textbf{3.13} & \textbf{7.52} & \textbf{1.63} & \underline{0.900} & \underline{3.62} \\
    XML & Single category & \underline{77.17} & 4.38 & 8.05 & 1.85 & 0.866 & \textbf{3.73} \\
    \bottomrule
  \end{tabular*}
  \par\vspace{9pt}
  {\sffamily\bfseries Category: Compositional editing}\par\smallskip
  \setlength{\tabcolsep}{1.7pt}
  \begin{tabular*}{\linewidth}{@{\extracolsep{\fill}}ll c c c c c c c c c c c @{}}
    \toprule[1pt]
    \tablehead{Instruction} & \tablehead{Editing\\data} & \tablehead{Component\\(\%)\,$\uparrow$} & \tablehead{All\\(\%)\,$\uparrow$} & \tablehead{Text\\(\%)\,$\uparrow$} & \tablehead{Emotion\\(\%)\,$\uparrow$} & \tablehead{Prosody\\(\%)\,$\uparrow$} & \tablehead{Pause\\(\%)\,$\uparrow$} & \tablehead{WER/CER\\(\%)\,$\downarrow$} & \tablehead{WDTW\\Dur (\%)\,$\downarrow$} & \tablehead{WDTW\\F0 (st)\,$\downarrow$} & SpkSim\,$\uparrow$ & UTMOS\,$\uparrow$ \\
    \midrule[0.3pt]
    \begin{tabular}[c]{@{}l@{}}XML\end{tabular} & \begin{tabular}[c]{@{}l@{}}All\\categories\end{tabular} & \begin{tabular}[c]{@{}c@{}}\textbf{68.12}\end{tabular} & \begin{tabular}[c]{@{}c@{}}\textbf{37.92}\end{tabular} & \begin{tabular}[c]{@{}c@{}}\textbf{77.54}\end{tabular} & \begin{tabular}[c]{@{}c@{}}\underline{21.01}\end{tabular} & \begin{tabular}[c]{@{}c@{}}\textbf{83.33}\end{tabular} & \begin{tabular}[c]{@{}c@{}}\textbf{90.58}\end{tabular} & \begin{tabular}[c]{@{}c@{}}\textbf{2.73}\end{tabular} & \begin{tabular}[c]{@{}c@{}}\textbf{6.88}\end{tabular} & \begin{tabular}[c]{@{}c@{}}\textbf{1.80}\end{tabular} & \begin{tabular}[c]{@{}c@{}}\textbf{0.908}\end{tabular} & \begin{tabular}[c]{@{}c@{}}\underline{3.37}\end{tabular} \\
    \addlinespace[3pt]
    \begin{tabular}[c]{@{}l@{}}Natural\\language\end{tabular} & \begin{tabular}[c]{@{}l@{}}All\\categories\end{tabular} & \begin{tabular}[c]{@{}c@{}}\underline{55.80}\end{tabular} & \begin{tabular}[c]{@{}c@{}}\underline{26.67}\end{tabular} & \begin{tabular}[c]{@{}c@{}}\underline{71.01}\end{tabular} & \begin{tabular}[c]{@{}c@{}}\textbf{21.74}\end{tabular} & \begin{tabular}[c]{@{}c@{}}\underline{73.19}\end{tabular} & \begin{tabular}[c]{@{}c@{}}\underline{57.25}\end{tabular} & \begin{tabular}[c]{@{}c@{}}\underline{3.21}\end{tabular} & \begin{tabular}[c]{@{}c@{}}\underline{7.08}\end{tabular} & \begin{tabular}[c]{@{}c@{}}\underline{1.85}\end{tabular} & \begin{tabular}[c]{@{}c@{}}\underline{0.907}\end{tabular} & \begin{tabular}[c]{@{}c@{}}\textbf{3.39}\end{tabular} \\
    \addlinespace[3pt]
    \begin{tabular}[c]{@{}l@{}}XML\end{tabular} & \begin{tabular}[c]{@{}l@{}}Single\\category\end{tabular} & \begin{tabular}[c]{@{}c@{}}---\end{tabular} & \begin{tabular}[c]{@{}c@{}}---\end{tabular} & \begin{tabular}[c]{@{}c@{}}---\end{tabular} & \begin{tabular}[c]{@{}c@{}}---\end{tabular} & \begin{tabular}[c]{@{}c@{}}---\end{tabular} & \begin{tabular}[c]{@{}c@{}}---\end{tabular} & \begin{tabular}[c]{@{}c@{}}---\end{tabular} & \begin{tabular}[c]{@{}c@{}}---\end{tabular} & \begin{tabular}[c]{@{}c@{}}---\end{tabular} & \begin{tabular}[c]{@{}c@{}}---\end{tabular} & \begin{tabular}[c]{@{}c@{}}---\end{tabular} \\
    \bottomrule
  \end{tabular*}
\end{table*}

\ifreportversion\else
\FloatBarrier
\fi

\paragraph{Experimental setup and results}

The ablations hold the training settings fixed. All ablation models share the same initialization,
optimizer, learning-rate schedule, and number of optimizer updates. The XML and
natural-language models use the same training data and sampling proportions,
differing only in instruction representation.

In \cref{tab:report-ablation-full}, the two representations achieve broadly
comparable results on Text, Emotion, and Prosody, with similar local preservation
and audio quality. XML's main advantages appear in compositional editing and
pause control: component success increases from 55.80\% to 68.12\%, all-operation
success from 26.67\% to 37.92\%, and Pause direction accuracy from 69.33\% to
92.00\%. Compositional requests require the model to distinguish multiple
operations, parameters, and target locations simultaneously. XML binds operations
directly to transcript spans or boundaries, providing an explicit structure that
helps maintain these associations.

Note that the natural-language instructions in this comparison already follow consistent
templates with complete edit specifications and explicit localization, making
their organization close to a structural representation. XML's compositional
advantage under these conditions supports the value of explicit
transcript-grounded operation-to-location binding for precise editing. More flexible natural-language
requests introduce additional variation in wording and potential ambiguity,
which can be expected to make both learning the intended mappings and executing
them reliably more demanding.

\subsection{Unified and Category-Specific Training}
\label{app:category-expert-ablations}
This ablation examines whether unified training can combine effective control
and local preservation across editing categories within one model. These experiments hold the training settings fixed.
The unified model and category experts share the same initialization, optimizer,
learning-rate schedule, and number of optimizer updates. Category experts concentrate editing samples on a single category while keeping
the proportion of TTS replay unchanged.

In \cref{tab:report-ablation-full}, the unified model reduces Text
Hard edit error from the expert's 15.42\% to 14.45\% and raises Pause direction
accuracy from 77.17\% to 92.00\%. In emotion editing, it exceeds the expert's accuracy on Neutral and matches it on Intense while better preserving content, timing, and pitch. The Prosody
expert has lower target pitch error, while the unified model better preserves
content, timing, pitch, and speaker identity. Unified training thus combines control
and local preservation across editing categories in one model, which also
supports their composition.

\section{doteBench}
\label{app:benchmark-metrics}
\subsection{Overview}

Speech creation requires targeted changes to an existing utterance, such as
replacing words, adjusting the emotion or prosody of a phrase, or changing a
pause. A speech editing model must control the operation, its parameters, and the location
of a change while preserving the remaining content and acoustic attributes and
maintaining natural speech. Evaluating precise speech editing therefore
requires evidence of instruction following, local preservation, and overall
audio quality.

doteBench makes these requirements explicit in a bilingual benchmark.
A semi-formal transcript-grounded edit instruction is the
sole authority for each request. It determines the source and target
transcripts, operation categories and parameters, and the transcript spans or
boundaries to which the operations apply. These specifications determine the
instruction-following measurements and their local scope, together with the
regions selected for preservation. Candidates may compile the instruction and
source audio into their supported inputs and execution plans. Evaluation
always uses the original complete request, allowing different interfaces to
be assessed against a common editing objective.

\subsection{Tasks and Data Composition}
\label{app:benchmark-construction}

doteBench contains 2,081 evaluation cases. \Cref{tab:benchmark-sources}
summarizes their languages and source origins: 1,487 cases use recorded speech
and 594 use synthesized speech. Recorded sources come from public test splits
of LibriTTS-R~\mbox{\citep{koizumi2023librittsr}}, Common Voice 20.0~\mbox{\citep{ardila2020commonvoice}},
FLEURS~\mbox{\citep{conneau2023fleurs}}, Multilingual LibriSpeech~\mbox{\citep{pratap2020mls}},
AISHELL-3~\mbox{\citep{shi2021aishell3}}, and AISHELL-1~\mbox{\citep{bu2017aishell1}},
covering audiobook and read speech. Synthesized sources provide
additional voice-designed utterances and strongly emotional starting
conditions. Several instructions can share a source utterance, so counts refer
to evaluation cases. Evaluation uses text units corresponding to English words
and Chinese characters. All categories also report speaker similarity
(SpkSim) and overall audio quality (UTMOS).

\begin{table}[!htbp]
\centering\small
\caption{doteBench source composition and evaluation focus. Counts refer to
evaluation cases. EN and ZH denote English and Chinese. Recorded and synthetic
describe source audio.}
\label{tab:benchmark-sources}
\renewcommand{\tabularxcolumn}[1]{m{#1}}
\begin{tabularx}{\textwidth}{@{}lccccc>{\raggedright\arraybackslash}X@{}}
\toprule
Category / shard & Cases & EN & ZH & Recorded & Synthetic & Evaluation focus \\
\midrule
\rowcolor{black!4}
Text / Easy & 209 & 125 & 84 & 209 & 0 & Small local text changes \\
Text / Hard & 360 & 180 & 180 & 234 & 126 & Longer contexts, larger replacements, multiple text operations \\
\rowcolor{black!4}
Emotion / Neutral & 312 & 156 & 156 & 312 & 0 & Introducing emotion into neutral speech \\
Emotion / Intense & 300 & 150 & 150 & 0 & 300 & Changing an established emotional delivery \\
\rowcolor{black!4}
Prosody & 360 & 180 & 180 & 360 & 0 & Pitch and rate control at different span granularities \\
Pause & 300 & 150 & 150 & 196 & 104 & Pause insertion and reduction at transcript boundaries \\
\rowcolor{black!4}
Compositional & 240 & 120 & 120 & 176 & 64 & Joint localization and execution of multiple edit families \\
\bottomrule
\end{tabularx}
\end{table}

\paragraph{Text editing}
Text editing tests lexical changes while retaining the surrounding speech.
Both Easy and Hard contain insertion, deletion, and substitution requests.
Easy emphasizes smaller local changes, while Hard includes longer contexts,
larger replacements, and multiple text operations. The instruction identifies
each edit's target-text interval or deletion boundary. We expand it by 3
text units on either side, clipped at the transcript boundaries, to form the
edit neighborhood. Instruction following uses WER/CER within this
neighborhood. Content preservation counts recognition substitutions, deletions,
and insertions on its complement. Acoustic preservation compares duration and
F0 over the units selected for preservation by the instruction.

\paragraph{Emotion editing}
Emotion editing tests the requested delivery over a whole utterance or local
span. Neutral contains 104 recorded source utterances with neutral delivery,
each paired with whole-utterance, single-span, and two-span requests
(312 cases). Intense uses 100 strongly emotional synthesized sources with the
same three scope patterns (300 cases). Its sources were generated with
IndexTTS2~\mbox{\citep{zhou2025indextts2}}, Qwen3-TTS~\mbox{\citep{hu2026qwen3tts}},
VoxCPM2~\mbox{\citep{zhou2026voxcpm2}}, and MiMo-Audio~\mbox{\citep{xiaomi2025mimoaudio}}, then screened for content,
audio quality, and emotional expression. The six source emotions are afraid,
angry, happy, melancholic, sad, and surprised. In Intense, requested targets
differ from the source emotion, and the two targets in a two-span case are
distinct, testing changes to an established emotional delivery. Shard names
describe the starting audio. Each operation explicitly specifies its target
category through the \texttt{type} attribute of the \texttt{<emo>} tag. The
classifier uses 8 labels: happy, angry, sad, afraid, disgusted, melancholic,
surprised, and calm. Whole-utterance requests are evaluated on the complete
output audio. Local requests use the aligned target interval with 250\,ms of
context on each side, clipped to the audio boundaries. Emotion accuracy is
the proportion of requested emotion operations whose predicted label matches
the target category. Each region in a multi-region request contributes one
operation. Content preservation uses full-utterance WER/CER, and acoustic
preservation uses the preserved units outside the requested regions.

\paragraph{Prosody editing}
Prosody editing tests pitch and speaking-rate control within specified spans.
It uses 120 recorded source utterances with three instruction variants each
(360 cases), covering different span granularities. Pitch L1 is the absolute
difference between the observed and requested pitch shifts in semitones.
Praat CLI~\mbox{\citep{boersma2001praat}} extracts voiced F0 for each aligned unit. We compute the shift from
the ratio of output to source median voiced F0, then average over source units
with positive aligned duration and valid median voiced F0 within the
operation's span. Duration L1 measures the absolute error in seconds between
the output-span duration and the source-span duration divided by the requested
rate factor. Content preservation uses full-utterance WER/CER. Acoustic
preservation covers unchanged regions and unrequested attributes within edited
regions: duration in pitch-only spans and F0 in rate-only spans.

\paragraph{Pause editing}
Pause editing tests insertion and reduction at transcript boundaries.
Evaluation compares the output gap with the frozen source gap and requires
a change of at least 160\,ms in the requested direction. Direction accuracy
is aggregated over requested operations. A genuine zero gap is a valid
observation. Content preservation uses full-utterance WER/CER, and acoustic
preservation compares duration and F0 over preserved units.

\paragraph{Compositional editing}
\label{app:compositional-criteria}
Compositional editing tests joint localization and execution of disjoint
operations within one source utterance. Its 240 distinct source utterances
comprise 176 recordings and 64 voice-designed synthetic utterances. Six pairs
of operation families have 30 cases each, four three-family combinations have
12 each, and 12 cases combine all four families. This yields 180 two-operation,
48 three-operation, and 12 four-operation cases, balanced between English and
Chinese. Pitch requests use shifts of $\pm4$ semitones, and rate requests use
factors of 0.8 or 1.25, corresponding to duration increases of 25\% or decreases
of 20\%, respectively. To provide a compact comparison across heterogeneous
operation families, compositional instruction following uses binary success criteria. Text requires zero
local recognition error, and emotion requires the requested label. To assess
both the direction and magnitude of prosody changes, pitch and rate require
the actual signed change to be 0.6--1.6 times the requested change. For pitch,
the ratio is $\widehat{\Delta}/\Delta$, where $\Delta$ and $\widehat{\Delta}$
are the requested and measured shifts in semitones. For rate, the ratio is
$(\widehat{f}-1)/(f-1)$, where $f$ is the requested speed factor and
$\widehat{f}$ is source-span duration divided by output-span duration.
Pause requires a gap change of at least 160\,ms in the requested direction.
Component success is the fraction of all requested operations meeting their
criteria. All-operation success is the fraction of cases in which every
requested operation meets its criterion. Content preservation excludes the text-edit
neighborhoods specified by the complete instruction. Acoustic preservation
selects corresponding units wherever each attribute should remain unchanged.

\subsection{Evaluation Criteria}
\label{app:evaluation-criteria}

\subsubsection{Local Preservation Metrics}

\paragraph{Unit correspondence and preservation scope}
Preservation depends on both the location and the attribute being edited,
so that intended changes can be distinguished from unintended drift.
XML node identity fixes the correspondence of preserved source and target
units before alignment times are selected. Repeated lexical forms are
distinguished by occurrence rather than paired through subsequence matching.
Source alignments are frozen. One output alignment against the target
transcript is shared by the metrics. Acoustic preservation scopes are
attribute-specific and distinct from the 3-unit recognition neighborhood.
WDTW-Dur includes pitch-only regions, and WDTW-F0 includes rate-only regions,
so each measure covers units where its corresponding attribute should remain
unchanged.

\paragraph{WDTW-Dur}
WDTW-Dur builds on the word-level duration formulation introduced by
AST~\citep{lv2026ast}. Dynamic time warping~\citep{sakoe1978dtw} is applied to
the duration sequences of the selected source and output units. Let
$X=((x_i,d_i^{\mathrm{src}}))_{i=1}^{m}$ and
$Y=((y_j,d_j^{\mathrm{out}}))_{j=1}^{n}$, where $x_i$ and $y_j$ are normalized
text units, $d_i^{\mathrm{src}}$ and $d_j^{\mathrm{out}}$ their durations, and
$m$ and $n$ the source and output unit counts. The local cost is
\begin{equation}
  c_{ij} =
  \begin{cases}
    \lvert d_i^{\mathrm{src}}-d_j^{\mathrm{out}}\rvert,
      & x_i=y_j,\\
    d_i^{\mathrm{src}}+d_j^{\mathrm{out}}+\lambda_{\mathrm{mis}},
      & x_i\ne y_j,
  \end{cases}
  \qquad \lambda_{\mathrm{mis}}=1.0\,\mathrm{s}.
\end{equation}
Here $\lambda_{\mathrm{mis}}$ is the mismatch penalty. Let $D_{ij}$ be the
minimum accumulated alignment cost through source unit $i$ and output unit
$j$. With $D_{00}=0$ and $D_{i0}=D_{0j}=\infty$ for positive indices,
\begin{equation}
  D_{ij}=c_{ij}+\min\{D_{i-1,j},D_{i,j-1},D_{i-1,j-1}\}.
\end{equation}
The normalized score for one example is
\begin{equation}
  \operatorname{WDTW\text{-}Dur}(X,Y)=
  \frac{D_{mn}}
  {\sum_{i=1}^{m}d_i^{\mathrm{src}}+
   \sum_{j=1}^{n}d_j^{\mathrm{out}}}.
\end{equation}
AST supplies the forced-alignment, duration-DTW, and length-normalization
premise, while doteBench makes the instruction-derived selection and local
cost explicit. Lower values indicate better timing preservation.

\paragraph{WDTW-F0}
Duration preservation is insensitive to pitch drift. WDTW-F0 complements it
by measuring pitch changes in units whose F0 should remain unchanged.
It uses the same Praat CLI extraction procedure as pitch-editing measurements
to obtain the minimum, maximum, and mean voiced F0 from aligned source and
output intervals. Let $\mathcal V$ contain preserved source units with
positive aligned duration and finite positive values for all three F0
statistics. Source-defined eligibility retains units whose output measurements
are missing. For a corresponding output unit that also has positive aligned
duration and three finite positive F0 statistics, the error is
\begin{equation}
  e_r=\frac{1}{3}\sum_{k\in\mathcal K}
  \left|12\log_2
  \frac{F^{\mathrm{out}}_{0,r,k}}{F^{\mathrm{src}}_{0,r,k}}\right|,
  \qquad \mathcal K=\{\min,\max,\operatorname{mean}\},
\end{equation}
where $F^{\mathrm{src}}_{0,r,k}$ and $F^{\mathrm{out}}_{0,r,k}$ denote the
source and output statistics. Missing-output values of $e_r$ are defined in
\mbox{\Cref{app:aggregation}}. The score for one example is
\begin{equation}
  \operatorname{WDTW\text{-}F0}=\frac{\sum_{r\in\mathcal V}e_r}{|\mathcal V|}.
\end{equation}
 It is reported in semitones, with lower values indicating
better pitch preservation.

\subsubsection{Aggregation and Failure Handling}
\label{app:aggregation}

\paragraph{Aggregation}
WER/CER pools substitutions, deletions, insertions, and reference units over
the evaluated set. Emotion accuracy, Pitch L1, Duration L1, Pause direction
accuracy, and Component success weight requested operations equally.
SpkSim, UTMOS, and All-operation success weight examples equally. Category
results pool the underlying records across shards.
WDTW-Dur divides the sum of alignment costs by the sum of duration normalizers
and reports 100 times this ratio in \%. WDTW-F0 pools unit errors and
counts of source units that satisfy the eligibility conditions above. For whole-utterance
edits, there may be no interval in which timing or pitch should be preserved.
Such samples are excluded from aggregation of the corresponding WDTW metric.

\paragraph{Missing outputs}
Missing local outputs retain their scoring units so that evaluation covers
the source units selected for evaluation and all requested operations. For WDTW-F0, a
source unit that satisfies these eligibility conditions without a positive-duration output unit or valid output F0
statistics receives $e_r=8$ semitones and remains in the denominator. For a
pitch operation, each source unit with valid pitch measurements lacking a valid output unit or voiced
F0 receives 0 shift in the operation-level mean used to compute Pitch L1.
The operation fails its success criterion whenever a required output
measurement is missing. Rate error retains the measured output-span duration,
while a missing required output unit makes the operation fail its success
criterion. Pause evaluation requires positive aligned durations for its output
anchor units. A pause at an utterance boundary may use a single anchor. If a
required anchor is missing, the operation fails, direction accuracy is 0,
and the scored gap change is 0. A genuinely measured zero gap on either
side remains a valid observation.

\paragraph{Generation failures}
Generation failures remain scored so that aggregate results cover all
evaluation requests. Recognition error is 100\% for a nonempty reference.
SpkSim is 0, UTMOS 1, WDTW-Dur 100\%, and WDTW-F0 8 semitones.
Preservation penalties apply only when source units satisfy the eligibility conditions for the metric. All
requested operations fail. Pitch L1 uses 0 observed shift and therefore
equals the absolute requested shift. Duration L1 equals the requested target
duration, computed from source-span duration and rate factor.

\subsection{Candidate Adaptation}
\label{app:candidate-adaptation}

Most baseline interfaces, including MiMo-Audio-Instruct
\mbox{\citep{xiaomi2025mimoaudio}}, Step-Audio-EditX
\mbox{\citep{yan2025stepaudioeditx}}, and AuK \mbox{\citep{ma2026auk}},
lack native support for local acoustic edits or requests combining multiple
editing operations. Adapting them to doteBench therefore requires careful
interface design and validation to make effective use of their capabilities
for a fair comparison. A central difficulty is that the instruction
generalization of text LLMs cannot be assumed to
carry over to audio models. AHAMask and ISA-Bench show that audio understanding
models can suffer substantial performance degradation even when instructions
express the same task in different ways, as well as when tasks are combined
\mbox{\citep{guo2026ahamask,li2025isabench}}. In our adapter validation, giving editing models
unseen local or multi-operation editing requests beyond their supported
interfaces severely degraded generation quality (e.g., reading only the instruction,
mixing the instruction with the target transcript, producing semantically
incoherent speech, repeating content, or generating no speech), making semantically coherent
speech difficult to obtain reliably. We therefore prioritize coherent speech
with the intended linguistic content, then express edits through controls each
model can reliably use. For each baseline, we tried multiple simplification
strategies on a small validation set, seeking to make effective use of its editing
capabilities while maintaining basic semantic correctness. In summary, we
simplify requests where necessary, retaining supported controls through each model's
native interface. We
compile requests for two execution strategies: one-take uses a single model call, and
sequential uses successive calls.

\paragraph{One-take execution}
\dotsttsedit{} receives the complete instruction with only the syntactic
conversion required by its runtime, preserving spans and parameter values.
The other evaluated interfaces use natural-language requests, task templates,
or transcript arguments. All outputs are scored against the original complete
instruction, including operations not expressed in the selected single-call
request.

Step-Audio-EditX, Ming-UniAudio, and MiMo-Audio-Instruct use the complete
target transcript for text edits and select the first acoustic operation in
source order for compositional one-take requests. Local acoustic controls are
expressed as utterance-level requests. Step-Audio-EditX uses its native TTS
path for text, editing templates for emotion and speed, and free-form requests
for pitch and pause insertion. Pause reduction uses VAD editing or the
compositional silence-removal request. Ming-UniAudio retains its native
speech-plus-instruction format, with its editing reasoning path for text-only
requests. MiMo-Audio-Instruct receives the transcript and instruction as
separate arguments, with the current audio as voice reference.

FireRedTTS3-Instruct selects the first operation in source/XML order
for one-take evaluation. FireRedTTS3-Instruct uses local semantic editing for
text, disambiguating repeated words with the smallest unique lexical context,
and global requests for acoustic edits. Rate and pitch values are mapped to
the nearest supported settings in the requested direction.

AuK receives the complete XML request with tag explanations
through its prompt enhancer, which maps the request to native controls.
The resulting controls are  applied to the current source audio
through AuK's whole-utterance editing interface. Sequential execution uses
the same interface for each operation.

\paragraph{Sequential execution}
Localizing an individual edit and executing multiple edits are distinct
requirements. Sequential execution lets each requested operation be attempted
separately, reducing the need to interpret multiple controls jointly. Each
call retains the candidate's adaptation of local scope. Some baselines have
limited preservation across successive edits, making their results sensitive to
operation order. To support baseline performance, we selected a common execution
order through small-scale pilot tests. Text edits precede
acoustic edits, with each output becoming the next source. This establishes
the intended transcript before delivery changes and avoids a later
text-generation step replacing earlier acoustic edits.

\section{Data Pipeline Details}
\label{app:data-pipeline-details}
\label{app:data}

\subsection{Construction Challenges and Shared Principles}
\label{app:pipeline-shared}

Precisely controlled speech editing requires paired utterances that differ in
the requested content or attribute while preserving the remaining speech. Such
pairs are rarely available in natural recordings. Constructing them introduces
two coupled challenges. First, a local intervention must realize the requested
change and join naturally with its acoustic context. Independently synthesizing
both utterances can introduce unrelated differences in timing, voice, and
rendering, while direct segment replacement can expose alignment errors and
audible discontinuities. Second, the data must cover a deliberately chosen
range of operation types, locations, granularities, and control strengths.
These properties must be controlled during construction so that difficult forms
of editing receive sufficient supervision.
During training, concatenating independently sampled segments augments these
examples with longer, compositional inputs.

\paragraph{An instruction-centered construction process}
We organize construction around a semi-formal transcript-grounded edit instruction, denoted by \(u\). The language model
generates this instruction under programmatically specified constraints. Once
validated, it is the authoritative description of the edit: operation
categories and parameters are read from its tags, and edit locations are read
from their positions in the transcript. Alignment queries, synthesis requests,
and paired transcripts are derived from this same representation. This common
interface connects semantic planning to precise tool execution.
A sampler selects the original utterance, the planner specifies the edit,
and the synthesizer constructs its generated counterpart using speech models
and signal-processing tools. Original/generated denote construction provenance;
source/target denote the direction of an editing example. For text, prosody,
and pause, exchanging the two waveforms and inverting the instruction yields
reverse supervision while leaving their provenance unchanged. When the original waveform is human-recorded speech, the reverse direction uses it as the target.

Every training example follows the shared definition
\begingroup
\[
(T_{\mathrm{src}}, A_{\mathrm{src}}, u, T_{\mathrm{tgt}}, A_{\mathrm{tgt}}),
\qquad T_{\mathrm{src}}=g_{\mathrm{src}}(u),\quad
T_{\mathrm{tgt}}=g_{\mathrm{tgt}}(u).
\]
\endgroup

Here, \(A\) denotes a waveform and \(T\) its transcript. The source and target
projections determine the text before and after the operation. The pipeline
checks these projections and the requested control values before invoking the
acoustic tools. All constructed training pairs undergo quality filtering for
edit realization, intelligibility, speaker consistency, local preservation, and
audio quality. The saver retains accepted pairs with their alignments,
intervention parameters, validation results, and source provenance.
Construction procedures and parameter settings for each editing task are
selected through evaluation using objective metrics and subjective preferences.

\paragraph{Controlling coverage before semantic planning}
Programmatic sampling determines the axes on which the construction
distribution should be controlled, including span granularity, operation count,
emotion condition, and pitch, rate, or pause parameters. The sampled
specification is passed to the language model, which chooses semantically
appropriate spans, generates replacement content, or describes the intended
vocal delivery. Validation returns feedback when a proposal violates the
specification, allowing the model to revise its output.

This division is particularly useful for multi-operation text editing. Several
changes within one utterance require coordinated generation and provide more
demanding training examples. Unconstrained language-model proposals need not
allocate enough examples to these cases. We therefore sample the operation
count explicitly and ask the planner to realize that count. For example, one-,
two-, and three-operation proposals use weights of \(2:3:5\), favoring
multi-operation examples. The weights govern the proposal distribution.
Linguistic validity and acoustic quality checks subsequently determine which
examples are retained. Within the sampled constraints, the language model
retains the flexibility to select meaningful edits.

\paragraph{A common local replacement operation}
The acoustic constructions share a simple local operation. Let \(I=[s,e)\)
denote an interval in a base waveform \(A\), and let \(A_{\mathrm{patch}}\) be
the waveform that will replace it. We write
\begingroup
\[
\operatorname{Replace}(A,I,A_{\mathrm{patch}})
= A[0:s]\;\Vert\;A_{\mathrm{patch}}\;\Vert\;A[e:],
\]
\endgroup

where \(\Vert\) denotes concatenation. The patch may contain newly generated
words, a different emotional realization, transformed prosody, or silence. An
insertion uses \(s=e\). Text infilling implements the same local organization
by generating a masked patch within the surrounding waveform. Emotion
construction uses a complete target realization as the base and constructs the
paired source through replacement, as detailed in \Cref{app:pipeline-emotion}.

Alignment uncertainty can place a cut inside a word, and acoustic differences
between a patch and its context can produce an audible seam. We use
context-conditioned F5-TTS~\citep{chen2024f5tts} regeneration around the affected boundaries to
repair these transitions. Boundary repair permits limited changes to
nominally unedited speech near a join; the repaired pair must still pass the
preservation and quality checks. Exact waveform reuse applies outside the replacement
and boundary-repair regions, accounting for any shift in the subsequent
timeline. The following sections describe how each category chooses its patch
and the constraints needed to preserve the requested control.

\subsection{Instruction-Derived Alignment}
\label{app:pipeline-alignment}

Local construction requires the waveform interval corresponding to a particular
transcript span or boundary. The semi-formal transcript-grounded edit instruction already identifies the intended occurrence
through its position in the full transcript. We preserve this positional
information when constructing the alignment request. This is useful for
repeated words: a request containing only the isolated word leaves its
occurrence unspecified, whereas marking it in its full transcript identifies
the intended occurrence directly.

For the waveform being aligned, the pipeline first renders the corresponding
transcript projection. It then replaces the operation-specific markup with a
query marker around the same character span. For example, selecting the second
occurrence of ``again'' yields

\begin{quote}
\ttfamily Say again, then say <q>again</q>.
\end{quote}

The query is derived from the parsed instruction span, so repeated occurrences
are distinguished before acoustic alignment. For an insertion or pause at a
boundary, the query marks the point between adjacent transcript characters. The
full transcript provides the surrounding linguistic context, while
Qwen3-ForcedAligner-0.6B~\citep{shi2026qwen3asr} maps the specified occurrence to waveform time.
Character positions remove ambiguity about the requested occurrence. The
returned acoustic boundary remains an estimate and motivates the subsequent
local repair. The same interface supports lexical spans, emotion and prosody regions, and
pause boundaries.

\begin{pipelinealgorithm}{Localize an instruction-defined edit}{alg:pipeline-alignment}
\textbf{Input:} waveform \(A\), instruction \(u\), side
\(\in\{\mathrm{source},\mathrm{target}\}\).\par
\textbf{Output:} edit interval \(I\), or a boundary for a point operation.\par
\begin{enumerate}[label=\arabic*.,leftmargin=1.7em,topsep=3pt,itemsep=1pt,parsep=0pt]
\item Render the transcript \(T\) for side from \(u\).
\item Read the selected edit's character span from the parsed instruction.
\item Mark that span with \texttt{<q>\ldots</q>} inside the complete \(T\). Use an
empty marked span for a point operation.
\item Query the forced aligner with \(A\) and the marked transcript.
\item Return the aligned start and end. Collapse a point operation to its boundary.
\end{enumerate}
\end{pipelinealgorithm}

\subsection{Text Editing}
\label{app:pipeline-text}

Text editing changes lexical content and can change the duration of an
utterance. The construction must therefore supply enough time for the new words
while retaining the original context. We support insertion, deletion, and
substitution. Their locations and replacement text are specified by a
semi-formal transcript-grounded edit instruction,
from which the source and target transcripts are rendered.
\Cref{alg:pipeline-alignment} localizes each source-side span or insertion
boundary.

\paragraph{Duration planning and the target timeline}
The replacement duration is estimated from the source speaking rate. Let
\(\ell(T)\) denote the text-length measure used by the duration estimator and
\(d_{\mathrm{src}}\) the source duration. For replacement text
\(T_{\mathrm{new}}\), the estimate is
\begingroup
\[
r_{\mathrm{src}}=\frac{\ell(T_{\mathrm{src}})}{d_{\mathrm{src}}},
\qquad d_{\mathrm{new}}=\frac{\ell(T_{\mathrm{new}})}{r_{\mathrm{src}}}.
\]
\endgroup

Deletion has an empty replacement and \(d_{\mathrm{new}}=0\). Insertion uses a
point interval. For a replacement at \(I=[s,e)\), the change in duration is
\begingroup
\[
\Delta d=d_{\mathrm{new}}-(e-s).
\]
\endgroup

The prefix retains its original position, and the unchanged suffix moves by
\(\Delta d\) on the target timeline. For example, replacing a 0.6-second phrase
with a 0.9-second phrase moves all subsequent speech 0.3 seconds later while
reusing its waveform samples. For multiple non-overlapping edits, the pipeline
processes spans in temporal order and accumulates these shifts.

\paragraph{Context-conditioned local generation}
The aligned edit windows are expanded by 0.4 seconds on each side, bounded by
the utterance endpoints. The extra context allows regeneration to accommodate
imperfect cuts and the transition between old and new content. Windows
separated by less than 0.1 seconds are merged so that closely spaced changes
are synthesized together. These settings were selected through listening and
inspection of construction failures.

Using the estimated durations, the pipeline places preserved speech on the
target timeline and leaves the expanded edit regions masked. F5-TTS fills the
masks using the target transcript and the surrounding waveform as acoustic
context. Its output is assembled with the preserved intervals, with a
20-millisecond source-side cross-fade at the joins. The generation step can
adapt local articulation and coarticulation while the surrounding speech
remains directly anchored to the original speech.

\begin{pipelinealgorithm}{Construct a lexical editing pair}{alg:pipeline-text}
\textbf{Input:} original waveform \(A_{\mathrm{ori}}\), validated instruction \(u\), expansion
0.4 s, merge gap 0.1 s.\par
\textbf{Output:} source and target waveforms with their instruction and transcripts.\par
\begin{enumerate}[label=\arabic*.,leftmargin=1.7em,topsep=3pt,itemsep=1pt,parsep=0pt]
\item Render source and target transcripts from \(u\). Localize each edit with
\Cref{alg:pipeline-alignment}.
\item Estimate source speaking rate and the duration of each replacement text. Assign
0 duration to deletions.
\item Expand the source windows and include the duration of their added context.
Merge nearby windows.
\item Build the target timeline, copying preserved samples and reserving masks for
replacement content and transition context.
\item Run F5-TTS with the target transcript, waveform condition, and masks.
\item Assemble the result with the copied context and apply the source-side
cross-fade.
\item Return the original waveform as source and the locally generated waveform as
target.
\end{enumerate}
\end{pipelinealgorithm}

The planner covers word, phrase, clause, sentence, and arbitrary spans.
 The inverse instruction
exchanges insertion and deletion and reverses substitutions, enabling the
constructed waveform pair to support the opposite editing direction as well.

\subsection{Emotion Editing}
\label{app:pipeline-emotion}

Emotion changes several correlated properties of speech, including pitch,
energy, rhythm, and voice quality. A strongly expressive speaker prompt can
carry an emotional bias that interferes with synthesis under another strong
emotion. We address this by drawing speaker prompts from a neutral library and
generating both emotional realizations from the same prompt. The
speaker-balanced library is built from open-source speech datasets and
covers 4,000 speakers and approximately 1 million
utterances. Holding its selected prompt fixed provides a common identity
reference for both sides of the pair.

\paragraph{Emotion conditions and local instructions}
Categorical construction uses eight emotion categories---happy, angry, sad,
afraid, disgusted, melancholic, surprised, and calm.
The planner specifies the desired emotional delivery.
IndexTTS2~\citep{zhou2025indextts2} renders the same text under the
selected conditions, so the intended contrast is determined by the emotion
controls while the lexical content and identity reference are shared.

Description-conditioned construction expresses the desired vocal delivery in
natural language. The planner supplies a concise description of the emotion and
its audible realization, which conditions the IndexTTS2-based synthesis
procedure following the description-conditioning approach of TED-TTS~\citep{liang2026tedtts}. This form
accommodates finer descriptions of delivery than a category label alone. For
both forms, the local change is serialized as a semi-formal transcript-grounded edit instruction. The span specifies the
affected text, and its attributes specify the target emotional control.
Categorical and descriptive emotion values are both operation parameters,
encoded in XML attributes.

\paragraph{Constructing the source around a complete target}
For local supervision, directly pairing two independent syntheses would
introduce differences outside the desired emotion region. Instead, let \(A_1\)
and \(A_2\) be the complete realizations under the source and target emotion
conditions, and let \(I_1\) and \(I_2\) be the corresponding aligned intervals.
We retain the complete second realization as the target and construct the
source from it:
\begingroup
\[
A_{\mathrm{tgt}}=A_2,\qquad
A_{\mathrm{src}}=
\operatorname{Repair}\!\left(
\operatorname{Replace}(A_2,I_2,A_1[I_1])
\right).
\]
\endgroup

Here, \(\operatorname{Repair}\) denotes local F5-TTS regeneration around the
replacement boundaries. The source contains the first emotion in the selected
region and otherwise shares the target realization's context. The target
retains the coherent utterance-level rendering produced by the emotion
synthesizer. This arrangement both avoids teaching assembly artifacts on the
target side and preserves the target's overall emotional and prosodic
continuity.

\begin{pipelinealgorithm}{Construct an ordered local emotion pair}{alg:pipeline-emotion}
\textbf{Input:} transcript \(T\), source and target emotion conditions, selected transcript
span.\par
\textbf{Output:} a locally assembled source and a complete target realization.\par
\begin{enumerate}[label=\arabic*.,leftmargin=1.7em,topsep=3pt,itemsep=1pt,parsep=0pt]
\item Select one neutral speaker prompt and keep it fixed.
\item Synthesize \(A_1\) and \(A_2\) from \(T\) under the two emotion conditions.
\item Align the selected transcript span in each waveform, obtaining \(I_1\) and
\(I_2\).
\item Set \(A_{\mathrm{tgt}}\leftarrow A_2\).
\item Set \(A_{\mathrm{src}}\leftarrow\operatorname{Replace}(A_2,I_2,A_1[I_1])\).
\item Repair the source's splice boundaries with F5-TTS.
\item Associate the ordered pair with the corresponding local emotion instruction.
\item For the opposite direction, exchange the emotional realizations and repeat
Steps 3--7.
\end{enumerate}
\end{pipelinealgorithm}

Repeating the assembly for the opposite direction keeps a complete synthesis as
the target in each ordered pair. The construction uses the same replacement
operation as the other categories, with the assembled waveform occupying the
source role. For an utterance-level emotion change, the two complete
realizations directly provide the pair. Local and utterance-level examples
therefore share the emotion controls and neutral identity reference while
differing in how the pair is assembled.

\subsection{Prosody Editing}
\label{app:pipeline-prosody}

Prosody control requires a prescribed pitch or rate change within a selected
region while maintaining the speaker's identity and the surrounding speech.
Signal processing provides an explicit acoustic intervention, but the
transformed segment can contain artifacts or a changed perceived timbre. We
combine the intervention with neural resynthesis that uses the original speech
as an identity reference.

The semi-formal transcript-grounded edit instruction supplies the pitch shift in semitones or the speech-rate factor and
identifies the span to localize with \Cref{alg:pipeline-alignment}. Pitch
conversion uses Praat PSOLA~\citep{boersma2001praat,moulines1990psola}. For a requested shift of \(\Delta\) semitones, the
ideal pitch multiplier is \(2^{\Delta/12}\). Quantitative calibration of the
subsequent resynthesis gives a compensation factor of 1.6, which we apply to
the requested shift before signal-level conversion:
\begingroup
\[
\Delta_{\mathrm{pre}}=1.6\Delta,\qquad
F_{0,\mathrm{pre}}(t)=2^{\Delta_{\mathrm{pre}}/12}F_0(t).
\]
\endgroup

The instruction retains the requested value \(\Delta\). The larger intermediate
shift compensates for attenuation during resynthesis. The converted segment
provides semantic, speaker-conditioning, and style features to IndexTTS2. The
original segment provides the base speaker prompt and mel reference. This
conditioning arrangement is designed to retain the pitch intervention while
recovering the original identity and naturalness.

Rate conversion uses SoX WSOLA~\citep{verhelst1993wsola}. A factor \(r>1\) speeds up speech, while
\(r<1\) slows it down. For an interval of duration \(e-s\), the intended
transformed duration is approximately \((e-s)/r\). IndexTTS2 extracts content
from the time-stretched segment and uses the original utterance as its speaker
reference for resynthesis.

For either control, the resulting patch is inserted using
\(\operatorname{Replace}\). A 40-millisecond in-window cross-fade blends the
patch with its context, and F5-TTS locally repairs the boundaries. The rest of
the waveform is reused with the timeline correspondence described in
\Cref{app:pipeline-text}. Pitch proposals cover shifts from \(-9\) to \(+9\)
semitones excluding 0, with separate near-neutral, moderate, and wide
ranges. Rate proposals cover 0.50--0.99 and 1.01--1.99. Programmatic sampling
controls coverage across these strengths and across word, short-phrase, clause,
and sentence spans. The original waveform is the source and the repaired
waveform is the target for the forward example. Swapping the waveforms gives
the reverse example: a pitch shift of $\Delta$ becomes $-\Delta$, and a rate
factor of $r$ becomes $1/r$. The reverse instruction localizes the transformed
span on the generated waveform's timeline.

\ifreportversion
\begin{table}[!htbp]
\centering
\small
\caption{Principal controls and acoustic construction parameters.}
\label{tab:data-construction-summary}
\begin{tabular}{@{}>{\raggedright\arraybackslash}m{0.19\textwidth}
  >{\raggedright\arraybackslash}m{0.42\textwidth}
  >{\raggedright\arraybackslash}m{0.33\textwidth}@{}}
\toprule Component & Principal setting & Purpose \\ \midrule
\rowcolor{black!4}
Shared planning &
Operation, span granularity, operation count, and parameter ranges &
Set proposal coverage before semantic planning \\[1.5pt]
Alignment &
Qwen3-ForcedAligner-0.6B with full-transcript queries and a marked span or point &
Identify the intended occurrence and locate its waveform interval \\[1.5pt]
\rowcolor{black!4}
Text sampling &
1/2/3-operation proposal weights: 2:3:5 &
Increase coverage of multi-operation edits \\[1.5pt]
Text duration &
Estimate replacement duration from source speaking rate &
Allocate time for the new content \\[1.5pt]
\rowcolor{black!4}
Text context &
0.4 s window expansion, merging gaps below 0.1 s &
Accommodate cut uncertainty and closely spaced edits \\[1.5pt]
Text assembly &
20 ms source-side cross-fade &
Blend generated content with retained speech \\[1.5pt]
\rowcolor{black!4}
Emotion identity reference &
Neutral library: 4,000 speakers, approximately 1 million utterances &
Share a neutral identity reference across emotional realizations \\[1.5pt]
Emotion controls &
8 categories, description-conditioned delivery &
Specify the intended emotional contrast \\[1.5pt]
\rowcolor{black!4}
Prosody pitch &
\(\pm\)1--\(\pm\)9 semitones, 1.6\(\times\) intermediate compensation &
Control pitch strength and compensate for resynthesis attenuation \\[1.5pt]
Prosody rate &
0.50--0.99 and 1.01--1.99 &
Cover slower and faster delivery \\[1.5pt]
\rowcolor{black!4}
Prosody assembly &
40 ms cross-fade, local F5-TTS boundary repair &
Restore continuity after patch insertion \\[1.5pt]
Pause duration &
200--800 ms &
Set the inserted silence duration \\[1.5pt]
\rowcolor{black!4}
Pause repair &
Separate left/right repair windows, fixed central silence &
Repair cut errors while retaining the intended gap \\[1.5pt]
F5-TTS infilling &
32 sampling steps, guidance 2, sway \(-1\) &
Shared local waveform regeneration \\[1.5pt]
\bottomrule
\end{tabular}
\end{table}

\fi

\subsection{Pause Editing}
\label{app:pipeline-pause}

Pause editing introduces a prescribed gap at a transcript boundary while
retaining the surrounding lexical content. The semi-formal transcript-grounded edit instruction specifies the boundary and
pause level. Its transcript projections leave the words unchanged, and
\Cref{alg:pipeline-alignment} maps the marked boundary to a waveform time
\(s\). The construction inserts silence lasting 200--800\,ms.

Using the common replacement notation, the initial construction is
\begingroup
\[
\widetilde A=
\operatorname{Replace}\!\left(A_{\mathrm{ori}},[s,s),\mathbf{0}_d\right),
\qquad
d\in[0.2,0.8]\;\mathrm{s},
\]
\endgroup

where \(\mathbf{0}_d\) denotes a zero-valued waveform segment of duration
\(d\). This operation establishes the gap explicitly. The source waveform
supplies both the preserved speech and the acoustic condition for subsequent
local regeneration.

\paragraph{Repairing the two boundaries while preserving the gap}
An estimated alignment boundary can fall inside a word. Inserting silence there
may split that word across the two sides of the gap, so the transitions require
local reconstruction. At the same time, unrestricted regeneration across the
entire inserted interval can extend speech into the intended silence and
substantially shorten the pause. We therefore restrict F5-TTS regeneration to
separate neighborhoods of the left and right boundaries and retain the
intervening silence.

Let \(\mathcal R_{\mathrm{left}}\) and \(\mathcal R_{\mathrm{right}}\) denote
these two repair windows. Their extents are constrained to leave a nonempty
central silence interval fixed. Regeneration uses the transcript and
surrounding speech to restore the word boundaries and transitions, while
samples outside the repair windows are copied from \(\widetilde A\). This
combines explicit duration control with local acoustic repair. Exchanging the
resulting waveform pair and reversing the instruction supplies pause-reduction
supervision.

\subsection{Principal Construction Parameters}
\label{app:pipeline-parameters}

\ifreportversion\else

\fi

\Cref{tab:data-construction-summary} collects the parameters that directly
determine control strength, affected extent, or acoustic reconstruction.  Sampling controls determine the
coverage of candidate plans, while quality filtering retains the pairs used for
training.

Together, the pipelines expose a common instruction-centered construction
process while choosing category-specific interventions for lexical, emotional,
prosodic, and temporal control.

\fi

\ifreportversion\else
\input{sections/journal_declarations}
\section*{Declaration of generative AI and AI-assisted technologies in the manuscript preparation process}
During the preparation of this manuscript, the authors used OpenAI Codex to
assist with drafting and revising the text and preparing LaTeX materials.
The authors reviewed and edited the resulting content and take full
responsibility for the content of the published article.
\fi

\FloatBarrier
\begingroup
\setstretch{1.0}
\bibliography{references,references_related,references_report}

@inproceedings{koizumi2023librittsr,
  title = {{LibriTTS-R}: A Restored Multi-Speaker Text-to-Speech Corpus},
  author = {Koizumi, Yuma and Zen, Heiga and Karita, Shigeki and Ding, Yifan
            and Yatabe, Kohei and Morioka, Nobuyuki and Bacchiani, Michiel
            and Zhang, Yu and Han, Wei and Bapna, Ankur},
  booktitle = {Proceedings of Interspeech},
  year = {2023},
  pages = {5496--5500},
  doi = {10.21437/Interspeech.2023-1584},
  url = {https://www.isca-archive.org/interspeech_2023/koizumi23_interspeech.html}
}

@inproceedings{ardila2020commonvoice,
  title = {{Common Voice}: A Massively-Multilingual Speech Corpus},
  author = {Ardila, Rosana and Branson, Megan and Davis, Kelly and Kohler, Michael
            and Meyer, Josh and Henretty, Michael and Morais, Reuben
            and Saunders, Lindsay and Tyers, Francis and Weber, Gregor},
  booktitle = {Proceedings of the Twelfth Language Resources and Evaluation Conference},
  publisher = {European Language Resources Association},
  year = {2020},
  pages = {4218--4222},
  url = {https://aclanthology.org/2020.lrec-1.520/}
}

@inproceedings{conneau2023fleurs,
  title = {{FLEURS}: Few-Shot Learning Evaluation of Universal Representations of Speech},
  author = {Conneau, Alexis and Ma, Min and Khanuja, Simran and Zhang, Yu
            and Axelrod, Vera and Dalmia, Siddharth and Riesa, Jason
            and Rivera, Clara and Bapna, Ankur},
  booktitle = {2022 IEEE Spoken Language Technology Workshop (SLT)},
  year = {2023},
  doi = {10.1109/SLT54892.2023.10023141},
  url = {https://ieeexplore.ieee.org/document/10023141/}
}

@inproceedings{pratap2020mls,
  title = {{MLS}: A Large-Scale Multilingual Dataset for Speech Research},
  author = {Pratap, Vineel and Xu, Qiantong and Sriram, Anuroop
            and Synnaeve, Gabriel and Collobert, Ronan},
  booktitle = {Proceedings of Interspeech},
  year = {2020},
  pages = {2757--2761},
  doi = {10.21437/Interspeech.2020-2826},
  url = {https://www.isca-archive.org/interspeech_2020/pratap20_interspeech.html}
}

@inproceedings{shi2021aishell3,
  title = {{AISHELL-3}: A Multi-Speaker Mandarin {TTS} Corpus},
  author = {Shi, Yao and Bu, Hui and Xu, Xin and Zhang, Shaoji and Li, Ming},
  booktitle = {Proceedings of Interspeech},
  year = {2021},
  doi = {10.21437/Interspeech.2021-755},
  url = {https://www.isca-archive.org/interspeech_2021/shi21c_interspeech.html}
}

@inproceedings{bu2017aishell1,
  title = {{AISHELL-1}: An Open-Source Mandarin Speech Corpus and a Speech Recognition Baseline},
  author = {Bu, Hui and Du, Jiayu and Na, Xingyu and Wu, Bengu and Zheng, Hao},
  booktitle = {2017 20th Conference of the Oriental Chapter of the International
               Coordinating Committee on Speech Databases and Speech I/O Systems
               and Assessment (O-COCOSDA)},
  year = {2017},
  doi = {10.1109/ICSDA.2017.8384449},
  url = {https://ieeexplore.ieee.org/abstract/document/8384449}
}

@misc{dotstts2026,
  title         = {{dots.tts} Technical Report},
  author        = {Lian, Shi and Li, Changtao and Li, Bohan and Wang, Hankun
                   and Zheng, Da and Tian, Junfeng and Ma, Yufeng
                   and Zhang, Colin and Yu, Kai},
  year          = {2026},
  eprint        = {2606.07080},
  archiveprefix = {arXiv},
  primaryclass  = {cs.SD},
  url           = {https://arxiv.org/abs/2606.07080}
}

@inproceedings{li2026holitok,
  title         = {{HoliTok}: A Continuous Holistic Tokenization with Robust
                   Dual Capabilities of Speech Generation and Understanding},
  author        = {Li, Bohan and Lian, Shi and Wang, Hankun and Guo, Yiwei
                   and Xi, Yu and Li, Zhihan and Zheng, Da and Zhang, Colin
                   and Yu, Kai},
  year          = {2026},
  booktitle     = {Proceedings of the 2026 Conference on Empirical Methods in Natural Language Processing},
  publisher     = {Association for Computational Linguistics},
  pages         = {29781--29795},
  url           = {https://aclanthology.org/2026.emnlp-main.1308/}
}

@misc{seedtts2024,
  title         = {{Seed-TTS}: A Family of High-Quality Versatile Speech
                   Generation Models},
  author        = {Anastassiou, Philip and Chen, Jiawei and Chen, Jitong
                   and Chen, Yuanzhe and Chen, Zhuo and Chen, Ziyi
                   and Cong, Jian and Deng, Lelai and Ding, Chuang
                   and Gao, Lu and others},
  year          = {2024},
  eprint        = {2406.02430},
  archiveprefix = {arXiv},
  primaryclass  = {eess.AS},
  url           = {https://arxiv.org/abs/2406.02430}
}

@misc{qwen2024qwen25,
  title         = {{Qwen2.5} Technical Report},
  author        = {{Qwen Team}},
  year          = {2024},
  eprint        = {2412.15115},
  archiveprefix = {arXiv},
  primaryclass  = {cs.CL},
  url           = {https://arxiv.org/abs/2412.15115}
}

@inproceedings{wang2023campplus,
  title     = {{CAM++}: A Fast and Efficient Network for Speaker Verification
               Using Context-Aware Masking},
  author    = {Wang, Hui and Zheng, Siqi and Chen, Yafeng and Cheng, Luyao
               and Chen, Qian},
  booktitle = {Proceedings of Interspeech},
  pages     = {5301--5305},
  year      = {2023},
  doi       = {10.21437/Interspeech.2023-1513}
}

@misc{hu2026qwen3tts,
  title         = {{Qwen3-TTS} Technical Report},
  author        = {Hu, Hangrui and Zhu, Xinfa and He, Ting and Guo, Dake
                   and Zhang, Bin and Wang, Xiong and Guo, Zhifang
                   and Jiang, Ziyue and Hao, Hongkun and Guo, Zishan
                   and Zhang, Xinyu and Zhang, Pei and Yang, Baosong
                   and Xu, Jin and Zhou, Jingren and Lin, Junyang},
  year          = {2026},
  eprint        = {2601.15621},
  archiveprefix = {arXiv},
  primaryclass  = {cs.SD},
  url           = {https://arxiv.org/abs/2601.15621}
}

@misc{zhou2026voxcpm2,
  title         = {{VoxCPM2} Technical Report},
  author        = {Zhou, Yixuan and Zeng, Guoyang and Liu, Xin and Li, Xiang
                   and Yu, Renjie and Gui, Jiancheng and Wu, Jiaheng
                   and Wang, Ziyang and Shen, Xudong and Ye, Runchuan
                   and others},
  year          = {2026},
  eprint        = {2606.06928},
  archiveprefix = {arXiv},
  primaryclass  = {cs.SD},
  url           = {https://arxiv.org/abs/2606.06928}
}

@inproceedings{saeki2022utmos,
  title     = {{UTMOS}: {UTokyo-SaruLab} System for {VoiceMOS} Challenge 2022},
  author    = {Saeki, Takaaki and Xin, Detai and Nakata, Wataru
               and Koriyama, Tomoki and Takamichi, Shinnosuke
               and Saruwatari, Hiroshi},
  booktitle = {Proceedings of Interspeech},
  pages     = {4521--4525},
  year      = {2022},
  doi       = {10.21437/Interspeech.2022-439}
}

@misc{shi2026qwen3asr,
  title         = {{Qwen3-ASR} Technical Report},
  author        = {Shi, Xian and Wang, Xiong and Guo, Zhifang and Wang, Yongqi
                   and Zhang, Pei and Zhang, Xinyu and Guo, Zishan
                   and Hao, Hongkun and Xi, Yu and Yang, Baosong and Xu, Jin
                   and Zhou, Jingren and Lin, Junyang},
  year          = {2026},
  eprint        = {2601.21337},
  archiveprefix = {arXiv},
  primaryclass  = {cs.CL},
  url           = {https://arxiv.org/abs/2601.21337}
}

@inproceedings{chen2024f5tts,
  title         = {{F5-TTS}: A Fairytaler that Fakes Fluent and Faithful
                   Speech with Flow Matching},
  author        = {Chen, Yushen and Niu, Zhikang and Ma, Ziyang and Deng, Keqi
                   and Wang, Chunhui and Zhao, Jian and Yu, Kai and Chen, Xie},
  booktitle     = {Proceedings of the 63rd Annual Meeting of the Association for Computational Linguistics},
  pages         = {6255--6271},
  year          = {2025},
  doi           = {10.18653/v1/2025.acl-long.313},
  url           = {https://aclanthology.org/2025.acl-long.313/}
}

@inproceedings{zhou2025indextts2,
  title         = {{IndexTTS2}: A Breakthrough in Emotionally Expressive and
                   Duration-Controlled Auto-Regressive Zero-Shot
                   Text-to-Speech},
  author        = {Zhou, Siyi and Zhou, Yiquan and He, Yi and Zhou, Xun
                   and Wang, Jinchao and Deng, Wei and Shu, Jingchen},
  booktitle     = {Proceedings of the AAAI Conference on Artificial Intelligence},
  volume        = {40},
  pages         = {35139--35148},
  year          = {2026},
  doi           = {10.1609/aaai.v40i41.40820},
  url           = {https://ojs.aaai.org/index.php/AAAI/article/view/40820}
}

@inproceedings{liang2026tedtts,
  title         = {{TED-TTS}: Training-Free Intra-Utterance Emotion and
                   Duration Control for Text-to-Speech Synthesis},
  author        = {Liang, Qifan and Liu, Yuansen and Wei, Ruixin and Lu, Nan
                   and Zhao, Junchuan and Wang, Ye},
  booktitle     = {Proceedings of the 64th Annual Meeting of the Association for Computational Linguistics},
  pages         = {23485--23508},
  year          = {2026},
  doi           = {10.18653/v1/2026.acl-long.1077},
  url           = {https://aclanthology.org/2026.acl-long.1077/}
}

@article{boersma2001praat,
  title   = {{Praat}, a System for Doing Phonetics by Computer},
  author  = {Boersma, Paul},
  journal = {Glot International},
  volume  = {5},
  number  = {9/10},
  pages   = {341--345},
  year    = {2001},
  url     = {https://www.fon.hum.uva.nl/paul/papers/speakUnspeakPraat_glot2001.pdf}
}

@article{moulines1990psola,
  title   = {Pitch-Synchronous Waveform Processing Techniques for
             Text-to-Speech Synthesis Using Diphones},
  author  = {Moulines, Eric and Charpentier, Francis},
  journal = {Speech Communication},
  volume  = {9},
  number  = {5--6},
  pages   = {453--467},
  year    = {1990},
  doi     = {10.1016/0167-6393(90)90021-Z}
}

@inproceedings{verhelst1993wsola,
  title     = {An Overlap-Add Technique Based on Waveform Similarity
               ({WSOLA}) for High Quality Time-Scale Modification of Speech},
  author    = {Verhelst, Werner and Roelands, Marc},
  booktitle = {Proceedings of the IEEE International Conference on Acoustics,
               Speech, and Signal Processing},
  volume    = {2},
  pages     = {554--557},
  year      = {1993},
  doi       = {10.1109/ICASSP.1993.319366}
}

@misc{ma2026mmae,
  title         = {{MMAE}: A Massive Multitask Audio Editing Benchmark},
  author        = {Ma, Ziyang and Yan, Ruiqi and Xu, Ruiyang and Fang, Jie
                   and Niu, Zhikang and Chao, Yi-Wen and Tu, Wenming
                   and Wang, Tianrui and {Auden} and Chen, Qi and Chen, Wenxi
                   and Chi, Jiaying and Huo, Yanru and Jiang, Zixuan
                   and Li, Xiquan and Li, Yalin and Liu, Junxi and Liu, Minghao
                   and Qiang, Binghao and Shan, Yijia and Song, Zheshu
                   and Tan, Tian and Wang, Zixiang and Xie, Zeyu and Xie, Zhifei
                   and Xing, Xiaoyu and Xu, Qixiang and Yang, Chen
                   and Yang, Guanrou and Yang, Shan and Yang, Yifan
                   and Yves, Steve and Zhang, Haotian and Zhu, Haina and Yu, Kai
                   and Bo, Liefeng and Chng, Eng-Siong and Chen, Xie},
  year          = {2026},
  eprint        = {2606.07229},
  archiveprefix = {arXiv},
  primaryclass  = {cs.SD},
  url           = {https://arxiv.org/abs/2606.07229}
}

@misc{xu2025qwen3omni,
  title         = {{Qwen3-Omni} Technical Report},
  author        = {Xu, Jin and Guo, Zhifang and Hu, Hangrui and Chu, Yunfei
                   and Wang, Xiong and He, Jinzheng and Wang, Yuxuan
                   and Shi, Xian and He, Ting and Zhu, Xinfa and others},
  year          = {2025},
  eprint        = {2509.17765},
  archiveprefix = {arXiv},
  primaryclass  = {cs.CL},
  url           = {https://arxiv.org/abs/2509.17765}
}

@misc{xiaomi2025mimoaudio,
  title         = {{MiMo-Audio}: Audio Language Models are Few-Shot Learners},
  author        = {{Xiaomi LLM-Core Team}},
  year          = {2025},
  eprint        = {2512.23808},
  archiveprefix = {arXiv},
  primaryclass  = {cs.SD},
  url           = {https://arxiv.org/abs/2512.23808}
}

@article{chen2022wavlm,
  title   = {{WavLM}: Large-Scale Self-Supervised Pre-Training for Full Stack
             Speech Processing},
  author  = {Chen, Sanyuan and Wang, Chengyi and Chen, Zhengyang and Wu, Yu
             and Liu, Shujie and Chen, Zhuo and Li, Jinyu and Kanda, Naoyuki
             and Yoshioka, Takuya and Xiao, Xiong and Wu, Jian and Zhou, Long
             and Ren, Shuo and Qian, Yanmin and Qian, Yao and Wu, Jian
             and Zeng, Michael and Yu, Xiangzhan and Wei, Furu},
  journal = {IEEE Journal of Selected Topics in Signal Processing},
  volume  = {16},
  number  = {6},
  pages   = {1505--1518},
  year    = {2022},
  doi     = {10.1109/JSTSP.2022.3188113}
}

@inproceedings{desplanques2020ecapa,
  title     = {{ECAPA-TDNN}: Emphasized Channel Attention, Propagation and
               Aggregation in {TDNN} Based Speaker Verification},
  author    = {Desplanques, Brecht and Thienpondt, Jenthe
               and Demuynck, Kris},
  booktitle = {Proceedings of Interspeech},
  pages     = {3830--3834},
  year      = {2020},
  doi       = {10.21437/Interspeech.2020-2650}
}

@inproceedings{lipman2023flow,
  title     = {Flow Matching for Generative Modeling},
  author    = {Lipman, Yaron and Chen, Ricky T. Q. and Ben-Hamu, Heli
               and Nickel, Maximilian and Le, Matt},
  booktitle = {International Conference on Learning Representations},
  year      = {2023},
  url       = {https://arxiv.org/abs/2210.02747}
}

@inproceedings{loshchilov2019adamw,
  title     = {Decoupled Weight Decay Regularization},
  author    = {Loshchilov, Ilya and Hutter, Frank},
  booktitle = {International Conference on Learning Representations},
  year      = {2019},
  eprint    = {1711.05101},
  archiveprefix = {arXiv},
  primaryclass  = {cs.LG},
  url       = {https://arxiv.org/abs/1711.05101}
}

@inproceedings{hu2024minicpm,
  title         = {{MiniCPM}: Unveiling the Potential of Small Language
                   Models with Scalable Training Strategies},
  author        = {Hu, Shengding and Tu, Yuge and Han, Xu and Cui, Ganqu
                   and He, Chaoqun and Zhao, Weilin and Long, Xiang
                   and Zheng, Zhi and Fang, Yewei and Huang, Yuxiang
                   and Zhang, Xinrong and Thai, Zhen Leng and Wang, Chongyi
                   and Yao, Yuan and Zhao, Chenyang and Zhou, Jie and Cai, Jie
                   and Zhai, Zhongwu and Ding, Ning and Jia, Chao
                   and Zeng, Guoyang and Li, Dahai and Liu, Zhiyuan and Sun, Maosong},
  year          = {2024},
  booktitle     = {Proceedings of the First Conference on Language Modeling},
  url           = {https://openreview.net/forum?id=3X2L2TFr0f}
}

@inproceedings{peebles2023dit,
  title     = {Scalable Diffusion Models with Transformers},
  author    = {Peebles, William and Xie, Saining},
  booktitle = {Proceedings of the IEEE/CVF International Conference on
               Computer Vision},
  pages     = {4195--4205},
  year      = {2023},
  eprint    = {2212.09748},
  archiveprefix = {arXiv},
  primaryclass  = {cs.CV},
  url       = {https://arxiv.org/abs/2212.09748}
}

@misc{ho2022classifierfree,
  title         = {Classifier-Free Diffusion Guidance},
  author        = {Ho, Jonathan and Salimans, Tim},
  year          = {2022},
  eprint        = {2207.12598},
  archiveprefix = {arXiv},
  primaryclass  = {cs.LG},
  url           = {https://arxiv.org/abs/2207.12598}
}

@article{deepseekai2026v41flash,
  title   = {{DeepSeek-V4.1-Flash}: Pushing the Limits of {KV} Cache Compression},
  author  = {{DeepSeek-AI}},
  journal = {arXiv preprint arXiv:2609.19969},
  year    = {2026},
  doi     = {10.48550/arXiv.2609.19969},
  url     = {https://arxiv.org/abs/2609.19969}
}

@article{sakoe1978dtw,
  title   = {Dynamic Programming Algorithm Optimization for Spoken Word
             Recognition},
  author  = {Sakoe, Hiroaki and Chiba, Seibi},
  journal = {IEEE Transactions on Acoustics, Speech, and Signal Processing},
  volume  = {26},
  number  = {1},
  pages   = {43--49},
  year    = {1978},
  doi     = {10.1109/TASSP.1978.1163055}
}

@article{guo2026ahamask,
  title = {{AHAMask}: Reliable Task Specification for Large Audio Language Models Without Instructions},
  author = {Guo, Yiwei and Li, Bohan and Wang, Hankun and Li, Zhihan and Wang, Shuai and Chen, Xie and Yu, Kai},
  journal = {Proceedings of the AAAI Conference on Artificial Intelligence},
  volume = {40},
  number = {26},
  pages = {21513--21521},
  year = {2026},
  doi = {10.1609/aaai.v40i26.39299}
}

@article{li2025isabench,
  title = {{ISA-Bench}: Benchmarking Instruction Sensitivity for Large Audio Language Models},
  author = {Li, Bohan and Huang, Wenbin and Qiu, Yuhang and Guo, Yiwei and Wang, Hankun and Li, Zhihan and Peng, Jing and Ma, Ziyang and Chen, Xie and Yu, Kai},
  journal = {arXiv preprint arXiv:2510.23558},
  year = {2025},
  doi = {10.48550/arXiv.2510.23558}
}

@article{ghosh2026audioflamingonext,
  title = {{Audio Flamingo Next}: Next-Generation Open Audio-Language Models for Speech, Sound, and Music},
  author = {Ghosh, Sreyan and Goel, Arushi and Jayakumar, Kaousheik and Koroshinadze, Lasha and Anand, Nishit and Kong, Zhifeng and Gururani, Siddharth and Lee, Sang-gil and Kim, Jaehyeon and Aljafari, Aya and Yang, Chao-Han Huck and Kim, Sungwon and Duraiswami, Ramani and Manocha, Dinesh and Shoeybi, Mohammad and Catanzaro, Bryan and Liu, Ming-Yu and Ping, Wei},
  journal = {arXiv preprint arXiv:2604.10905},
  year = {2026},
  eprint = {2604.10905},
  archiveprefix = {arXiv},
  url = {https://arxiv.org/abs/2604.10905}
}

@inproceedings{sun2026spotsound,
  title = {{SpotSound}: Enhancing Large Audio-Language Models with Fine-Grained Temporal Grounding},
  author = {Sun, Luoyi and Zhou, Xiao and Li, Zeqian and Zhang, Ya and Wang, Yanfeng and Xie, Weidi},
  booktitle = {Proceedings of the 34th ACM International Conference on Multimedia},
  year = {2026},
  eprint = {2604.13023},
  archiveprefix = {arXiv},
  url = {https://arxiv.org/abs/2604.13023}
}

@inproceedings{tan2021editspeech,
  title         = {{EditSpeech}: A Text Based Speech Editing System Using Partial Inference and Bidirectional Fusion},
  author        = {Tan, Daxin and Deng, Liqun and Yeung, Yu Ting and Jiang, Xin and Chen, Xiao and Lee, Tan},
  booktitle     = {2021 IEEE Automatic Speech Recognition and Understanding Workshop (ASRU)},
  pages         = {626--633},
  year          = {2021},
  doi           = {10.1109/ASRU51503.2021.9688051},
  eprint        = {2107.01554},
  archiveprefix = {arXiv},
  primaryclass  = {eess.AS},
  url           = {https://arxiv.org/abs/2107.01554}
}

@inproceedings{liu2023fluenteditor,
  title         = {{FluentEditor}: Text-based Speech Editing by Considering Acoustic and Prosody Consistency},
  author        = {Liu, Rui and Xi, Jiatian and Jiang, Ziyue and Li, Haizhou},
  booktitle     = {Proceedings of Interspeech},
  pages         = {3435--3439},
  year          = {2024},
  doi           = {10.21437/Interspeech.2024-2320},
  url           = {https://www.isca-archive.org/interspeech_2024/liu24p_interspeech.html}
}

@inproceedings{peng2024voicecraft,
  title     = {{VoiceCraft}: Zero-Shot Speech Editing and Text-to-Speech in the Wild},
  author    = {Peng, Puyuan and Huang, Po-Yao and Li, Shang-Wen and Mohamed, Abdelrahman and Harwath, David},
  booktitle = {Proceedings of the 62nd Annual Meeting of the Association for Computational Linguistics},
  year      = {2024},
  url       = {https://aclanthology.org/2024.acl-long.673/}
}

@misc{chen2026cosyedit,
  title         = {{CosyEdit}: Unlocking End-to-End Speech Editing Capability from Zero-Shot Text-to-Speech Models},
  author        = {Chen, Junyang and Jia, Yuhang and Wang, Hui and Zhou, Jiaming and Qin, Yong},
  year          = {2026},
  eprint        = {2601.05329},
  archiveprefix = {arXiv},
  primaryclass  = {cs.SD},
  url           = {https://arxiv.org/abs/2601.05329}
}

@misc{lv2026ast,
  title         = {{AST}: Adaptive, Seamless, and Training-Free Precise Speech Editing},
  author        = {Lv, Sihan and Jin, Yechen and Li, Zhen and Chen, Jintao and Zhang, Jinshan and Li, Ying and Yin, Jianwei and Xi, Meng},
  year          = {2026},
  eprint        = {2604.16056},
  archiveprefix = {arXiv},
  primaryclass  = {cs.SD},
  url           = {https://arxiv.org/abs/2604.16056}
}

@misc{yan2025stepaudioeditx,
  title         = {{Step-Audio-EditX} Technical Report},
  author        = {Yan, Chao and Wu, Boyong and Yang, Peng and Tan, Pengfei and Hu, Guoqiang and Xie, Li and Zhang, Yuxin and Zhang, Xiangyu (Tony) and Tian, Fei and Yang, Xuerui and Zhang, Xiangyu and Jiang, Daxin and Zhou, Shuchang and Yu, Gang},
  year          = {2025},
  eprint        = {2511.03601},
  archiveprefix = {arXiv},
  primaryclass  = {cs.CL},
  url           = {https://arxiv.org/abs/2511.03601}
}

@misc{yan2025minguniaudio,
  title         = {{Ming-UniAudio}: Speech {LLM} for Joint Understanding, Generation and Editing with Unified Representation},
  author        = {Yan, Canxiang and Jin, Chunxiang and Huang, Dawei and Yu, Haibing and Peng, Han and Zhan, Hui and Gao, Jie and Peng, Jing and Chen, Jingdong and Zhou, Jun and Ren, Kaimeng and Yang, Ming and Yang, Mingxue and Xu, Qiang and Zhao, Qin and Xiong, Ruijie and Lin, Shaoxiong and Wang, Xuezhi and Yuan, Yi and Wu, Yifei and Lyu, Yongjie and He, Zhengyu and Qiu, Zhihao and Fang, Zhiqiang and Huang, Ziyuan},
  year          = {2025},
  eprint        = {2511.05516},
  archiveprefix = {arXiv},
  primaryclass  = {cs.CL},
  url           = {https://arxiv.org/abs/2511.05516}
}

@misc{tao2025mmedit,
  title         = {{MMEDIT}: A Unified Framework for Multi-Type Audio Editing via Audio Language Model},
  author        = {Tao, Ye and Wu, Wen and Zhang, Chao and Wu, Mengyue and Wang, Shuai and Xu, Xuenan},
  year          = {2025},
  eprint        = {2512.20339},
  archiveprefix = {arXiv},
  primaryclass  = {cs.SD},
  url           = {https://arxiv.org/abs/2512.20339}
}

@inproceedings{tian2026audioomni,
  title         = {{Audio-Omni}: Extending Multi-modal Understanding to Versatile Audio Generation and Editing},
  author        = {Tian, Zeyue and Yang, Binxin and Liu, Zhaoyang and Zhang, Jiexuan and Yuan, Ruibin and Yin, Hubery and Chen, Qifeng and Li, Chen and Lyu, Jing and Xue, Wei and Guo, Yike},
  booktitle     = {Proceedings of the ACM SIGGRAPH Conference Papers},
  articleno     = {193},
  numpages      = {10},
  year          = {2026},
  doi           = {10.1145/3799902.3811191},
  url           = {https://doi.org/10.1145/3799902.3811191}
}

@inproceedings{li2026unison,
  title         = {{UNISON}: A Unified Sound Generation and Editing Framework via Deep {LLM} Fusion},
  author        = {Li, Zhaoqing and Xu, Haoning and Su, Jingran and Liu, Yaofang and Rao, Zhefan and Wang, Huimeng and Deng, Jiajun and Wang, Tianzi and Jin, Zengrui and Liu, Rui and Che, Haoxuan and Liu, Xunying},
  year          = {2026},
  booktitle     = {Proceedings of the 2026 Conference on Empirical Methods in Natural Language Processing},
  publisher     = {Association for Computational Linguistics},
  pages         = {45902--45922},
  url           = {https://aclanthology.org/2026.emnlp-main.2015/}
}

@misc{zhu2026unisae,
  title         = {{UniSAE}: Unified Speech Attribute Editing on Speaker, Emotion and Low-Level Content via Discrete Phonetic Posteriorgram Modelling},
  author        = {Zhu, Chuanbo and Zhou, Wuyou and Zhong, Rongxiu and Zhang, Shilei and Qian, Kun and Guo, Yike and Xue, Wei},
  year          = {2026},
  eprint        = {2606.31128},
  archiveprefix = {arXiv},
  primaryclass  = {cs.SD},
  url           = {https://arxiv.org/abs/2606.31128}
}

@inproceedings{zhang2026speecheditbench,
  title         = {{SpeechEditBench}: A Bilingual Multi-Attribute Benchmark for Instruction-Guided Speech Editing},
  author        = {Zhang, Hanlin and Tan, Daxin and Tao, Dehua and Chen, Xiao and Tan, Haochen and Song, Linqi},
  year          = {2026},
  booktitle     = {Proceedings of the 2026 Conference on Empirical Methods in Natural Language Processing},
  publisher     = {Association for Computational Linguistics},
  pages         = {26394--26414},
  url           = {https://aclanthology.org/2026.emnlp-main.1151/}
}

@article{jin2017voco,
  title   = {{VoCo}: Text-based Insertion and Replacement in Audio Narration},
  author  = {Jin, Zeyu and Mysore, Gautham J. and DiVerdi, Stephen and Lu, Jingwan and Finkelstein, Adam},
  journal = {ACM Transactions on Graphics},
  volume  = {36},
  number  = {4},
  pages   = {1--13},
  year    = {2017},
  doi     = {10.1145/3072959.3073702},
  url     = {https://doi.org/10.1145/3072959.3073702}
}

@article{wang2022campnet,
  title         = {{CampNet}: Context-Aware Mask Prediction for End-to-End Text-Based Speech Editing},
  author        = {Wang, Tao and Yi, Jiangyan and Fu, Ruibo and Tao, Jianhua and Wen, Zhengqi},
  journal       = {IEEE/ACM Transactions on Audio, Speech, and Language Processing},
  volume        = {30},
  pages         = {2241--2254},
  year          = {2022},
  doi           = {10.1109/TASLP.2022.3190717},
  eprint        = {2202.09950},
  archiveprefix = {arXiv},
  primaryclass  = {eess.AS},
  url           = {https://arxiv.org/abs/2202.09950}
}

@inproceedings{bai2022a3t,
  title     = {{$A^3T$}: Alignment-Aware Acoustic and Text Pretraining for Speech Synthesis and Editing},
  author    = {Bai, He and Zheng, Renjie and Chen, Junkun and Ma, Mingbo and Li, Xintong and Huang, Liang},
  booktitle = {Proceedings of the 39th International Conference on Machine Learning},
  series    = {Proceedings of Machine Learning Research},
  volume    = {162},
  pages     = {1399--1411},
  year      = {2022},
  url       = {https://proceedings.mlr.press/v162/bai22d.html}
}

@article{liu2024fluenteditor2,
  title         = {{FluentEditor2}: Text-based Speech Editing by Modeling Multi-Scale Acoustic and Prosody Consistency},
  author        = {Liu, Rui and Xi, Jiatian and Jiang, Ziyue and Li, Haizhou},
  journal       = {IEEE Transactions on Audio, Speech and Language Processing},
  volume        = {33},
  pages         = {4500--4514},
  year          = {2025},
  doi           = {10.1109/TASLPRO.2025.3624913},
  url           = {https://doi.org/10.1109/TASLPRO.2025.3624913}
}

@inproceedings{du2023unicats,
  title         = {{UniCATS}: A Unified Context-Aware Text-to-Speech Framework with Contextual {VQ}-Diffusion and Vocoding},
  author        = {Du, Chenpeng and Guo, Yiwei and Shen, Feiyu and Liu, Zhijun and Liang, Zheng and Chen, Xie and Wang, Shuai and Zhang, Hui and Yu, Kai},
  booktitle     = {Proceedings of the AAAI Conference on Artificial Intelligence},
  volume        = {38},
  pages         = {17924--17932},
  year          = {2024},
  doi           = {10.1609/aaai.v38i16.29747},
  url           = {https://ojs.aaai.org/index.php/AAAI/article/view/29747}
}

@article{liang2024e3tts,
  title         = {{E$^3$TTS}: End-to-End Text-Based Speech Editing {TTS} System and Its Applications},
  author        = {Liang, Zheng and Ma, Ziyang and Du, Chenpeng and Yu, Kai and Chen, Xie},
  journal       = {IEEE/ACM Transactions on Audio, Speech, and Language Processing},
  volume        = {32},
  pages         = {4810--4821},
  year          = {2024},
  doi           = {10.1109/TASLP.2024.3485466},
  url           = {https://ieeexplore.ieee.org/document/10731477/}
}

@inproceedings{le2023voicebox,
  title         = {{Voicebox}: Text-Guided Multilingual Universal Speech Generation at Scale},
  author        = {Le, Matthew and Vyas, Apoorv and Shi, Bowen and Karrer, Brian and Sari, Leda and Moritz, Rashel and Williamson, Mary and Manohar, Vimal and Adi, Yossi and Mahadeokar, Jay and Hsu, Wei-Ning},
  booktitle     = {Advances in Neural Information Processing Systems},
  volume        = {36},
  year          = {2023},
  eprint        = {2306.15687},
  archiveprefix = {arXiv},
  primaryclass  = {cs.CL},
  url           = {https://arxiv.org/abs/2306.15687}
}

@article{wang2023speechx,
  title         = {{SpeechX}: Neural Codec Language Model as a Versatile Speech Transformer},
  author        = {Wang, Xiaofei and Thakker, Manthan and Chen, Zhuo and Kanda, Naoyuki and Eskimez, Sefik Emre and Chen, Sanyuan and Tang, Min and Liu, Shujie and Li, Jinyu and Yoshioka, Takuya},
  journal       = {IEEE/ACM Transactions on Audio, Speech, and Language Processing},
  volume        = {32},
  pages         = {3355--3364},
  year          = {2024},
  doi           = {10.1109/TASLP.2024.3419418},
  url           = {https://ieeexplore.ieee.org/document/10577150/}
}

@misc{pei2026speechedit,
  title         = {A Unified Neural Codec Language Model for Selective Editable Text to Speech Generation},
  author        = {Pei, Hanchen and Liu, Shujie and Liu, Yanqing and Yu, Jianwei and Qian, Yuanhang and Huang, Gongping and Zhao, Sheng and Lu, Yan},
  year          = {2026},
  eprint        = {2601.12480},
  archiveprefix = {arXiv},
  primaryclass  = {cs.SD},
  url           = {https://arxiv.org/abs/2601.12480}
}

@article{chen2025valle,
  title = {{Neural Codec Language Models are Zero-Shot Text to Speech Synthesizers}},
  author = {Chen, Sanyuan and Wang, Chengyi and Wu, Yu and Zhang, Ziqiang and Zhou, Long and Liu, Shujie and Chen, Zhuo and Liu, Yanqing and Wang, Huaming and Li, Jinyu and He, Lei and Zhao, Sheng and Wei, Furu},
  year = {2025},
  journal = {IEEE Transactions on Audio, Speech and Language Processing},
  doi = {10.1109/taslpro.2025.3530270},
  volume = {33},
  pages = {705--718},
}

@inproceedings{eskimez2024e2tts,
  title = {{E2 TTS: Embarrassingly Easy Fully Non-Autoregressive Zero-Shot TTS}},
  author = {Eskimez, Sefik Emre and Wang, Xiaofei and Thakker, Manthan and Li, Canrun and Tsai, Chung-Hsien and Xiao, Zhen and Yang, Hemin and Zhu, Zirun and Tang, Min and Tan, Xu and Liu, Yanqing and Zhao, Sheng and Kanda, Naoyuki},
  year = {2024},
  booktitle = {2024 IEEE Spoken Language Technology Workshop (SLT)},
  doi = {10.1109/slt61566.2024.10832320},
  pages = {682--689},
}

@inproceedings{guo2023prompttts,
  title = {{PromptTTS}: Controllable Text-to-Speech with Text Descriptions},
  author = {Guo, Zhifang and Leng, Yichong and Wu, Yihan and Zhao, Sheng and Tan, Xu},
  year = {2023},
  booktitle = {ICASSP 2023 - 2023 IEEE International Conference on Acoustics, Speech and Signal Processing (ICASSP)},
  doi = {10.1109/icassp49357.2023.10096285},
  pages = {1--5},
}

@inproceedings{wang2017tacotron,
  title = {{Tacotron: Towards End-to-End Speech Synthesis}},
  author = {Wang, Yuxuan and Skerry-Ryan, R.J. and Stanton, Daisy and Wu, Yonghui and Weiss, Ron J. and Jaitly, Navdeep and Yang, Zongheng and Xiao, Ying and Chen, Zhifeng and Bengio, Samy and Le, Quoc and Agiomyrgiannakis, Yannis and Clark, Rob and Saurous, Rif A.},
  year = {2017},
  booktitle = {Interspeech 2017},
  doi = {10.21437/interspeech.2017-1452},
  pages = {4006--4010},
}

@inproceedings{du2022vqtts,
  title = {{VQTTS: High-Fidelity Text-to-Speech Synthesis with Self-Supervised VQ Acoustic Feature}},
  author = {Du, Chenpeng and Guo, Yiwei and Chen, Xie and Yu, Kai},
  year = {2022},
  booktitle = {Interspeech 2022},
  doi = {10.21437/interspeech.2022-489},
  pages = {1596--1600},
}

@misc{touvron2023llama,
  title = {{LLaMA: Open and Efficient Foundation Language Models}},
  author = {Touvron, Hugo and Lavril, Thibaut and Izacard, Gautier and Martinet, Xavier and Lachaux, Marie-Anne and Lacroix, Timothée and Rozière, Baptiste and Goyal, Naman and Hambro, Eric and Azhar, Faisal and Rodriguez, Aurelien and Joulin, Armand and Grave, Edouard and Lample, Guillaume},
  year = {2023},
  eprint = {2302.13971},
  archiveprefix = {arXiv},
  url = {https://arxiv.org/abs/2302.13971},
}

@misc{du2024cosyvoice,
  title = {{CosyVoice: A Scalable Multilingual Zero-shot Text-to-speech Synthesizer based on Supervised Semantic Tokens}},
  author = {Du, Zhihao and Chen, Qian and Zhang, Shiliang and Hu, Kai and Lu, Heng and Yang, Yexin and Hu, Hangrui and Zheng, Siqi and Gu, Yue and Ma, Ziyang and Gao, Zhifu and Yan, Zhijie},
  year = {2024},
  eprint = {2407.05407},
  archiveprefix = {arXiv},
  url = {https://arxiv.org/abs/2407.05407},
}

@misc{wang2025sparktts,
  title = {{Spark-TTS: An Efficient LLM-Based Text-to-Speech Model with Single-Stream Decoupled Speech Tokens}},
  author = {Wang, Xinsheng and Jiang, Mingqi and Ma, Ziyang and Zhang, Ziyu and Liu, Songxiang and Li, Linqin and Liang, Zheng and Zheng, Qixi and Wang, Rui and Feng, Xiaoqin and Bian, Weizhen and Ye, Zhen and Cheng, Sitong and Yuan, Ruibin and Zhao, Zhixian and Zhu, Xinfa and Pan, Jiahao and Xue, Liumeng and Zhu, Pengcheng and Chen, Yunlin and Li, Zhifei and Chen, Xie and Xie, Lei and Guo, Yike and Xue, Wei},
  year = {2025},
  eprint = {2503.01710},
  archiveprefix = {arXiv},
  url = {https://arxiv.org/abs/2503.01710},
}

@misc{liu2024ardit,
  title = {{Autoregressive Diffusion Transformer for Text-to-Speech Synthesis}},
  author = {Liu, Zhijun and Wang, Shuai and Inoue, Sho and Bai, Qibing and Li, Haizhou},
  year = {2024},
  eprint = {2406.05551},
  archiveprefix = {arXiv},
  url = {https://arxiv.org/abs/2406.05551},
}

@misc{yu2026das,
  title = {{Distributed Agent System: Fault-Tolerant Collaboration Among Embodied Agents}},
  author = {Yu, Kai and Chen, Lu and Li, Hanqi},
  year = {2026},
  eprint = {2607.10811},
  archiveprefix = {arXiv},
  url = {https://arxiv.org/abs/2607.10811},
}

@misc{li2026xflow,
  title = {{XFlow: An Executable Protocol Programming System for Reliable Multi-Agent Workflows}},
  author = {Li, Hanqi and Peng, Jing and Wang, Zijian and Chen, Lu and Yu, Kai},
  year = {2026},
  eprint = {2606.14790},
  archiveprefix = {arXiv},
  url = {https://arxiv.org/abs/2606.14790},
}

@InProceedings{jia2025ditar,
  title = 	 {{D}i{TAR}: Diffusion Transformer Autoregressive Modeling for Speech Generation},
  author =       {Jia, Dongya and Chen, Zhuo and Chen, Jiawei and Du, Chenpeng and Wu, Jian and Cong, Jian and Zhuang, Xiaobin and Li, Chumin and Wei, Zhen and Wang, Yuping and Wang, Yuxuan},
  booktitle = 	 {Proceedings of the 42nd International Conference on Machine Learning},
  pages = 	 {27255--27270},
  year = 	 {2025},
  editor = 	 {Singh, Aarti and Fazel, Maryam and Hsu, Daniel and Lacoste-Julien, Simon and Berkenkamp, Felix and Maharaj, Tegan and Wagstaff, Kiri and Zhu, Jerry},
  volume = 	 {267},
  series = 	 {Proceedings of Machine Learning Research},
  month = 	 {13--19 Jul},
  publisher =    {PMLR},
  url = 	 {https://proceedings.mlr.press/v267/jia25a.html}
}

@inproceedings{brown2020gpt3,
 author = {Brown, Tom and Mann, Benjamin and Ryder, Nick and Subbiah, Melanie and Kaplan, Jared D and Dhariwal, Prafulla and Neelakantan, Arvind and Shyam, Pranav and Sastry, Girish and Askell, Amanda and Agarwal, Sandhini and Herbert-Voss, Ariel and Krueger, Gretchen and Henighan, Tom and Child, Rewon and Ramesh, Aditya and Ziegler, Daniel and Wu, Jeffrey and Winter, Clemens and Hesse, Chris and Chen, Mark and Sigler, Eric and Litwin, Mateusz and Gray, Scott and Chess, Benjamin and Clark, Jack and Berner, Christopher and McCandlish, Sam and Radford, Alec and Sutskever, Ilya and Amodei, Dario},
 booktitle = {Advances in Neural Information Processing Systems},
 editor = {H. Larochelle and M. Ranzato and R. Hadsell and M.F. Balcan and H. Lin},
 pages = {1877--1901},
 publisher = {Curran Associates, Inc.},
 title = {Language Models are Few-Shot Learners},
 url = {https://proceedings.neurips.cc/paper_files/paper/2020/file/1457c0d6bfcb4967418bfb8ac142f64a-Paper.pdf},
 volume = {33},
 year = {2020}
}

@misc{openai2025audio,
  title = {{Introducing Next-Generation Audio Models in the API}},
  author = {{OpenAI}},
  year = {2025},
  url = {https://openai.com/index/introducing-our-next-generation-audio-models/},
}

@techreport{itu2019bs1284,
  author = {{ITU-R}},
  title = {General methods for the subjective assessment of sound quality},
  institution = {International Telecommunication Union},
  type = {Recommendation},
  number = {BS.1284-2},
  year = {2019},
  url = {https://www.itu.int/rec/R-REC-BS.1284-2-201901-I/en}
}

@techreport{itu2021p808,
  author = {{ITU-T}},
  title = {Subjective evaluation of speech quality with a crowdsourcing approach},
  institution = {International Telecommunication Union},
  type = {Recommendation},
  number = {P.808},
  year = {2021},
  url = {https://www.itu.int/rec/T-REC-P.808-202106-I/en}
}

@article{shen2026fireredtts3,
  title={{FireRedTTS3}: Unified Speech Generation and Editing with Semantically Enriched Speech Representations},
  author={Shen, Feiyu and Xie, Kun and Wu, Yichen and Dai, Ziqi and Han, Yichen and Li, Junjie and Geng, Xuelong and Xie, Fenglong and Xie, Lei and Tang, Xu and Hu, Yao},
  journal={arXiv preprint arXiv:2608.17492},
  year={2026}
}

@article{ma2026auk,
  title={{AuK} Technical Report: An Open-Source Foundational Model for Speech Generation and Editing},
  author={Ma, Ziyang and Niu, Zhikang and Tu, Wenming and Wang, Tianrui and Yan, Ruiqi and Liu, Junxi and Huo, Yanru and Huang, Nickk and Liu, Yang and Xie, Qicong and Xie, Zeyu and Wang, Hui and Li, Haitao and Jiang, Zixuan and Li, Yalin and Fang, Jie and Duan, Yifan and Tian, Zeyue and Li, Guangzheng and Zhu, Haina and Wang, Shuyi and Wang, Jinwen and Cui, Mingyu and Tan, Tian and Auden and Liang, Sen and Yves, Steve and Yang, Shan and Bo, Liefeng and Zheng, Zilong and Yu, Kai and Chng, Eng-Siong and Chen, Xie},
  journal={arXiv preprint arXiv:2609.08936},
  year={2026}
}
\endgroup

\end{document}